\documentclass[twocolumn]{aastex63}

\usepackage{graphicx}	% Including figure files
\usepackage{amsmath}	% Advanced maths commands
\usepackage{amssymb}	% Extra maths symbols
\usepackage{natbib}
\usepackage{scalerel}
\graphicspath{{./}{images/}}

\def\msol{$M_{\odot}$}

\def\lum{erg~s$^{-1}$}

\def\Lx{$L_{\mbox{\scriptsize{X}}}$}

\def\chandra{{\itshape Chandra\/}}

\def\hst{{\itshape HST\/}}

\def\nustar{{\it NuSTAR\/}}

\def\xmm{{\it XMM-Newton\/}}

\def\ltsima{$\; \buildrel < \over \sim \;$}
\def\simlt{\lower.5ex\hbox{\ltsima}}
\def\gtsima{$\; \buildrel > \over \sim \;$}
\def\simgt{\lower.5ex\hbox{\gtsima}}
\def\kms{\ifmmode{~{\rm km~s^{-1}}}\else{~km s$^{-1}$}\fi}
\def\lsim{\lower0.3em\hbox{$\,\buildrel <\over\sim\,$}}
\def\gsim{\lower0.3em\hbox{$\,\buildrel >\over\sim\,$}}          
\renewcommand*{\thefootnote}{\fnsymbol{footnote}}

\DeclareMathAlphabet{\mathsc}{OT1}{cmr}{m}{sc}
\def\testbx{bx}%
\DeclareRobustCommand{\ion}[2]{%
\relax\ifmmode
\ifx\testbx\f@series
{\mathbf{#1\,\mathsc{#2}}}\else
{\mathrm{#1\,\mathsc{#2}}}\fi
\else\textup{#1\,{\mdseries\textsc{#2}}}%
\fi}

\received{July 13, 2020}
\revised{September 8th, 2020}
\accepted{September 17th, 2020}
\submitjournal{ApJ}

\shorttitle{X-ray SED of VV~114}
\shortauthors{Garofali et al.}

\begin{document}
\title{On the X-ray Spectral Energy Distributions of Star-Forming Galaxies: the 0.3--30 keV Spectrum of the Low-Metallicity Starburst Galaxy VV~114}

\correspondingauthor{Kristen Garofali}
\email{kristen.garofali@nasa.gov}

\author[0000-0002-9202-8689]{Kristen Garofali}
\affil{Department of Physics, University of Arkansas, 825 West Dickson St, Fayetteville, AR 72701, USA}
 
 \author{Bret D. Lehmer}
\affil{Department of Physics, University of Arkansas, 825 West Dickson St, Fayetteville, AR 72701, USA}
 
 \author{Antara Basu-Zych}
\affil{NASA Goddard Space Flight Center, Code 662, Greenbelt, MD 20771, USA}
\affil{Center for Space Science and Technology, University of Maryland Baltimore County, 1000 Hilltop Circle, Baltimore, MD 21250, USA}

 \author{Lacey A. West}
\affil{Department of Physics, University of Arkansas, 825 West Dickson St, Fayetteville, AR 72701, USA}

 \author{Daniel Wik}
\affil{Department of Physics and Astronomy, University of Utah, 201 James Fletcher Building, Salt Lake City, UT 84112, USA}

 \author{Mihoko Yukita}
\affil{The Johns Hopkins University, Homewood Campus, Baltimore, MD 21218, USA}

 \author{Neven Vulic}
\affil{NASA Goddard Space Flight Center, Code 662, Greenbelt, MD 20771, USA}
\affil{Department of Astronomy and Center for Space Science and Technology (CRESST), University of Maryland, College Park, MD 20742-2421, USA}

\author{Andrew Ptak}
\affil{NASA Goddard Space Flight Center, Code 662, Greenbelt, MD 20771, USA}

\author{Ann Hornschemeier}
\affil{NASA Goddard Space Flight Center, Code 662, Greenbelt, MD 20771, USA}

\begin{abstract}
Binary population synthesis combined with cosmological models suggest that X-ray emission from star-forming galaxies, consisting primarily of emission from X-ray binaries (XRBs) and the hot interstellar medium (ISM), could be an important, and perhaps dominant, source of heating of the intergalactic medium prior to the epoch of reionization. However, such models rely on empirical constraints for the X-ray spectral energy distributions (SEDs) of star-forming galaxies, which are currently lacking for low-metallicity galaxies. Using a combination of  \chandra, \xmm, and \nustar\ observations, we present new constraints on the 0.3--30 keV SED of the low-metallicity starburst galaxy VV~114, which is known to host several ultra-luminous X-ray sources (ULXs) with luminosities above 10$^{40}$ erg s$^{-1}$. We use an archival \chandra\ observation of VV~114 to constrain the contributions to the X-ray SED from the major X-ray emitting components of the galaxy, and newly acquired, nearly simultaneous \xmm\ and \nustar\ observations to extend the spectral model derived from \chandra\ to cover the 0.3--30 keV range. Using our best-fit galaxy-wide spectral model, we derive the 0.3--30 keV SED of VV~114, which we find is dominated by emission from the XRB population, and in particular ULXs, at energies $>$ 1.5 keV, and which we find to have an elevated galaxy-integrated X-ray luminosity per unit star formation rate relative to higher-metallicity star-forming galaxies. We discuss our results in terms of the effect of metallicity on XRB populations and the hot ISM, and the importance of X-ray emission from star-forming galaxies in the high redshift Universe. 
\end{abstract}


%%%%%%%%%%%%%%%%%%%%%%%%%%%%%%%%%%%%%%%%%%%%%%%%%%

%%%%%%%%%%%%%%%%% BODY OF PAPER %%%%%%%%%%%%%%%%%%

\section{Introduction}

X-ray binaries (XRBs), systems in which a compact object (a black hole [BH] or neutron star [NS]) accretes material from a less evolved stellar companion, are important probes of stellar and binary evolution, compact object populations, and physical accretion mechanisms. Studies of XRB populations in nearby galaxies have revealed important scalings between XRB populations and host galaxy properties, including star formation rate (SFR), stellar mass, and metallicity \citep[e.g.,][]{Prestwich2013,Lehmer2019}. In particular, the galaxy-integrated emission from high-mass XRBs (HMXBs), systems in which the donor star is massive ($>$ 8 \msol), has been shown to scale with SFR \citep[e.g.,][]{Grimm2003,Fabbiano2006,Mineo2012,Lehmer2010,Lehmer2019,Kou2020}, while the emission from low-mass XRBs (LMXBs), systems with low-mass donor stars, has been observed to scale with stellar mass \citep[e.g.,][]{Colbert2004,Gilfanov2004,Boroson2011,Lehmer2010,Lehmer2019,Lehmer2020}. These scalings can be explained by stellar evolution timescales: the high-mass donor stars in HMXBs die off rapidly ($\lesssim$ 40 Myr) following a star forming episode, while the low-mass donors in LMXBs will live for billions of years following an episode of star formation. 

These locally-derived scaling relations for galaxy-integrated \Lx\ with SFR and mass have also been shown empirically to evolve with redshift \citep{BZ13LBG,Lehmer2016,Aird2017}, and very recently \citet{Fornasini2019,Fornasini2020} demonstrated that the increase of galaxy-integrated \Lx\ per unit SFR with increasing redshift is likely tied to the metallicity evolution of the Universe. This metallicity dependence of \Lx\ per unit SFR (\Lx/SFR) is supported by studies of HMXBs and ultra-luminous X-ray sources (ULXs), XRBs with \Lx\ $>$ 10$^{39}$ \lum, whose galaxy-integrated \Lx/SFR has been shown to increase with decreasing metallicity \citep[e.g.,][]{Prestwich2013,Douna2015,Brorby2014,Brorby2016,Basu-Zych2013,Basu-Zych2016}. This metallicity scaling for HMXBs and ULXs is corroborated by theoretical binary population synthesis models \citep[e.g.,][]{Linden2010,Fragos2013,Wiktor2017,Wiktor2019}, which find order-of-magnitude differences in galaxy-integrated \Lx/SFR between environments at solar and 0.1 $Z_{\odot}$ metallicities. The inverse correlation between XRB \Lx/SFR and metallicity is due to the effects of metallicity on stellar and binary evolution, namely the production of more massive BHs and/or more compact binaries, and therefore brighter systems, at lower metallicities \citep[e.g.,][]{Mapelli2010,Linden2010}. 

A key implication of these empirically derived and theoretically corroborated scalings of \Lx\ with host galaxy properties is the importance of XRBs to normal galaxy emissivity across cosmic time. Theoretical binary population synthesis models, when coupled with prescriptions for the cosmic star formation history (SFH) and metallicity evolution of the Universe predict that HMXBs will begin to dominate normal galaxy emissivity over LMXBs at $z$ $\gtrsim$ 1--2, and further that the normal galaxy emissivity due to XRBs may begin to dominate over active galactic nuclei (AGN) at $z$ $\gtrsim$ 5--6 \citep[e.g.,][]{Fragos2013,Madau2017}. 
 
The rising dominance of XRBs in the early Universe coupled with cosmological models suggests that emission from XRBs may provide a non-negligible heating source to the intergalactic medium (IGM) during the ``epoch of heating" at $z \approx$ 10--20, prior to reionization \citep[e.g.,][]{Mesinger2013,Mesinger2014,Pacucci2014,Mirocha2014,Fialkov2014,Fialkov2017}. This further suggests that XRBs could have a significant imprint on the 21-cm signal observed from these redshifts \citep[e.g.,][]{Das2017}. In the near future, 21-cm interferometers like the Hydrogen Epoch of Reionisation Array (HERA) and Square Kilometre Array (SKA) are expected to observe the signals from this epoch of heating, thus providing constraints on the ionizing properties of X-ray sources during this epoch \citep[e.g.,][]{Greig2017,Park2019}. However, interpreting the 21-cm results in the context of XRB populations, and further refining predictions from binary population synthesis models for the importance of XRBs at different epochs requires empirically constraining the metallicity dependence of the X-ray spectral energy distribution (SED). In particular,  it is critical to constrain the X-ray SED for {\it low-metallicity}, star-forming galaxies, which serve as better analogs to the first galaxies, and in the rest frame soft band (0.5--2 keV), which is the energy band of interest for the photons that most strongly interact with the IGM at high redshift  \citep[e.g.,][]{McQuinn2012}. In this work, we present the 0.3--30 keV SED of the low-metallicity, starburst galaxy VV~114, providing an important empirical benchmark for the metallicity dependence of the X-ray SED in both the soft (0.5--2 keV) and hard (2--30 keV) bands. 

\begin{deluxetable*}{ccccc}
\centering
\tablewidth{\textwidth}
 \tablecaption{Archival and New Observations Used in This Work \label{tab:obs}}
\tablehead{
 \colhead{Obs. Start Date} & \colhead{Obs. ID} & \colhead{Inst.} & \colhead{Eff. Exposure (ks)} & \colhead{PI} \\
  \colhead{(1)} & \colhead{(2)} & \colhead{(3)} & \colhead{(4)} & \colhead{(5)} }

\startdata 
\hline 
& & {\chandra} & & \\
\hline 
2005-10-20 & 7063 & ACIS-S & 59  & T. Heckman \\ 
\hline 
& & { \xmm} & & \\
\hline 
2019-01-10 & 0830440101 & EPIC-pn & 26 & B. Lehmer \\ 
2019-01-10 & 0830440101 & EPIC-MOS1 & 30  & B. Lehmer \\
2019-01-10 & 0830440101 & EPIC-MOS2 & 26 & B. Lehmer \\ 
\hline 
& & { \nustar} & & \\
\hline 
2019-01-19 & 50401001002 & FPMA & 205 & B. Lehmer \\
2019-01-19 & 50401001002 & FPMB & 204 & B. Lehmer \\
\enddata 
\tablecomments{Col. (1): observation start date. Col. (2): observation ID. Col. (3): instrument. Col. (4): good time interval effective exposure times in ks after removing flared intervals. Col. (4): observation PI.}
\end{deluxetable*}

VV~114 is a prime target for calibrating the metallicity dependence of the X-ray SED as it is a relatively nearby ($D$ = 88 Mpc\footnote[2]{We calculate the distance for VV~114 taking $z = 0.02$ from NED and assuming $H_{0}$ = 70 km s$^{-1}$ Mpc$^{-1}$, $\Omega_{M}$ = 0.3, and $\Omega_{\Lambda}$ = 0.7}) Lyman break analog (LBA). LBAs are highly star-forming, yet relatively dust- and metal-poor galaxies at $z < 0.3$ that resemble higher redshift ($z > 2$) Lyman break galaxies \citep[e.g.,][]{Heckman2005,Hoopes2007,BZ2009}. With VV~114's gas phase metallicity of 12+log(O/H) = 8.4\footnote[3]{We calculate the gas-phase metallicity for VV~114 from the [\ion{O}{iii}]~$\lambda$5007 and [\ion{N}{ii}]~$\lambda$6584 emission line ratios taken from \citet{Moustakas2006}, and using the method outlined in \citet[][``PP04 {\it O3N2}"]{PP04}. In what follows, we adopt a global metallicity of 0.51 $Z_{\odot}$ for VV~114, assuming $Z_{\odot}$ corresponds to 12+log(O/H) = 8.69 \citep{Asplund2009}.}, global UV + IR SFR of $\sim$ 38 $M_{\odot}$ yr$^{-1}$ as measured from {\it GALEX} and {\it WISE}, and stellar mass of log $M_{\star}$ = 10.65 $M_{\odot}$ \citep{Basu-Zych2013}, scaling relations dictate that it should host a substantial XRB population. With a specific star formation rate (SFR/$M_{\star}$) $>$ 10$^{-10}$ yr$^{-1}$, VV~114 is further expected to be dominated by HMXBs or ULXs, as opposed to LMXBs \citep[e.g.,][]{Lehmer2010}. Indeed, previous X-ray studies of VV~114 have revealed a galaxy with a well-populated X-ray luminosity function (XLF), comprised of six ULXs, which put VV~114 above the ``normal" \Lx/SFR derived from nearby star-forming galaxies \citep{Basu-Zych2013,Basu-Zych2016}. Thus, VV~114 offers a unique environment for studying the X-ray SED in that it is highly star forming, relatively low metallicity, and is known to host six ULXs \citep{Basu-Zych2016}.

In this paper, we use new nearly-simultaneous observations of VV~114 from \xmm\ and \nustar\ coupled with archival \chandra\ data to characterize its 0.3--30 keV SED in terms of both the galaxy-wide X-ray emission and the resolved X-ray source population. This paper is organized as follows: Section~\ref{sec:reduce} discusses the data sets in use, as well as the reduction procedures and spectral fitting techniques employed in the analysis of these datasets. Section~\ref{sec:results} provides the results of the custom spectral modeling of both the point source population and galaxy-wide X-ray emission of VV~114 using all three data sets. Section~\ref{sec:discuss} presents a discussion and interpretation of these results for the low-metallicity SED in the context of previous works, the theoretical scalings of XRB emission with SFR and metallicity, and future 21-cm measurements. Finally, in Section~\ref{sec:conclude} we summarize this work, and discuss future directions. 

Throughout this paper we assume a \citet{Kroupa} initial mass function (IMF) and, when comparing to any previous works, correct all SFRs following this assumption. Furthermore, we standardize all quoted gas-phase metallicities to values determined using the {\it O3N2} \citet{PP04} calibration. 

\section{Observations \& Data Reduction}\label{sec:reduce}

In this section we describe the observations, both archival (\chandra) and new (\xmm\ and \nustar), used as part of this analysis, as well as an overview of the data reduction procedures.

\subsection{{\itshape Chandra} Imaging \& Spectra}\label{sec:chobs}

We used archival \chandra\ data to assess the point source population in VV~114, as only \chandra\ has the required spatial resolution to resolve the galaxy-wide emission of VV~114 into individual point sources. The archival \chandra\ observation is listed in Table~\ref{tab:obs}, where the observation was performed with ACIS-S in ``Very Faint" (VFAINT) mode, and the listed effective exposure time includes only good time intervals (GTIs). 

We reduced the archival \chandra\ observation using the standard reduction tools included with {\tt CIAO} version 4.10  and {\tt CALDB} version 4.8.1\footnote[7]{\url{http://cxc.harvard.edu/ciao/download/}}. The {\tt level=1} event files were reprocessed to {\tt level=2} event files using the latest calibration and script {\tt chandra$\_$repro}. We subsequently 
filtered the {\tt level=2} event files on GTIs determined from background light curves filtered with the task {\tt lc$\_$clean} with default filtering parameters. 

\begin{figure}
\vspace{-5pt}
\includegraphics[width=\linewidth]{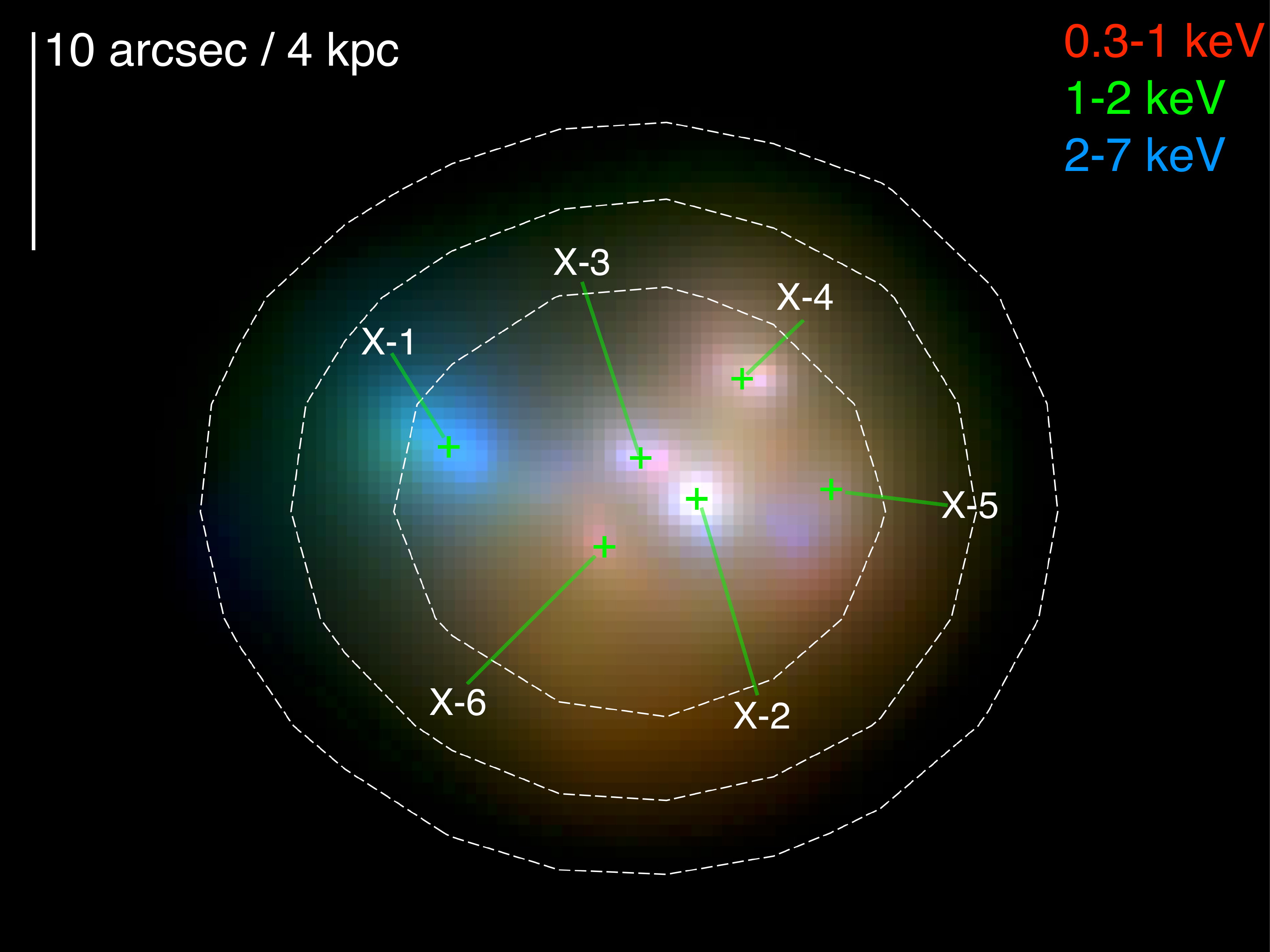}
\caption{Adaptively smoothed three-color (red: 0.3--1 keV, green: 1--2 keV, and blue: 2--7 keV) \chandra\ image of VV~114 showing the six \chandra-detected point sources as green crosses, annotated in order of decreasing brightness, as well as the hot, diffuse gas which permeates the galaxy. The white dashed curves overlaid represent the 4--25 keV \nustar\ intensity contours (1.2 $\times$ 10$^{-5}$ counts s$^{-1}$, 1.0 $\times$ 10$^{-5}$ counts s$^{-1}$, 8.6 $\times$ 10$^{-6}$ counts s$^{-1}$), which are comparable to a single source PSF for VV~114.}
\label{fig:chandra_rgb}
\end{figure}

\begin{figure}
\vspace{-5pt}
\includegraphics[width=\linewidth]{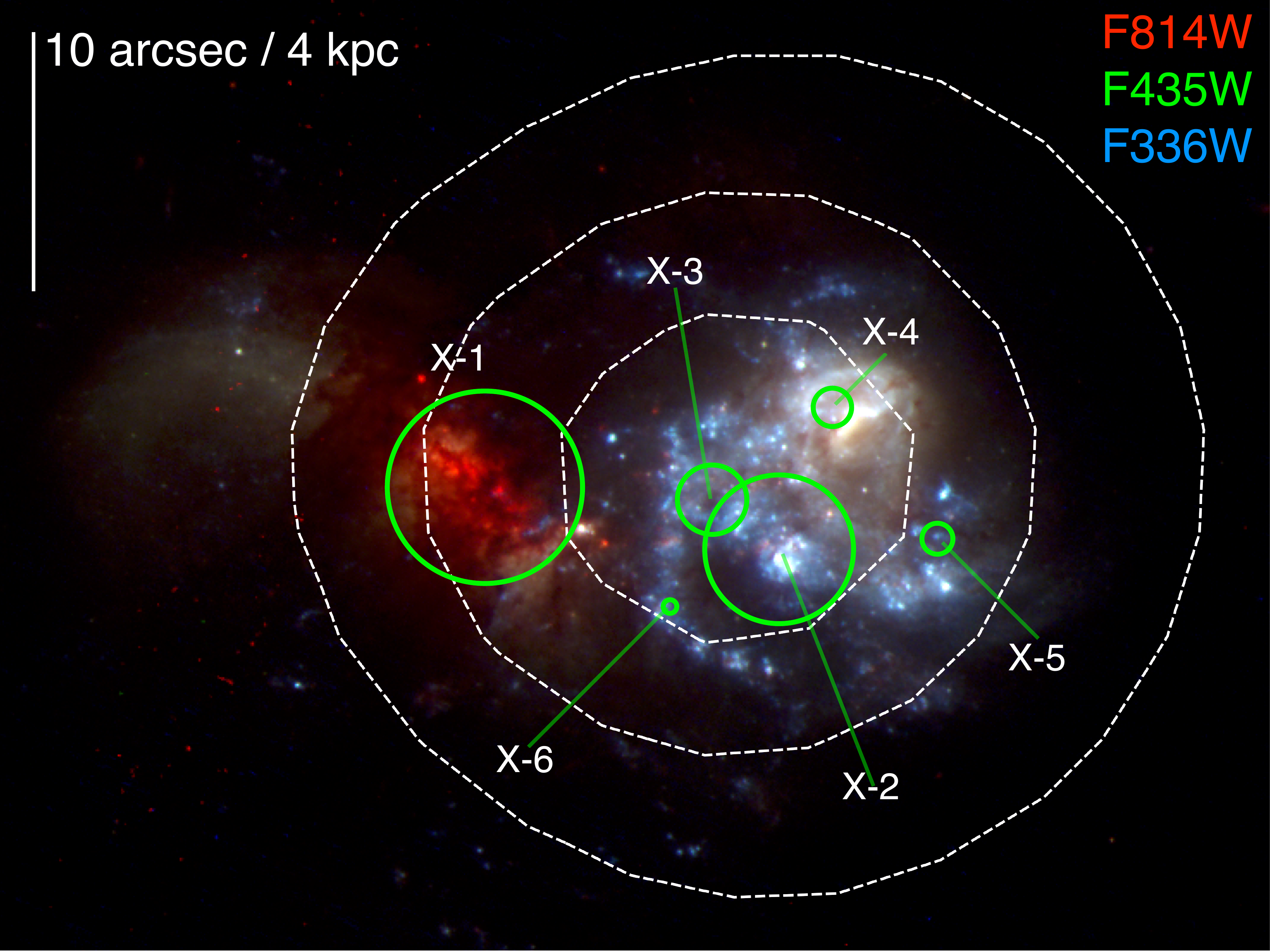}
\caption{Three-color (red: F814W, green: F435W, and blue: F336W) \hst\ image of VV~114 with the positions of the six \chandra-detected point sources overlaid as green circles, where the region size is scaled by the point source luminosity. The white dashed curves are the 0.3--12 keV \xmm\ intensity contours (2.3 $\times$ 10$^{-3}$ counts s$^{-1}$, 1.2 $\times$ 10$^{-3}$ counts s$^{-1}$, 4.0 $\times$ 10$^{-4}$ counts s$^{-1}$), comparable to a single point source as for the \nustar\ observations.}\label{fig:hst_rgb}
\end{figure}

We then created an exposure map and point spread function (PSF) map using regions that encompassed 90\% of the encircled energy of the PSF using the {\tt CIAO} tools {\tt fluximage} and {\tt mkpsfmap}. We used the images from these procedures as input to {\tt wavdetect} to determine positions and PSF-corrected extraction regions for each of the six point sources in the galaxy. We then extracted spectral files for each of the six point sources using the task {\tt specextract} with the {\tt wavdetect}-determined source positions and extraction regions and the flag {\tt psfcorr = yes}. The {\tt specextract} task produces not only source spectra, but also response matrix files (RMFs), ancillary response files (ARFs), and background spectra when provided with a background extraction region. We use this task for extracting all subsequent \chandra\ spectral products, as described next. 

For extracting background spectra we chose regions encircling VV~114 free of point sources and uncontaminated by diffuse emission from the galaxy. We estimated the diffuse extent ($\sim$ 13 kpc) of the galaxy visually using a combination of soft-band (0.3--1 keV and 1--2 keV) \chandra\ and optical \hst\ images. We also extracted spectral products for the point-source-free diffuse emission using a region encompassing the diffuse extent as described above, but excluding the 90\% encircled energy fraction regions for all six detected point sources. We likewise extracted spectral products for the galaxy-wide emission, including all six detected point sources, the diffuse emission, and any unresolved component. These \chandra\ spectral products are used in constraining the components of the galaxy-wide emission and compared with the \xmm\ and \nustar\ data, for which the galaxy appears as a single source, as described in Section~\ref{sec:point_src_decomp}.

Finally, we created exposure-corrected images in the soft (0.3--1 keV), medium (1--2 keV), and hard (2--7 keV) bands using the task {\tt fluximage}. We imposed a $\sim$ 1$^{\prime \prime}$ $\times$ 1$^{\prime \prime}$ pixel binning (2 $\times$ 2 native pixels) on the resultant images, which we subsequently used to create a three-color, adaptively smoothed image using {\tt csmooth}. The three-color, adaptively smoothed image is shown in Figure~\ref{fig:chandra_rgb}, and highlights the locations of each of the six \chandra-detected point sources (VV~114 X-1 to X-6), along with the hot, diffuse gas, which suffuses the galaxy.

\subsection{{\itshape XMM-Newton} Imaging \& Spectra}\label{sec:xmmobs}

In addition to using the archival \chandra\ observation to measure the resolved components of VV~114, we obtained new \xmm\ observations of the galaxy to provide additional constraints on the galaxy-wide emission. Observational data files (ODFs) for these new observations were processed using the \xmm\ Science Analysis System ({\tt SAS} version 17.0)\footnote[8]{\url{https://www.cosmos.esa.int/web/xmm-newton/sas}}. We created event lists from the ODFs for the EPIC-MOS \citep{Turner2001} and EPIC-pn detectors \citep{Struder2001} using the {\tt SAS} tasks {\tt emchain} and {\tt epchain}, respectively. We applied standard filters for the EPIC-MOS detectors to include single, double, and quadruple events (PATTERN $\leq$ 12 \&\& flag== \#XMMEA$\_$EM), and similarly applied standard criteria to include single and double events with conservative flagging for the EPIC-pn detector (PATTERN $\leq$ 4 \&\& FLAG == 0). 

With the filtered event lists for each detector, we next constructed X-ray light curves from the entire field, from which we determined the rate thresholds for filtering the event lists for background flaring events. For the MOS detectors we created $>$ 10 keV light curves, creating GTIs by filtering out periods with count rates $>$ 0.2 counts s$^{-1}$, and for the pn camera we created a 10-12 keV light curve, filtering out periods with count rates $>$ 0.5 counts s$^{-1}$. The effective exposures for each observation after filtering on these GTIs are listed in Table~\ref{tab:obs}. 

Following GTI correction, we performed source detection on images in five bands for each detector using the task {\tt edetect$\_$chain}. We cross-correlated 29 of the detected point sources with counterparts from the \chandra\ observation to determine the translation shift between the observations, finding shifts of +0\farcs25 in R.A. and +1\farcs66 uncertainty in decl. necessary to bring the images into astrometric alignment. 

As VV~114 is consistent with being a single source in \xmm, we determined the appropriate galaxy-wide spectral extraction region for VV~114 in the \xmm\ observations by simulating the combination of the \chandra-detected point source PSFs on each \xmm\ detector, using the point source physical positions determined from source detection after astrometric correction was applied. To determine the overall expected PSF of VV~114 in each \xmm\ exposure we used the {\tt SAS} task {\tt psfgen} to simulate the PSF for each point source at its physical position on the \xmm\ detectors, and then combined the simulated PSFs for each of the six point sources, accounting for the physical offsets between each source. The 80\% encircled energy fraction spectral extraction regions in each EPIC exposure determined from this procedure were found to be in good agreement with the optical extent of VV~114 from \hst\ imaging. In Figure~\ref{fig:hst_rgb}, we show a three-color \hst\ image of VV~114 with 0.3--12 keV intensity contours from \xmm\ overlaid in white, where the contours approximate the extent of the galaxy-wide spectral extraction region constructed from the PSF simulation procedure. 

We extracted source spectra for VV~114 for each detector with the task {\tt evselect} using the source regions described above and a spectral bin size of five for the pn and 15 for the MOS exposures. For the pn detector we extracted a background spectrum using {\tt evselect} from a source-free region at a similar RAWY position as VV~114 on the detector, while for both MOS detectors we chose background regions from source-free areas on the same CCD as VV~114. We produced the associated RMFs and ARFs for each spectrum using the tasks {\tt rmfgen} and {\tt arfgen}. These \xmm\ spectral products are used in the analysis of the galaxy-wide spectrum of VV~114 as described in Section~\ref{sec:gal_wide_spec}.

\subsection{{\itshape NuSTAR} Imaging \& Spectra}\label{sec:nuobs}

The \nustar\ data were reduced using {\tt HEASoft} v6.24 and the \nustar\ Data Analysis Software  {\tt NuSTARDAS} v1.8.0 with {\tt CALDB} version 20190627. 
We produced level 2 data products by processing the level 1 data files through {\tt nupipeline}, which filters out bad pixels, screens for cosmic rays and high background intervals, and projects the detected events to proper sky coordinates.

The \nustar\ PSF has an 18$^{\prime \prime}$ FWHM core and a 58$^{\prime \prime}$ half power diameter \citep{Harrison2013} resulting in all six point sources in VV~114 appearing blended as one source in the \nustar\ observations (see white, dashed contours, Figure~\ref{fig:chandra_rgb}). Given the extent of the \nustar\ PSF, we chose a 30$^{\prime \prime}$ region for extracting the galaxy-wide source spectra to encompass most of the emission from VV~114 while minimizing background contamination. We defined a region for extraction of background spectra from a source-free area on the same detector as VV~114, but separated by at least 20$^{\prime \prime}$ from the galaxy. We produced source and background spectra using these regions, as well as RMFs and ARFs for both the FPMA and FPMB using the task {\tt nuproducts}, ensuring no background subtraction was performed on the source spectra during extraction. These spectral products were used in our analysis of the 3--30 keV spectrum of VV~114, as described in Section~\ref{sec:gal_wide_spec}. 

\subsection{X-ray Spectral Fitting Technique}\label{sec:fit_techniques}

All spectral fitting was performed with {\tt XSPEC} v12.10.0c \citep{Arnaud1996} using the Cash statistic \citep{Cash1979} as the fit statistic. Because the Cash statistic does not yield a straightforward way to evaluate goodness-of-fit (gof), we evaluated the gof of spectral models using the Anderson-Darling (ad) test statistic and the {\tt XSPEC} Monte Carlo command {\tt goodness}. For each model, we ran the {\tt goodness} command for 1000 realizations with the {\tt ``nosim''} and {\tt ``fit''} options. This procedure simulates spectra based on the current best-fit model, fits these simulated spectra to the model, and then calculates the new test statistic. The {\tt goodness} command returns the distribution of test statistics for the simulated data, which can then be compared to the test statistic for the actual data. Our reported ``gof'' for each model fit is the fraction of simulations returned from {\tt goodness} with a test statistic as large, or larger (i.e., statistically worse fits) than the test statistic for the actual data \citep[e.g.,][]{Maccarone2016}. Therefore, gof = 0.5 is expected for data consistent with the model, and gof $\sim$ 1 can be interpreted as overfitting the data, since it implies nearly all simulations produced worse fits than the data itself. If all simulations returned smaller test statistics (better fits) than the actual data, the model is rejected (gof $<$ 10$^{-3}$). It is important to note that the gof calculated in this way provides a measure of the confidence level with which a model can be rejected, not a probability for whether the model is correct. Errors on all free model parameters are reported as 90\% confidence intervals, and are computed using the {\tt XSPEC error} command using the output of the {\tt XSPEC mcmc} routine. In all models we set abundances relative to solar using the \citet{Asplund2009} abundance tables. 

All spectral fits were performed on the unbinned source spectra, without any background subtraction. To perform such fits, we must define a model for the background for each instrument. For each observation, we modeled the background with both a sky component, representing the contribution from the diffuse background and the unresolved cosmic X-ray background \citep[e.g.,][]{Kuntz2000,Lumb2002}, as well as an instrumental component, representing the contribution from the instrumental continuum and detector lines. We describe the details of the sky and instrumental background components for each instrument below. 

We modeled the sky background for both \chandra\ and \xmm\ as an absorbed two-temperature thermal plasma ({\tt APEC}) plus power-law. These model components represent the diffuse Galactic and extragalactic cosmic X-ray background, respectively, where we fix the photon index for the cosmic X-ray background to $\Gamma$ = 1.42 \citep{Lumb2002}. For \nustar, we modeled the sky background as an absorbed single temperature thermal plasma plus power-law, accounting for the ``Solar" and cosmic X-ray background components, respectively \citep{Wik2014back}. For each sky background model we fixed the foreground Galactic absorption component ({\tt Tbabs}) to $N_{H}$ = 1.20 $\times$ 10$^{20}$ cm$^{-2}$ \citep{nHpimms}. From our best-fit background models to the background spectra we found sky background {\tt APEC} temperatures of $kT_{1} = 0.1$ keV and $kT_{2} = 0.24$ keV for \chandra, and $kT_{1} = 0.1$ keV and $kT_{2} = 0.27$ keV for \xmm. We set $kT = 0.27$ keV for \nustar\ following \xmm.

The instrumental background for \chandra\ was modeled as a power-law continuum superposed with Gaussians, representing detector lines, with the energies and widths of the Gaussian lines fixed following \citet{Bartalucci2014}, but line normalizations allowed to vary relative to the continuum. For all \xmm\ detectors, the instrumental background was composed of a broken power-law for the continuum with detector fluorescence lines as described in \citet{Garofali2017}. The \nustar\ instrumental background was modeled as a broken power-law overlaid with 29 Lorentzians following \citet{Wik2014back}, where the line normalizations are allowed to vary relative to the continuum, and FPMA and FPMB were handled independently.

We fit the above described background models to the background spectra for each observation to determine the shape of the background at the location of VV~114 for each observation and detector. In subsequent fits to the source region, which includes source plus background data, we include this background as a model component, albeit with nearly all free parameters fixed to their best-fit values (e.g., plasma temperatures listed above), and the normalization fixed to the best-fit normalization for the background spectrum scaled by the ratio of the source to background extraction region areas. In this way, we fix the {\it shape} of the background in the fits to the source region  based on the best-fit models for the background spectra, and constrain the {\itshape contribution} of the background to the source spectra via the known source and background spectral extraction region areas. Thus, we fit the spectra for the source region without performing any background subtraction while still minimizing the number of free parameters.

\begin{deluxetable*}{cccccccccccccc}
\tabletypesize{\tiny}
\tablewidth{\textwidth} 
 \tablecaption{Spectral Fit Results for \chandra-detected Point Sources \label{tab:point_src_fits}}
\tablehead{ \colhead{} & \colhead{} & \colhead{} & \colhead{$kT_{\scaleto{1}{3pt}}$}  & \colhead{$A_{\scaleto{kT_{1}}{3pt}}$} & \colhead{$N_{\scaleto{\rm H,2}{3pt}}$} & \colhead{$kT_{\scaleto{2}{3pt}}$} & \colhead{$A_{\scaleto{kT_{2}}{3pt}}$} & \colhead{$N_{\scaleto{\rm H, 3}{3pt}}$} & \colhead{} &\colhead{$A_{\scaleto{\Gamma}{3pt}}$} & \colhead{log($L_{\scaleto{{\rm 0.5-8~keV}}{3pt}}$)} & \colhead{log($L_{\scaleto{{\rm 2-10~keV}}{3pt}}$)} & \colhead{} \\ 
\colhead{Source} & \colhead{Counts} & \colhead{C$_{\scaleto{kT}{3pt}}$} & \colhead{(keV)} & \colhead{(10$^{-5}$)} & \colhead{(10$^{22}$ cm$^{-2}$)} & \colhead{(keV)} & \colhead{(10$^{-4}$)} & \colhead{(10$^{22}$ cm$^{-2}$)} & \colhead{$\Gamma$} & \colhead{(10$^{-5}$)} & \colhead{(erg s$^{-1}$)} & \colhead{(erg s$^{-1}$)} & \colhead{gof} \\
\colhead{(1)} & \colhead{(2)} & \colhead{(3)} & \colhead{(4)} & \colhead{(5)} & \colhead{(6)} & \colhead{(7)} & \colhead{(8)} & \colhead{(9)} & \colhead{(10)} & \colhead{(11)} & \colhead{(12)} & \colhead{(13)} & \colhead{(14)}
 }
\startdata 
diffuse$^{{\tt a}}$ & 2477 &  $\cdots$ & 0.36$^{+0.26}_{-0.05}$ & 4.67$^{+1.47}_{-2.33}$ &  0.44$^{+0.23}_{-0.12}$ & 0.80$^{+0.11}_{-0.06}$ & 1.36$^{+0.18}_{-0.37}$ & $\cdots$ & 1.78$^{+0.27}_{-0.17}$  & 1.55$^{+0.50}_{-0.30}$ & 41.23$^{+0.02}_{-0.02}$ & 40.76$^{+0.08}_{-0.08}$ & 0.65 \\ 
X-1$^{{\tt b}}$ &  519 & $\cdots$ & $\dagger$ & 0.18$^{+0.12}_{-0.08}$ & 2.11$^{+0.36}_{-0.23}$ & $\dagger$ & 2.20$^{+0.93}_{-0.80}$ & $\cdots$ & 1.01$^{+0.59}_{-0.24}$ & 0.97$^{+1.24}_{-0.30}$ & 41.01$^{+0.05}_{-0.05}$ & 41.05$^{+0.07}_{-0.07}$ & 0.06 \\ 
X-2$^{{\tt c}}$ & 744 & 0.04$^{+0.06}_{-0.03}$ & $\dagger$ & $\dagger$ & $\dagger$ & $\dagger$& $\dagger$ & 0.22$^{+0.10}_{-0.06}$ & 2.02$^{+0.22}_{-0.22}$ & 2.44$^{+0.50}_{-0.48}$  & 40.92$^{+0.04}_{-0.04}$  & 40.75$^{+0.08}_{-0.08}$ & 0.14 \\ 
X-3$^{{\tt c}}$ & 293 & 0.03$^{+0.03}_{-0.02}$ & $\dagger$ & $\dagger$ & $\dagger$ & $\dagger$ & $\dagger$  & 0.03$^{+0.16}_{-0.02}$ & 1.53$^{+0.45}_{-0.23}$ & 0.55$^{+0.35}_{-0.12}$ & 40.57$^{+0.07}_{-0.07}$ & 40.45$^{+0.11}_{-0.12}$ & 0.27 \\  
X-4$^{{\tt c}}$ & 545 & 0.14$^{+0.05}_{-0.05}$ & $\dagger$ & $\dagger$ & $\dagger$ & $\dagger$ & $\dagger$ & 0.18$^{+0.14}_{-0.09}$ & 2.50$^{+0.53}_{-0.40}$ & 1.09$^{+0.69}_{-0.36}$ & 40.61$^{+0.04}_{-0.04}$ & 40.14$^{+0.05}_{-0.05}$ & 0.98 \\ 
X-5$^{{\tt c}}$ & 178 & 0.06$^{+0.05}_{-0.02}$ & $\dagger$& $\dagger$ & $\dagger$ & $\dagger$ & $\dagger$ & 0.25$^{+6.99}_{-0.08}$ & 2.17$^{+1.78}_{-0.79}$ & 0.28$^{+2.98}_{-0.18}$ & 40.16$^{+0.09}_{-0.08}$ &  39.76$^{+0.24}_{-0.23}$ & 0.72 \\ 
X-6$^{{\tt c}}$ & 196 &  0.09$^{+0.04}_{-0.02}$ & $\dagger$ & $\dagger$ & $\dagger$ & $\dagger$ & $\dagger$ & $>$ 0.10  &  2.34$^{+3.38}_{-0.46}$  & 0.19$^{+0.94}_{-0.08}$ & 40.16$^{+0.07}_{-0.07}$ & 39.55$^{+0.25}_{-0.27}$ & 0.09 \\ 
\enddata 
\tablecomments{Best-fit model parameters from spectral fits to the \chandra\ observation of each component of VV~114: hot gas and point sources VV~114 X-1 to X-6. Quoted uncertainties are 90\% confidence intervals. Col. (1): source name, footnote describes the spectral model employed in {\tt XSPEC}. Col. (2): total number of counts used in spectral fit. Col. (3): multiplicative constant modifying fixed diffuse model component. Col. (4): plasma temperature in keV of lower temperature {\tt APEC} model component. Col. (5): normalization for lower temperature {\tt APEC} component. Col. (6): column density in units of 10$^{22}$ cm$^{-2}$ for higher temperature {\tt APEC} component. Col. (7):  plasma temperature in keV for higher temperature {\tt APEC} component. Col. (8): normalization for higher temperature {\tt APEC} component. Col. (9): column density in units of 10$^{22}$ cm$^{-2}$ for power-law component. Col. (10): photon index for power-law component. Col. (11): normalization for power-law component. Col. (12): 0.5--8 keV luminosity, corrected for foreground Galactic absorption and assuming $D$ = 88 Mpc. Col. (13): 2--10 keV luminosity, corrected for foreground Galactic absorption and assuming $D$ = 88 Mpc. Col. (14): goodness-of-fit measure (see Section~\ref{sec:fit_techniques}).}
\tablenotetext{\dagger}{\footnotesize Parameters fixed to best-fit values from the fit to the point-source-free spectrum (``diffuse").}
\tablenotetext{{\tt a}}{\footnotesize {\tt XSPEC} model: \texttt{tbabs$_{\scaleto{\rm Gal}{4pt}}$*(apec$_{\scaleto{1}{4pt}}$ + tbabs$_{\scaleto{2}{4pt}}$*apec$_{\scaleto{2}{4pt}}$ + pow}), where the foreground Galactic absorption was fixed (\texttt{tbabs$_{\scaleto{\rm Gal}{4pt}}$}; $N_{\scaleto{\rm H}{4pt}}$ = 1.2 $\times$ 10$^{\scaleto{20}{4pt}}$ cm$^{\scaleto{-2}{4pt}}$), and the thermal models assumed $Z$ = 0.51 $Z_{\odot}$.}
\tablenotetext{{\tt b}}{\footnotesize {\tt XSPEC} model: {\tt tbabs$_{\scaleto{\rm Gal}{4pt}}$*(apec$_{\scaleto{1}{4pt}}$ + tbabs$_{\scaleto{2}{4pt}}$*(apec$_{\scaleto{2}{4pt}}$ + pow))}, where the foreground Galactic absorption was fixed (\texttt{tbabs$_{\scaleto{\rm Gal}{4pt}}$}; $N_{\scaleto{\rm H}{4pt}}$ = 1.2 $\times$ 10$^{\scaleto{20}{4pt}}$ cm$^{\scaleto{-2}{4pt}}$).}
  \tablenotetext{{\tt c}}{\footnotesize {\tt XSPEC} model: {\tt tbabs$_{\scaleto{\rm Gal}{4pt}}$*(constant$_{\scaleto{kT}{4pt}}$*(apec$_{\scaleto{1}{4pt}}$ + tbabs$_{\scaleto{2}{4pt}}$*apec$_{\scaleto{2}{4pt}}$) + tbabs$_{\scaleto{3}{4pt}}$*pow}), where the foreground Galactic absorption was fixed (\texttt{tbabs$_{\scaleto{\rm Gal}{4pt}}$}; $N_{\scaleto{\rm H}{4pt}}$ = 1.2 $\times$ 10$^{\scaleto{20}{4pt}}$ cm$^{\scaleto{-2}{4pt}}$).}
\end{deluxetable*}

\section{Results}\label{sec:results}

To construct the galaxy-wide X-ray SED of VV~114 and estimate the XRB contribution, we use the archival \chandra\ observation of VV~114 in conjunction with the newly obtained, nearly-simultaneous \xmm\ and \nustar\ observations. In Section~\ref{sec:point_src_decomp}, we present our overall spectral modeling approach and fit results for each major component of VV~114, including the point source population and the hot, diffuse gas component of the galaxy. As the \chandra\ observation was taken $\sim$ 13~yr prior to the newly obtained \xmm\ and \nustar\ observations, we consider the potential impact of variability from the sources of compact emission in VV~114 (ULXs and possible AGN) on this analysis. To mitigate the impact of variability, we analyze separately the \chandra\ spectra from the \xmm\ and \nustar\ observations, applying the constraints from the \chandra\ spectral fits, particularly for the non-time variable hot gas component of the galaxy, to the newly obtained \xmm\ and \nustar\ data. In Section~\ref{sec:gal_wide_spec} we present a comparison of results between the different epochs, as well as the galxy-wide spectrum and associated X-ray SED for VV~114. 

\subsection{\chandra\ Point Source Decomposition}\label{sec:point_src_decomp}

We began our investigation of the galaxy-wide SED of VV~114 through a spectral decomposition of the major galaxy components using the archival \chandra\ observation. With \chandra, VV~114 is resolved into six discrete point sources embedded in hot, diffuse gas (see Figure~\ref{fig:chandra_rgb}). All six point sources are ULXs with $L_{\rm 2-10~keV}$ $\approx$ (3--110) $\times$ 10$^{39}$ erg~s$^{-1}$, and are detected with sufficient counts ($\gtrsim$ 200) for simple spectral fitting (Table~\ref{tab:point_src_fits}). The brightest point source in the eastern region of the galaxy (VV~114 X-1) is a possible AGN \citep{Iono2013,Saito2015}, and we devote more attention to discussion of possible AGN contamination in Section~\ref{sec:agn_contrib}. For each of the six point sources, as well as the point-source-free diffuse emission component we fit the unbinned source spectra with an appropriately scaled background model as described in Section~\ref{sec:fit_techniques}. Below we describe the spectral models for each component resolved with \chandra. 

We first assessed the point-source-free hot, diffuse gas component using an absorbed (foreground Galactic and intrinsic) two-temperature thermal plasma ({\tt APEC}) model along with a power-law continuum to account for any unresolved XRB emission from undetected XRBs and the wings of the PSFs of X-ray detected point sources. Given the measured gas-phase metallicity for VV~114 (12 + log (O/H) = 8.4), we fixed the abundance for the {\tt APEC} model components to 0.51 $Z_{\odot}$, a direct conversion assuming \citet{Asplund2009} abundances, where $Z_{\odot}$ corresponds to 12 + log(O/H) = 8.69. 

The choice of the two-temperature thermal plasma model with intrinsic absorption is physically motivated assuming the diffuse emission detected with \chandra\ is produced via a hot disk seen through an intrinsic obscuring column (higher temperature, absorbed {\tt APEC} component), as well as a relatively unobscured hot halo (lower temperature, unabsorbed {\tt APEC} component) \citep[e.g.,][]{Martin2002,Strickland2004}. Such a model is consistent with hot, diffuse emission produced by feedback from supernovae and stellar winds \citep[e.g.,][]{Strickland2004,Grimes2005}. The model choice is further motivated by previous empirical studies, which have found that a two-temperature thermal plasma with intrinsic absorption well-describes the diffuse emission in star-forming galaxies across a range of SFRs \citep[e.g.,][]{Summers2003,Hartwell2004,MineoGas,Lehmer2015,Smith2018}. The choice to fix the abundances of the {\tt APEC} components to the measured gas-phase metallicity for VV~114 is supported by previous X-ray investigations of the hot interstellar medium (ISM) in star-forming galaxies, which have found spectral degeneracies when attempting to fit for metal abundances using X-ray spectra \citep[e.g.,][]{Weaver2000,Dahlem2000}, and further that the gas-phase metallicity is a good proxy for the metal abundance of the hot ISM  \citep[e.g.,][]{Ott2005I,Grimes2005,Grimes2006}. 

The best-fit values for the free parameters from this diffuse component spectral model are listed in the first row of Table~\ref{tab:point_src_fits}, with the {\tt XSPEC} description of the model listed in the table notes. The diffuse gas in VV~114 is well described by $\sim$ 0.4 keV and $\sim$ 0.8 keV components, consistent with plasma temperatures measured for the hot ISM in other star-forming galaxies \citep[e.g.,][]{Ott2005I,Grimes2005,MineoGas,Smith2018}. Previous X-ray analysis of VV~114 using the same \chandra\ data was performed by \citet{Grimes2006}, finding $kT = 0.3^{+0.75}_{-0.20}$ and $kT = 0.62 \pm 0.03$ keV. While the lower temperature component from \citet{Grimes2006} is consistent with our findings, the high temperature component of their model is inconsistent with our $kT = 0.80^{+0.11}_{-0.06}$ keV component. This inconsistency despite the same dataset may be due to differences in the spectral extraction and modeling approach. In particular, the \citet{Grimes2006} analysis did not explicitly separate the diffuse emission from the point source emission as we do here, and further employed a {\tt vmekal} model for the hot gas, fitting for the abundances using the {\tt angr} abundance tables \citep{Anders1989} in {\tt XSPEC}, while in this work we have employed the {\tt APEC} model with fixed abundances relative to the \citet{Asplund2009} abundance tables. 

In subsequent modeling of the six point sources, we included the diffuse gas component listed in Table~\ref{tab:point_src_fits} as a fixed component modified by a free multiplicative constant to account for any residual hot gas in the point source extraction regions. For the point sources themselves, we employed simple absorbed power-law models, appropriate for either XRBs or AGN. Each point source was therefore fit with four freely varying components for the source (i.e., diffuse gas normalization, intrinsic column density, photon index, and power-law normalization).

\begin{deluxetable*}{ccccccccccccc}
\tabletypesize{\tiny}
\tablewidth{\textwidth}
 \tablecaption{Spectral Fit Results for Galaxy-Wide Models \label{tab:gal_wide_fits}}
\tablehead{
 \colhead{} & \colhead{} & \colhead{} &  \colhead{} & \colhead{}  & \colhead{$N_{\scaleto{\rm H, XRB}{3pt}}$} & \colhead{}  & \colhead{$E_{\scaleto{\rm break}{3pt}}$} &  \colhead{} & \colhead{$A_{\scaleto{\Gamma_{\rm XRB}}{3pt}}$} & \colhead{log $L^{\scaleto{\rm gal}{3pt}}_{\scaleto{{\rm 0.5-8~keV}}{3pt}}$} & \colhead{log $L^{\scaleto{\rm gal}{3pt}}_{\scaleto{{\rm 2-10~keV}}{3pt}}$} &  \colhead{} \\ 
\colhead{Model} & \colhead{Inst.} & \colhead{C$_{\scaleto{\it XMM}{3pt}}$} & \colhead{C$_{\scaleto{\nustar}{3pt}}$}  & \colhead{C$_{\scaleto{kT}{3pt}}$} & \colhead{(10$^{22}$ cm$^{-2}$)} & \colhead{$\Gamma_{\scaleto{\rm XRB,1}{3pt}}$} & \colhead{(keV)} & \colhead{$\Gamma_{\scaleto{\rm XRB,2}{3pt}}$} & \colhead{(10$^{-5}$)} & \colhead{(erg s$^{-1}$)} & \colhead{(erg s$^{-1}$)} & \colhead{gof} \\
\colhead{(1)} & \colhead{(2)} &  \colhead{(3)} & \colhead{(4)} & \colhead{(5)} & \colhead{(6)} & \colhead{(7)} & \colhead{(8)} & \colhead{(9)} & \colhead{(10)} & \colhead{(11)} & \colhead{(12)} & \colhead{(13)}
}
\startdata 
{\scriptsize {\tt pow$_{\scaleto{\rm XRB}{2pt}}$ + pow$_{\scaleto{\rm AGN}{2pt}}$}} & \chandra$^{{\tt a}}$ & $\cdots$ & $\cdots$ & 1.26$^{+0.14}_{-0.13}$ &  0.13$^{+0.05}_{-0.03}$ & 2.07$^{+0.19}_{-0.16}$ & $\cdots$ & $\cdots$ &  6.94$^{+1.60}_{-0.85}$ & 41.65$^{+0.02}_{-0.01}$ & 41.43$^{+0.03}_{-0.03}$ & 0.07 \\  
{\scriptsize {\tt pow$_{\scaleto{\rm XRB}{2pt}}$}} & \chandra$^{{\tt b}}$ & $\cdots$ & $\cdots$ & 1.28$^{+0.15}_{-0.09}$ & 0.08$^{+0.04}_{-0.04}$ & 1.69$^{+0.11}_{-0.14}$ & $\cdots$ & $\cdots$ & 6.37$^{+0.76}_{-0.94}$ & 41.64$^{+0.02}_{-0.02}$ & 41.41$^{+0.04}_{-0.04}$ & 0.43 \\
{\scriptsize {\tt pow$_{\scaleto{\rm XRB}{2pt}}$ + pow$_{\scaleto{\rm AGN}{2pt}}$}} & {\it XMM}+\nustar$^{{\tt a\dagger}}$ & 0.97$^{+0.05}_{-0.04}$ & 0.56$^{+0.05}_{-0.05}$ & $\cdots$ & $\dagger$ & $\dagger$ & $\cdots$ & $\cdots$ & $\dagger$ & 41.60$^{+0.02}_{-0.01}$ & 41.30$^{+0.03}_{-0.03}$ & $<$ 10$^{-3}$ \\
{\scriptsize {\tt pow$_{\scaleto{\rm XRB}{2pt}}$}} & {\it XMM}+\nustar$^{{\tt b\ddagger}}$ & 0.96$^{+0.05}_{-0.04}$ & 0.63$^{+0.06}_{-0.06}$ & $\cdots$ & $\ddagger$ &  $\ddagger$ &  $\cdots$ & $\cdots$ & $\ddagger$ & 41.59$^{+0.02}_{-0.02}$ & 41.30$^{+0.03}_{-0.03}$ & $<$ 10$^{-3}$ \\ 
{\scriptsize {\tt bknpow$_{\scaleto{\rm ULX}{2pt}}$}}  & {\it XMM} + {\it NuSTAR}$^{{\tt c}}$ & 1.39$^{+0.12}_{-0.04}$ & 1.62$^{+0.33}_{-0.17}$ & $\cdots$ & 0.01$^{+0.05}_{-0.01}$ & 1.44$^{+0.08}_{-0.07}$ & 4.03$^{+1.37}_{-0.50}$ & 2.51$^{+0.33}_{-0.19}$ & 2.61$^{+0.23}_{-0.32}$ & 41.54$^{+0.02}_{-0.02}$ & 41.27$^{+0.04}_{-0.04}$ & 0.06 \\
\enddata 
\tablecomments{Best-fit model parameters from spectral fits to galaxy-wide \chandra, \xmm, and \nustar\ spectra of VV~114. Quoted uncertainties are 90\% confidence interval. Col. (1): model descriptor. Col. (2): instrument(s), footnote describe the spectral model employed in {\tt XSPEC}. Col. (3): multiplicative constant for \xmm\ spectra (\xmm\ + \nustar\ fits only). Col. (4): multiplicative constant for \nustar\ spectra (\xmm\ + \nustar\ fits only). Col. (5): multiplicative constant modifying the diffuse model component (\chandra\ fits only). Col. (6): column density in units of 10$^{22}$ cm$^{-2}$ for XRB power-law or broken power-law component. Col. (7): photon index for XRB power-law or first photon index for broken power-law component. Col. (8): break energy in keV for XRB broken power-law component. Col. (9):  second photon index for XRB broken power-law component. Col. (10): normalization for XRB power-law or broken power-law component. Col. (11): galaxy-integrated 0.5--8 keV \Lx\ derived from the model, corrected for foreground Galactic absorption and assuming $D$ = 88 Mpc. Col. (12): galaxy-integrated 2--10 keV \Lx\ derived from the model, corrected for foreground Galactic absorption and assuming $D$ = 88 Mpc. Col. (13): goodness-of-fit measure (see Section~\ref{sec:fit_techniques}).}
\tablenotetext{{\tt a}}{\footnotesize {\tt XSPEC} model: {\tt tbabs$_{\scaleto{\rm Gal}{4pt}}$*(constant$_{\scaleto{kT}{4pt}}$*(apec$_{\scaleto{1}{4pt}}$ + tbabs$_{\scaleto{2}{4pt}}$*apec$_{\scaleto{2}{4pt}}$) + tbabs$_{\scaleto{3}{4pt}}$*pow$_{\scaleto{\rm AGN}{4pt}}$ +  tbabs$_{\scaleto{4}{4pt}}$*pow$_{\scaleto{\rm XRB}{4pt}}$)}, where foreground Galactic absorption ({\tt tbabs$_{\scaleto{\rm Gal}{4pt}}$}) was fixed to $N_{\scaleto{\rm H}{4pt}}$ =1.20 $\times$ 10$^{\scaleto{20}{4pt}}$ cm$^{\scaleto{-2}{4pt}}$, all ({\tt apec$_{\scaleto{1}{4pt}}$ + tbabs$_{\scaleto{2}{4pt}}$*apec$_{\scaleto{2}{4pt}}$}) components were fixed to values from the fit to the point-source free spectrum, the ({\tt tbabs$_{\scaleto{3}{4pt}}$*pow$_{\scaleto{\rm AGN}{4pt}}$}) components were fixed to the values from the fit to the spectrum of VV~114 X-1, and the ({\tt tbabs$_{\scaleto{4}{4pt}}$*pow$_{\scaleto{\rm XRB}{4pt}}$}) components are allowed to freely vary.}
\tablenotetext{{\tt b}}{\footnotesize {\tt XSPEC} model: {\tt tbabs$_{\scaleto{\rm Gal}{4pt}}$*(constant$_{\scaleto{kT}{4pt}}$*(apec$_{\scaleto{1}{4pt}}$ + tbabs$_{\scaleto{2}{4pt}}$*apec$_{\scaleto{2}{4pt}}$) +  tbabs$_{\scaleto{3}{4pt}}$*pow$_{\scaleto{\rm XRB}{4pt}}$)}, where foreground Galactic absorption ({\tt tbabs$_{\scaleto{\rm Gal}{4pt}}$}) was fixed to $N_{\scaleto{\rm H}{4pt}}$ =1.20 $\times$ 10$^{\scaleto{20}{4pt}}$ cm$^{\scaleto{-2}{4pt}}$, all ({\tt apec$_{\scaleto{1}{4pt}}$ + tbabs$_{\scaleto{2}{4pt}}$*apec$_{\scaleto{2}{4pt}}$}) components were fixed to values from the fit to the point-source free spectrum, and the ({\tt tbabs$_{\scaleto{3}{4pt}}$*pow$_{\scaleto{\rm XRB}{4pt}}$}) components were allowed to freely vary.}
\tablenotetext{{\tt a\dagger}}{\footnotesize {\tt XSPEC} model: {\tt tbabs$_{\scaleto{\rm Gal}{4pt}}$*(constant$_{\scaleto{\rm inst}{4pt}}$*(apec$_{\scaleto{1}{4pt}}$ + tbabs$_{\scaleto{2}{4pt}}$*apec$_{\scaleto{2}{4pt}}$ + tbabs$_{\scaleto{3}{4pt}}$*pow$_{\scaleto{\rm AGN}{4pt}}$ +  tbabs$_{\scaleto{4}{4pt}}$*pow$_{\scaleto{\rm XRB}{4pt}}$))}, where the only freely varying parameter is the instrumental constant ({\tt constant$_{\scaleto{\rm inst}{4pt}}$}). The foreground Galactic absorption ({\tt tbabs$_{\scaleto{\rm Gal}{4pt}}$}) was fixed to $N_{\scaleto{\rm H}{4pt}}$ =1.20 $\times$ 10$^{\scaleto{20}{4pt}}$ cm$^{\scaleto{-2}{4pt}}$, all ({\tt apec$_{\scaleto{1}{4pt}}$ + tbabs$_{\scaleto{2}{4pt}}$*apec$_{\scaleto{2}{4pt}}$}) components were fixed to values from the fit to the point-source free spectrum, the ({\tt tbabs$_{\scaleto{3}{4pt}}$*pow$_{\scaleto{\rm AGN}{4pt}}$}) components were fixed to the values from the fit to the spectrum of VV~114 X-1, and the ({\tt tbabs$_{\scaleto{4}{4pt}}$*pow$_{\scaleto{\rm XRB}{4pt}}$}) component was fixed to the best-fit values from the galaxy-wide fit to the \chandra\ observation (model {\tt a}).}
\tablenotetext{{\tt b\ddagger}}{\footnotesize {\tt XSPEC} model: {\tt tbabs$_{\scaleto{\rm Gal}{4pt}}$*(constant$_{\scaleto{\rm inst}{4pt}}$*(apec$_{\scaleto{1}{4pt}}$ + tbabs$_{\scaleto{2}{4pt}}$*apec$_{\scaleto{2}{4pt}}$ +  tbabs$_{\scaleto{3}{4pt}}$*pow$_{\scaleto{\rm XRB}{4pt}}$))}, where the only freely varying parameter is the instrumental constant ({\tt constant$_{\scaleto{\rm inst}{4pt}}$}). The foreground Galactic absorption ({\tt tbabs$_{\scaleto{\rm Gal}{4pt}}$}) was fixed to $N_{\scaleto{\rm H}{4pt}}$ =1.20 $\times$ 10$^{\scaleto{20}{4pt}}$ cm$^{\scaleto{-2}{4pt}}$, all ({\tt apec$_{\scaleto{1}{4pt}}$ + tbabs$_{\scaleto{2}{4pt}}$*apec$_{\scaleto{2}{4pt}}$}) components were fixed to values from the fit to the point-source free spectrum, and the ({\tt  tbabs$_{\scaleto{3}{4pt}}$*pow$_{\scaleto{\rm XRB}{4pt}}$}) component was fixed to the best-fit values from the galaxy-wide fit to the \chandra\ observation (model {\tt b}).} 
\tablenotetext{{\tt c}}{\footnotesize {\tt XSPEC} model: {\tt tbabs$_{\scaleto{\rm Gal}{4pt}}$*(constant$_{\scaleto{\rm inst}{4pt}}$*(apec$_{\scaleto{1}{4pt}}$ + tbabs$_{\scaleto{2}{4pt}}$*apec$_{\scaleto{2}{4pt}}$ +  tbabs$_{\scaleto{3}{4pt}}$*bknpow))}, where the only freely varying parameters are the instrumental constant ({\tt constant$_{\scaleto{\rm inst}{4pt}}$}) and the parameters of the ({\tt  tbabs$_{\scaleto{3}{4pt}}$*bknpow}) model component. The foreground Galactic absorption ({\tt tbabs$_{\scaleto{\rm Gal}{4pt}}$}) was fixed to $N_{\scaleto{\rm H}{4pt}}$ =1.20 $\times$ 10$^{\scaleto{20}{4pt}}$ cm$^{\scaleto{-2}{4pt}}$, and all ({\tt apec$_{\scaleto{1}{4pt}}$ + tbabs$_{\scaleto{2}{4pt}}$*apec$_{\scaleto{2}{4pt}}$}) components were fixed to values from the fit to the point-source-free \chandra\ spectrum.}
\end{deluxetable*}

Using the above described power-law-plus-hot-gas model for VV~114 X-2 to X-6, we find steep photon indices ($\Gamma > 1.5$), relatively low column densities modifying the power-law components, and minimal contributions from the surrounding hot gas, as indicated by the small values of the normalizations to the fixed diffuse gas components. The best-fit parameters and their associated uncertainties (90\% confidence intervals) along with the 0.5--8 and 2--10 keV luminosities for each source from this model are summarized in Table~\ref{tab:point_src_fits}. The models for point sources VV~114 X-2 to X-6 are consistent with their being either collections of unresolved XRBs or ULXs embedded in hot gas, indicative of recent star formation. 

We initially applied the default power-law-plus-hot-gas model with fixed parameters to VV~114 X-1, but found gof = 0.03, suggesting the model could be improved. We next attempted to fit VV~114 X-1 with a simple absorbed power-law, but found the fit left residuals at energies $<$ 0.5 keV and at the location of emission line complexes between 1--2 keV, indicating the need to include one or more thermal components. Given these results and that the multiwavelength data available for VV~114 (e.g., Figure~\ref{fig:hst_rgb}) indicates heavy obscuration in the eastern portion of the galaxy where VV~114 X-1 is located, we next adopted a slightly altered version of the default model. The new model for VV~114 X-1 consists of an unabsorbed {\tt APEC} component (unobscured hot halo), as well as an absorbed {\tt APEC}-plus-power-law component, representing the obscured emission from the hot disk and VV~114 X-1. In this model for VV~114 X-1 we fixed the {\tt APEC} temperatures to the values from the default model ($kT_{1}$ = 0.36 keV and $kT_{2}$ = 0.80 keV), but allowed the {\tt APEC} normalizations to freely vary. We likewise allowed the intrinsic column density and power-law parameters to freely vary. The fit to VV~114 X-1 using this model yields gof = 0.06, a slight improvement over the default model, returning a high column density ($N_{\rm H} = 2.11^{+0.36}_{-0.23} \times 10^{22}$ cm$^{-2}$) modifying the power-law and higher temperature {\tt APEC} components, and a photon index of $\Gamma = 1.01^{+0.59}_{-0.24}$. The best-fit values and associated uncertainties from this model are listed in the second row of Table~\ref{tab:point_src_fits}. Importantly, the values for the column density and photon index from this model are consistent with values from \citet{Grimes2006} (their source VV~114E) using the same \chandra\ data, albeit a slightly different source model than the one employed here (see discussion in Section~\ref{sec:agn_contrib}). This lends additional support for the adoption of this model for VV~114 X-1. We discuss the significance of this spectral fit result for VV~114 X-1 as a possible AGN in more detail in Section~\ref{sec:agn_contrib}.

\subsection{Galaxy-Wide Spectral Analysis}\label{sec:gal_wide_spec}

We extend the results from the spectral decomposition of VV~114 to construct a galaxy-wide spectral model and determine the dominant spectral component at higher energies. In fitting the \chandra\ observation of the galaxy-wide spectrum we consider all major spectral components as derived from the spectral decomposition described in Section~\ref{sec:point_src_decomp}, and then apply these spectral constraints as appropriate in building a galaxy-wide spectral model to be applied to the \xmm\ and \nustar\ observations, in which VV~114 is consistent with being a single source (see contours in Figures~\ref{fig:chandra_rgb}--\ref{fig:hst_rgb}). We note that it is not as straightforward to interpret the gof measure for the galaxy-wide fits as for the point source fits. This is because the {\tt goodness} command simulates spectra assuming variance due only to counting statistics, whereas in fitting the galaxy-wide spectra we may be dominated by systematics. Thus, for the galaxy-wide fits, we caution against interpreting very low gof values as indicating the need for additional model components or free parameters, or interpreting gof $\sim$ 1 as ``overfitting,'' as proper inclusion of systematic error would likely serve to widen the distribution of test statistics for the simulated spectra.

\begin{figure}
\centering
\includegraphics[width=0.5\textwidth,trim=0 0 0 0, clip]{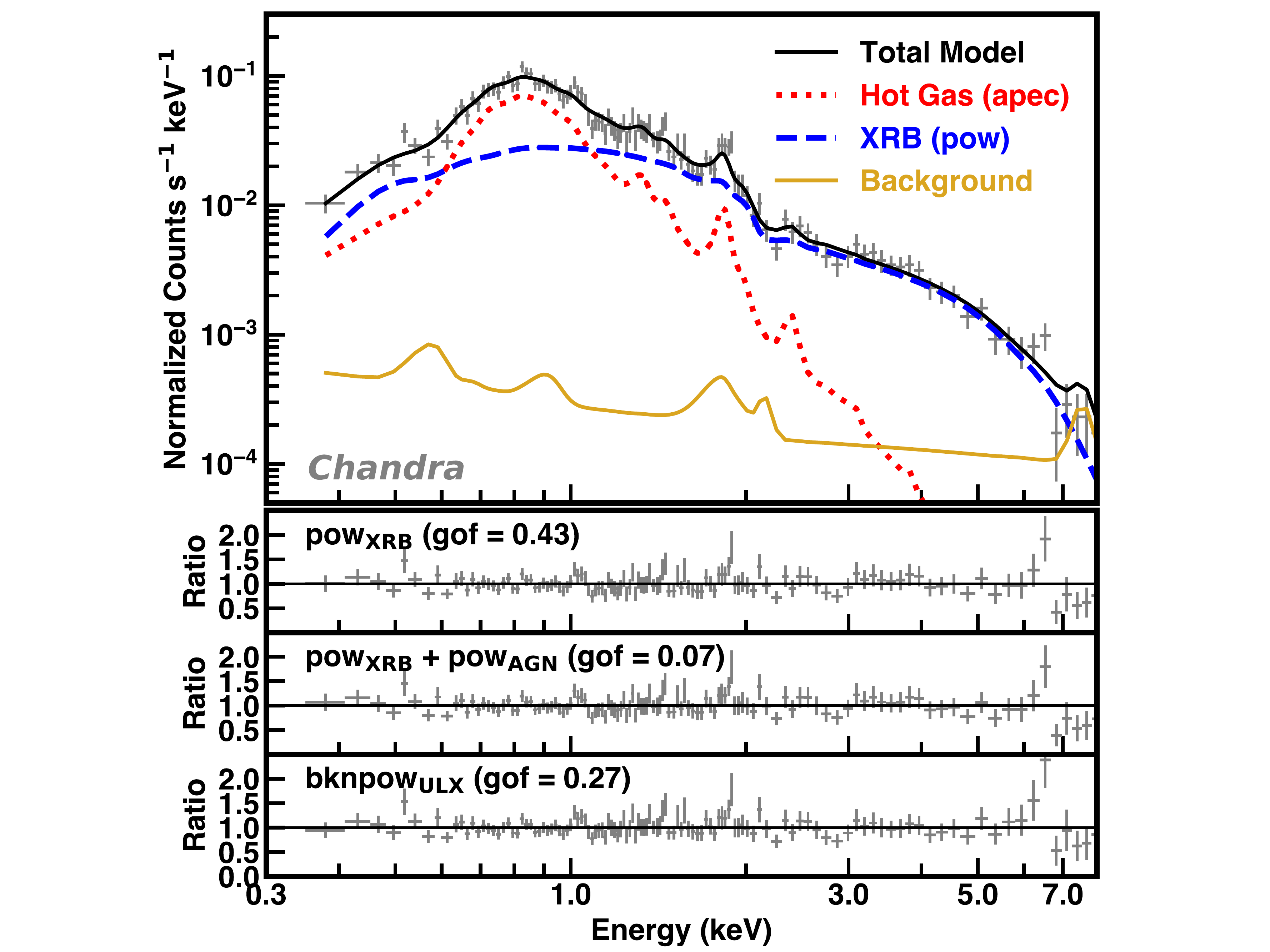}
\caption{Spectral fit to the \chandra\ spectrum (grey points), where the data points have been binned to a minimum significance of five per spectral bin for plotting purposes. The best-fit total model is shown as the solid black line. Each component of the model is also displayed: the hot gas component (two-temperature thermal plasma model) as a dotted red line, the XRB component (absorbed power-law model) as a dashed blue line, and the background, both sky and instrumental, as the solid gold line. We display the residuals for the {\tt pow$_{\rm XRB}$}, {\tt bknpow$_{\rm ULX}$}, and {\tt pow$_{\rm XRB}$ + pow$_{\rm AGN}$} models as applied to the \chandra\ observation in the bottom three panels, annotated with the gof for the fit. The model consisting of a single absorbed power-law is the most consistent with the \chandra\ data; however, as shown by the residuals, the quality of the \chandra\ data, particularly at energies $>$ 5 keV is not sufficient to effectively rule out any of the models tested here.}\label{fig:chandra_spec_resid}
\end{figure}

\begin{figure}
\centering
\includegraphics[width=0.5\textwidth,trim=0 0 0 0, clip]{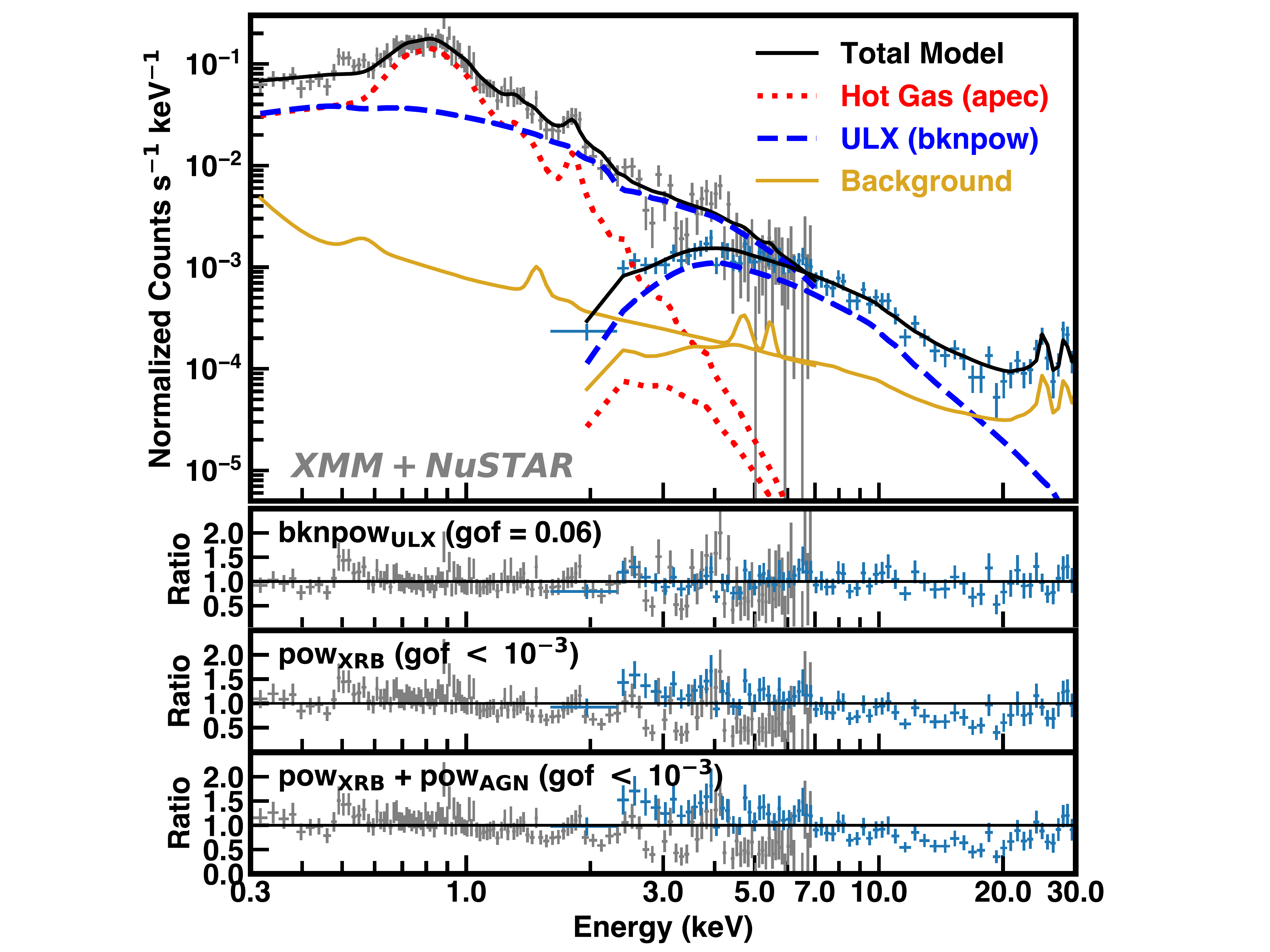}
\caption{Spectral fit to joint, nearly-simultaneous \xmm\ (pn: grey) and \nustar\ (FPMA + FPMB: cyan) spectra, where the data points have been grouped by instrument and to a minimum significance of five in {\tt XSPEC} for plotting purposes only. The total best-fit spectral model is displayed as a solid black line, with each major component of this model also labeled. The dotted red line shows the hot gas component (a two-temperature thermal plasma model), the dashed blue line represents the ULX component (an absorbed broken power-law), and the solid gold line shows the combined sky and instrumental background (described in Section~\ref{sec:fit_techniques}). Below the plotted spectra with best-fit model we show the residuals for all three models listed in the last three rows of Table~\ref{tab:gal_wide_fits} that were fit to the joint \xmm\ and \nustar\ spectra. In each residual panel we list a shorthand for the model type and the gof for the fit. Only the model with the broken power-law component is consistent with the data, indicating that the global emission of VV~114 is dominated by ULXs.}\label{fig:xmmnu_spec_resid}
\end{figure}

We first fit the galaxy-wide \chandra\ spectrum of VV~114 using a model comprised of a hot gas component, an obscured AGN-like component, and an XRB population component (model: {\tt pow$_{\rm XRB}$ + pow$_{\rm AGN}$}). In this {\tt pow$_{\rm XRB}$ + pow$_{\rm AGN}$} model we fixed the hot gas component to the best-fit model for the point-source-free spectrum (first row of Table~\ref{tab:point_src_fits}), allowing this component to be modified only by a freely varying multiplicative constant. We likewise fixed the obscured AGN-like component to the column density and power-law slope and normalization from the best-fit model to VV~114 X-1 (second row of Table~\ref{tab:point_src_fits}), under the assumption X-1 is an AGN candidate distinct from the other detected point sources. We consider the XRB population component of VV~114 to consist of the known ULXs VV~114 X-2 to X-6 as well as the unresolved XRBs (e.g., the power-law component in the diffuse-only spectrum), for which the best-fit models return varying levels of intrinsic absorption and a range of photon indices and power-law normalizations. In all galaxy-wide models for VV~114 we therefore model the XRB population with a single absorbed power-law component to account for the {\it combination} of all known ULXs and the unresolved XRBs. The assumption of a single absorbed power-law for the XRB population  is consistent with a fit to the stacked spectra of sources VV~114 X-2 to X-6. We allow all parameters associated with this absorbed power-law model (i.e., intrinsic absorption, photon index, normalization) to be freely varying in order to determine the best-fit parameters for the ensemble XRB population on the galaxy-wide emission. 

The results for the fit to the galaxy-wide \chandra\ spectrum using the {\tt pow$_{\rm XRB}$ + pow$_{\rm AGN}$} model are recorded in the first row of Table~\ref{tab:gal_wide_fits}, where we list the best-fit values for the free parameters and their associated uncertainties, as well as the gof value for the model. The {\tt pow$_{\rm XRB}$ + pow$_{\rm AGN}$} model produces an acceptable fit (gof = 0.07) to the galaxy-wide \chandra\ spectrum (residuals in the bottom panel of Figure~\ref{fig:chandra_spec_resid}), returning a photon index $\Gamma = 2.07^{+0.19}_{-0.16}$ for the XRB population power-law component, and a power-law normalization consistent with the summation of normalizations for the known ULXs and unresolved XRBs from the decomposition fits in Table~\ref{tab:point_src_fits}. We also tested a simpler model consisting of the same hot gas component, with parameters fixed to the best-fit values from the point-source-free spectrum, but only a single absorbed power-law component with freely varying absorption, photon index, and normalization (model: {\tt pow$_{\rm XRB}$}). In this model, the single power-law represents the combination of all six point sources (ULXs and possible AGN) and unresolved XRBs. The results from this {\tt pow$_{\rm XRB}$} model fit to the galaxy-wide \chandra\ spectrum are listed in the second row of Table~\ref{tab:gal_wide_fits} and shown in the top panel of Figure~\ref{fig:chandra_spec_resid}, with associated residuals in the panel just below. The {\tt pow$_{\rm XRB}$} model returns a photon index of $\Gamma = 1.69^{+0.11}_{-0.14}$, consistent with expectations for a population of XRBs. The gof = 0.43 further suggests that this {\tt pow$_{\rm XRB}$} model is a somewhat more acceptable fit to the galaxy-wide \chandra\ spectrum, indicating that an additional power-law component describing VV~114 X-1 is not {\it required} to model the galaxy-wide X-ray emission. 

We next applied these two models (i.e., {\tt pow$_{\rm XRB}$ + pow$_{\rm AGN}$}, and {\tt pow$_{\rm XRB}$} alone), to the \xmm\ and \nustar\ spectra. To start, we simply applied each model to the joint \xmm\ and \nustar\ spectra with the parameters of all model components fixed to the best-fit values from the fits to the \chandra\ spectrum, save for an overall multiplicative scaling constant for each instrument, which we allowed to vary to account for flux calibration differences or intrinsic variability. We record the results from applying these \chandra-derived models to \xmm\ and \nustar\ in the third and fourth rows of Table~\ref{tab:gal_wide_fits}, finding that neither model is consistent with the \xmm\ and \nustar\ data (gof $<$ 10$^{-3}$). We also attempted fitting the \xmm\ and \nustar\ spectra with {\tt pow$_{\rm XRB}$ + pow$_{\rm AGN}$} and {\tt pow$_{\rm XRB}$}  models with freely varying power-law parameters, i.e., without fixing these parameters to the \chandra-derived values, but found unacceptable fits in both cases (gof $<$ 10$^{-3}$).

The residuals from the \chandra-derived {\tt pow$_{\rm XRB}$ + pow$_{\rm AGN}$} and {\tt pow$_{\rm XRB}$} models (bottom two panels of Figure~\ref{fig:xmmnu_spec_resid}) indicate a reasonable fit to the \xmm\ + \nustar\ data at $E \lesssim$ 2--3 keV, but an overestimate of the $E >$ 2--3 keV emission (overall gof $<$ 10$^{-3}$). These results suggest that the extension of an XRB-like + AGN-like power-law or a single XRB-like power-law to higher energies using parameters derived from fits to the \chandra\ data is inconsistent with the observed \xmm\ and \nustar\ spectra. We discuss these results in the context of the potential AGN source VV~114 X-1 in Section~\ref{sec:agn_contrib}. 

\begin{deluxetable*}{cccccccc}
\tabletypesize{\tiny}
\tablewidth{\textwidth}
 \tablecaption{Luminosity of Galaxy Components from Best-Fit Spectral Models \label{tab:comp_lx}}
\tablehead{
\colhead{} & \colhead{} & \colhead{log $L^{\scaleto{\rm gas}{3pt}}_{\scaleto{{\rm 0.5-2~keV}}{3pt}}$ } & \colhead{log $L^{\scaleto{\rm XRB}{3pt}}_{\scaleto{{\rm 0.5-2~keV}}{3pt}}$} & \colhead{log $L^{\scaleto{\rm gas}{3pt}}_{\scaleto{{\rm 0.5-8~keV}}{3pt}}$} &  \colhead{log $L^{\scaleto{\rm XRB}{3pt}}_{\scaleto{{\rm 0.5-8~keV}}{3pt}}$} & \colhead{log $L^{\scaleto{\rm gas}{3pt}}_{\scaleto{{\rm 2-10~keV}}{3pt}}$}  & \colhead{log $L^{\scaleto{\rm XRB}{3pt}}_{\scaleto{{\rm 2-10~keV}}{3pt}}$} \\ 
\colhead{Inst.} & \colhead{Model} & \colhead{(erg s$^{-1}$)}  & \colhead{(erg s$^{-1}$)}  & \colhead{(erg s$^{-1}$)}  & \colhead{(erg s$^{-1}$)}  & \colhead{(erg s$^{-1}$)} & \colhead{(erg s$^{-1}$)} \\
\colhead{(1)} & \colhead{(2)} &  \colhead{(3)} & \colhead{(4)} & \colhead{(5)} & \colhead{(6)} & \colhead{(7)} & \colhead{(8)}
}
\startdata 
\chandra & {\scriptsize {\tt pow$_{\scaleto{\rm XRB}{2pt}}$}} & 41.06$^{+0.03}_{-0.03}$ & 41.04$^{+0.04}_{-0.04}$ &  41.09$^{+0.03}_{-0.03}$ & 41.50$^{+0.02}_{-0.02}$ &  39.98$^{+0.03}_{-0.03}$ & 41.39$^{+0.04}_{-0.04}$ \\
\xmm\ + \nustar\ & {\scriptsize {\tt bknpow$_{\scaleto{\rm ULX}{2pt}}$}} & 41.09$^{+0.04}_{-0.04}$ & 40.87$^{+0.04}_{-0.04}$ & 41.13$^{+0.04}_{-0.04}$ & 41.33$^{+0.04}_{-0.04}$ & 40.01$^{+0.04}_{-0.04}$ &  41.25$^{+0.04}_{-0.05}$ \\
\enddata 
\tablecomments{Luminosities in three different bands of the components (hot gas and XRB population) comprising the galaxy-integrated \Lx\ of VV~114 from the best-fit spectral models to the \chandra\ and \xmm\ and \nustar\ spectra. Col. (1): 0.5--2 keV luminosity of the hot gas component. Col. (2): 0.5--2 keV luminosity of the XRB population component. Col. (3): 0.5--8 keV luminosity of the hot gas component. Col. (4): 0.5--8 keV luminosity of the XRB population component. Col. (5): 2--10 keV luminosity of the hot gas component. Col. (6): 2--10 keV luminosity of the XRB population component.}
\end{deluxetable*}

We next swapped the power-law component in the {\tt pow$_{\rm XRB}$} model for a broken power-law (model: {\tt bknpow$_{\rm ULX}$}), a component which is physically and observationally motivated assuming a ULX-dominated population \citep[e.g.,][]{Gladstone2009, Wik2014,Walton2013,Walton2015,Rana2015,Lehmer2015,Yukita2016}. As in previous fits, we fixed all the diffuse gas model parameters to the best-fit values to the \chandra\ point-source-free spectrum, but allowed all broken power-law parameters as well as the overall multiplicative constant for each instrument to vary. This {\tt bknpow$_{\rm ULX}$} model yields a fit consistent with the \xmm\ and \nustar\ observations (gof  = 0.06), demonstrating that the data are consistent with a spectral turnover at $\sim$ 4 keV. We record the values for the free parameters and their associated uncertainties from the {\tt bknpow$_{\rm ULX}$} model in the last row of Table~\ref{tab:gal_wide_fits}.  

Given the success of this {\tt bknpow$_{\rm ULX}$} model in fitting the \xmm\ and \nustar\ spectra, we test this model on the \chandra\ observation as well, allowing only the broken power-law normalization and intrinsic absorption to freely vary. We show the residuals for the {\tt bknpow$_{\rm ULX}$} as applied to the \chandra\ data in the middle panel of Figure~\ref{fig:chandra_spec_resid}, demonstrating that this model provides an acceptable fit (gof = 0.27) to the \chandra\ observation. However, as shown by the residuals in Figure~\ref{fig:chandra_spec_resid}, the quality of the \chandra\ data, particularly at energies $>$ 5 keV, is not sufficient to distinguish between the different models tested here. In fact, in order to obtain an acceptable fit to the \chandra\ spectrum with the {\tt bknpow$_{\rm ULX}$} model we must freeze the majority of the model parameters to the best-fit values obtained from fits to the \xmm\ and \nustar\ spectra, indicating that the \chandra\ spectrum alone cannot constrain parameters such as the broken power-law photon indices and break energy. This underlines that, while \chandra\ is powerful for resolving point sources from the hot, diffuse emission in the galaxy, \xmm\ and \nustar\ are critical for high energy ($>$ 5 keV) constraints, where spectral turnover from a ULX population (or lack thereof) is more apparent. 

We display the best-fit {\tt bknpow$_{\rm ULX}$} model and its associated components (hot, diffuse gas, broken power-law, and background) as applied to the \xmm\ and \nustar\ spectra in the top panel of Figure~\ref{fig:xmmnu_spec_resid}, with the residuals for this model in the second panel from the top. The residuals from the \chandra-derived models ({\tt pow$_{\rm XRB}$ + pow$_{\rm AGN}$} and {\tt pow$_{\rm XRB}$}), which were poor fits to the \xmm\ and \nustar\ observations, are shown in the bottom two panels of the same figure for reference. The preference for a broken power-law over power-law component(s) as constrained by the nearly-simultaneous \xmm\ and \nustar\ spectra is consistent with the galaxy-wide emission of VV~114 being dominated by ULXs at energies $\gtrsim$ 2 keV \citep[e.g.,][]{Gladstone2009}. We discuss this ULX-dominated interpretation as it relates to the possible AGN in VV~114, as well as metallicity effects, in Sections~\ref{sec:agn_contrib}--\ref{sec:sed_metal}.

In columns 11--12 of Table~\ref{tab:gal_wide_fits} we list the galaxy-integrated {\it total} X-ray luminosity in the 0.5--8 keV and 2--10 keV bands (log $L^{\rm gal}_{\rm 0.5-8~keV}$ and log $L^{\rm gal}_{\rm 2-10~keV}$) corrected for foreground Galactic absorption from each of the spectral models applied to the \chandra\ and \xmm\ + \nustar\ spectra. In Table~\ref{tab:comp_lx} we likewise list the galaxy-integrated luminosities of the {\it components} which constitute $L^{\rm gal}_{\rm X}$, namely the luminosities of the hot, diffuse gas (log $L^{\rm gas}_{\rm X}$) and XRB population (log $L^{\rm XRB}_{\rm X}$), in the 0.5--2 keV, 0.5--8 keV, and 2--10 keV bands derived from the best-fit model to the \chandra\ spectrum ({\tt pow$_{\rm XRB}$}) and the best-fit model to the \xmm\ and \nustar\ spectra ({\tt bknpow$_{\rm ULX}$}). 

The discrepancy between the model XRB component luminosities listed in Table~\ref{tab:comp_lx} as measured with \chandra\ versus \xmm\ + \nustar\ can be attributed to multiple factors, including flux calibration differences between instruments, model differences, and the different integration times and epochs between observations. In Table~\ref{tab:comp_lx} we list XRB component luminosities derived from the {\it best-fit} models for each set of observations, thus some disagreement between \chandra\ and \xmm\ + \nustar\ is expected given these observations are fit with different best-fit models ({\tt pow$_{\rm XRB}$} and {\tt bknpow$_{\rm ULX}$}, respectively). However, when we calculate the XRB \Lx\ from the  {\tt bknpow$_{\rm ULX}$} model fit to the \chandra\ spectrum we find log $L^{\rm XRB}_{\rm 0.5-8~keV}$ = 41.47 $\pm$ 0.02 and log $L^{\rm XRB}_{\rm 2-10~keV}$ = 41.33 $\pm$ 0.02, still inconsistent with the \xmm\ and \nustar\ derived XRB component luminosities listed in Table~\ref{tab:comp_lx} using this same model. That the inconsistency is larger in the 0.5--8 keV band indicates that the depth of the observations is at least partly to blame for the discrepancy between luminosities. In particular, we use the \xmm\ spectra to constrain the luminosity in the 0.5--8 keV band given its increased sensitivity across the entirety of the bandpass relative to \nustar; however, \xmm\ is the shallowest of all observations used here, thus the XRB component luminosity in the 0.5--8 keV bandpass as constrained by \xmm\ is likely systematically low compared to \chandra. By contrast, the discrepancy in the 2--10 keV band, where we use both \nustar\ and \xmm\ to constrain the XRB component luminosity, is within the range expected due to flux calibration differences between instruments \citep{Madsen2015}. It is also possible that variability between epochs affects the derived luminosities, however, as we further demonstrate in Section~\ref{sec:compare_prev} using a set of archival \xmm\ observations, the inconsistency between \chandra\ and \xmm\ + \nustar\ derived luminosities does not necessarily suggest substantial variability between epochs.

All luminosities listed in Tables~\ref{tab:point_src_fits}--\ref{tab:comp_lx} are calculated assuming $D$ = 88 Mpc for VV~114, and using the {\tt cflux} convolution model in {\tt XSPEC} as a component modifying either the overall model (luminosities in Table~\ref{tab:point_src_fits} and Table~\ref{tab:gal_wide_fits}) or the appropriate model component (luminosities in Table~\ref{tab:comp_lx}). Therefore, all luminosities are based on fluxes corrected for Galactic extinction, but not intrinsic extinction. In the {\tt cflux} model component, we fixed the minimum and maximum energy parameters to return the flux in the appropriate band, and likewise fixed the normalizations of any model component modified by {\tt cflux} to the best-fit values from Tables~\ref{tab:point_src_fits}--\ref{tab:gal_wide_fits}. In refitting the models with {\tt cflux} to produce a galaxy-integrated total or component flux, we allowed only the flux parameter of the {\tt cflux} component and any free parameters excluding the component normalizations in the model to freely vary.

\begin{figure*}
\centering
\includegraphics[width=\textwidth,trim=0 0 0 0, clip]{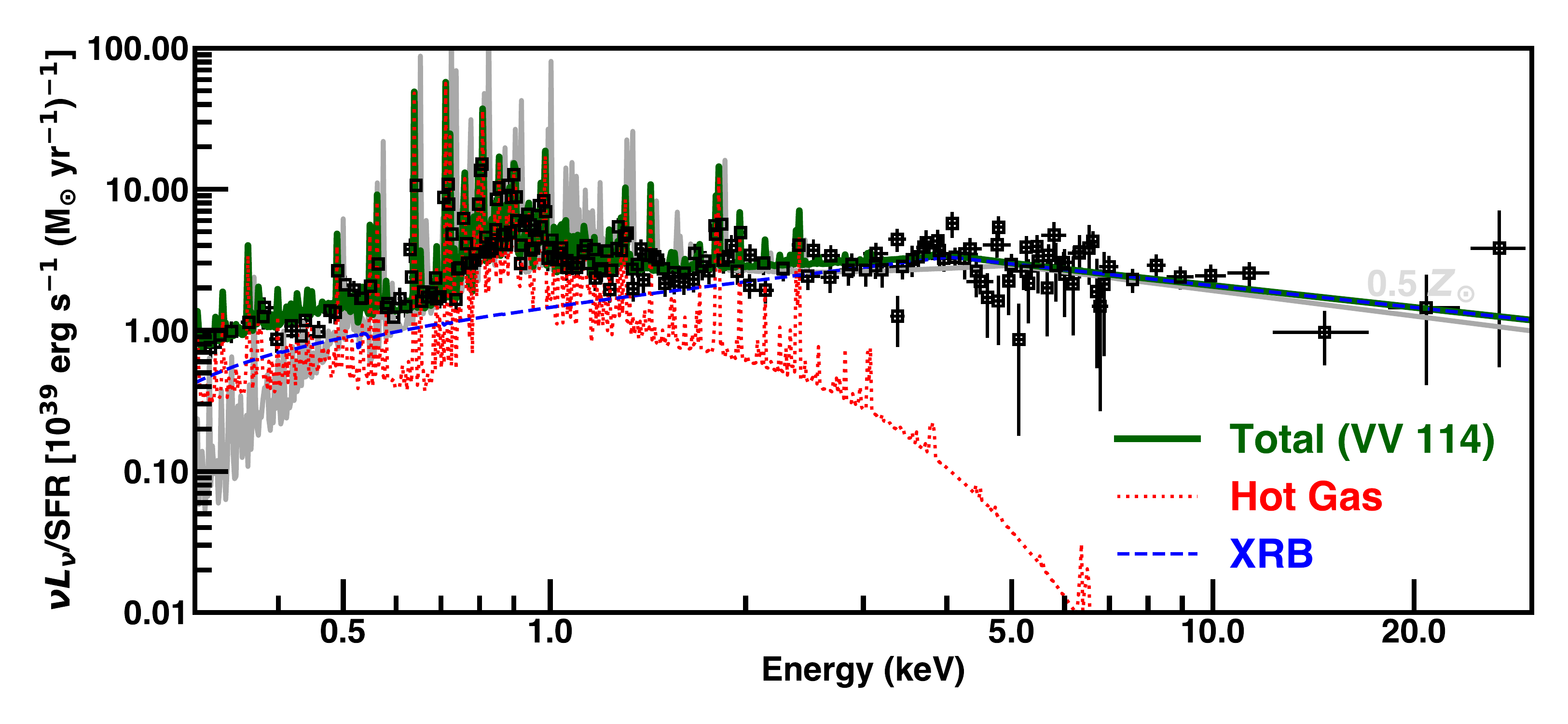}
\caption{The 0.3--30 keV SED of VV~114 (solid green line) from the best-fit model to the joint \xmm\ + \nustar\ spectra, with the major components of the SED included: hot gas component (dotted red line), and the XRB component (dashed blue line). The SED has been normalized by the SFR of VV~114 (38 $M_{\odot}$ yr$^{-1}$). We display the background-subtracted unfolded data points from \xmm\ and \nustar\ binned to a minimum significance of ten per spectral bin to roughly indicate how well the SED is constrained at different energies, although we note that we fit the unbinned spectra without any background subtraction to produce the SED shown in green. In the background as a light grey line we show the simulated SED for a star-forming galaxy at 0.5 $Z_{\odot}$ (see Figure~\ref{fig:sed_comp} for simulated SEDs at other metallicities, and Section~\ref{sec:sed_metal} for details on the construction of simulated SEDs). The SFR-normalized SED of VV~114 is consistent with the simulated SED for a 0.5 $Z_{\odot}$ galaxy, in line with its measured global, gas-phase metallicity of $\sim$ 0.5 $Z_{\odot}$.}\label{fig:sed_unfolded}
\end{figure*}

Finally, in Figure~\ref{fig:sed_unfolded} we present the SFR-normalized, unfolded 0.3--30 keV SED of VV~114 derived from the best-fit {\tt bknpow$_{\rm ULX}$} spectral model to the \xmm\ and \nustar\ spectra, where the solid green line in Figure~\ref{fig:sed_unfolded} represents the total X-ray SED of VV~114, and the dotted red and dashed blue lines represent the contributions from the hot, diffuse gas and XRB population in the galaxy, respectively. On this same plot, we overlay the unfolded data points from the \xmm\ and \nustar\ spectra, which have been binned to a minimum significance of ten per spectral bin and background subtracted for display purposes only. We display the data in this way to give a sense for the uncertainty associated with the galaxy SED, particularly at higher energies where the data become more highly background dominated.

\section{Discussion}\label{sec:discuss}

Previous studies of VV~114 have found an elevated galaxy-integrated \Lx/SFR relative to the average value of local star-forming galaxies \citep{Basu-Zych2013,Basu-Zych2016}. In this section, we discuss the possible explanations for the elevated \Lx/SFR in VV~114, namely the potential presence of an AGN and the sub-solar metallicity of the galaxy, ultimately ruling out a significant contribution from an AGN to the galaxy-integrated \Lx\ in VV~114. We present a further discussion of the SFR-scaled 0.3--30 keV SED of VV~114 in the context of metallicity effects on XRB populations and future 21-cm measurements. 

\subsection{Potential AGN Contribution}\label{sec:agn_contrib}

Previous X-ray analyses of VV~114 have explored the possibility that the galaxy contains an AGN in its heavily obscured eastern region, but have been inconclusive as to the definitive presence of an accreting supermassive BH \citep[e.g.,][]{Grimes2006,Basu-Zych2016}. Multiwavelength analysis beyond X-rays offers evidence in favor of the AGN interpretation for VV~114 X-1. Using ALMA, \citet{Iono2013} and \citet{Saito2015} showed evidence of compact and broad molecular emission component coincident with the eastern nucleus of VV~114, where the high detected HCN (3--4)/ HCO$^{+}$ (4--3) ratio is indicative of the presence of a dust-enshrouded AGN, possibly surrounded by compact star-forming regions. In the mid-IR, \citet{LeFloch2002} demonstrated that the strong continuum emission in the eastern portion of the galaxy may be indicative of a heavily obscured AGN. However, \citet{Basu-Zych2013} used optical line ratio diagnostics to demonstrate that VV~114 lies squarely in the star-forming region of the Baldwin, Phillips, and Terlevich diagram \citep{BPT}, indicating that an AGN is unlikely to be the {\it dominant} source of ionizing photons in VV~114. 

Using the same \chandra\ data as analyzed in this work, \citet{Grimes2006} performed a spectral fit to VV~114E (our source VV~114 X-1), finding that their fits required the addition of Gaussian line components centered at 1.39 and 1.83 keV superposed on the continuum emission, roughly line energies corresponding to enhanced Mg and Si, respectively. These authors note that such emission lines may be associated with the presence of a low-luminosity AGN, but are also consistent with emission features expected from a region of intense star formation. 

Whereas the \citet{Grimes2006} spectral model for VV~114E (our source VV~114 X-1) consists of a single thermal plasma, Gaussian lines, and an absorbed power-law, our model for this source does not require additional Gaussian lines at 1.39 and 1.83 keV to produce an acceptable fit to the source. Rather, our model consists of a two-temperature thermal plasma plus absorbed power-law as summarized in the second row of Table~\ref{tab:point_src_fits}. We find that adding the Gaussian lines at the energies included in \citet{Grimes2006} is degenerate with the features of our two-temperature thermal plasma with $kT = 0.36$ keV and $kT = 0.80$ keV. Furthermore, under the assumption that the hot gas component in VV 114 is associated with starburst activity, our model is strongly motivated by previous studies which have found that the hot gas component in star-forming galaxies is well-described by a two-temperature thermal plasma with characteristic temperatures $<$ 1 keV \citep[e.g.,][]{Strickland2004,Ott2005I,Ott2005II,Grimes2005,Tull2006I,Tull2006II,MineoGas,Smith2018}. Despite these model differences for the hot gas component, we find a photon index and column density for the power-law component of our model for VV~114 X-1 that is consistent with the values found by \citet{Grimes2006} for VV~114E using the same \chandra\ data. 

These fit results, both from this work as recorded in Table~\ref{tab:point_src_fits} and from \citet{Grimes2006} for the same \chandra\ data, demonstrate that VV~114 X-1 {\it is} unique relative to the other five point sources resolved by \chandra. In particular, VV~114 X-1 is consistent with a power-law spectrum with $\Gamma \sim 1.0$, while the point sources present in the western portion of the galaxy are consistent with power-law models with $\Gamma > 1.5$ . This harder spectrum is to be expected given that VV~114 X-1 sits behind a much higher column density than any sources in the western region of the galaxy (see Figure~\ref{fig:hst_rgb}). The photon index returned from our fit to the \chandra\ spectrum of VV~114 X-1 ($\Gamma = 1.01^{+0.59}_{-0.24}$) differs from expectations for a population of HMXBs or a ULX \citep[e.g.,][]{Remillard2006,Berghea2008,Gladstone2009}; however, within the upper range of the 90\% confidence interval on the best-fit value the photon index is consistent with the power-law slope for a population of more heavily obscured XRBs or perhaps a single ULX \citep[e.g.,][]{Lehmer2013}. 

Photon indices in the range $\Gamma \sim 1$, such as the best-fit value for VV~114 X-1, have been measured both for a subset of ULXs with \Lx $\gtrsim$ 10$^{40}$ erg s$^{-1}$, possibly indicative of ULXs in the power-law dominated very high state \citep[e.g.,][]{Berghea2008}, as well as for some Compton-thick AGN \citep[e.g.,][]{Winter2008}. We cannot distinguish between these possibilities based on the \chandra\ data alone, though our measured column density for VV~114 X-1 does not support the interpretation of this source as a Compton-thick AGN. We note that \citet{Prestwich2015} find a similarly hard spectrum for the highly luminous source Haro 11 X-1 ($\Gamma = 1.2$), a source which they report is consistent with being a single compact accretor. It is possible that the $\Gamma \sim 1$ photon index measured for VV~114 X-1 is a function of the limited data quality, where we are not sensitive to features such as a spectral turnover, which would point more definitively to a ULX versus AGN interpretation. In any case, in the absence of higher quality spectra or long-term monitoring to detect possible state transitions, we cannot distinguish between an AGN or ULX for VV~114 X-1 on the basis of the \chandra\ data alone. 

Although we cannot definitively determine the nature of VV~114 X-1, it is important to note that our spectral analysis using the newly obtained \xmm\ and \nustar\ data indicates that the galaxy-wide emission of VV~114 is not dominated by an AGN at energies $\gtrsim$ 2 keV. For an AGN-dominated galaxy we would expect a spectrum well-fit by a simple power-law \citep[e.g.,][]{Winter2008,Winter2009}. We find that the spectral fits to the joint \xmm\ and \nustar\ observations favor a broken power-law model with $E_{\rm break} \sim 4$~keV, and that the inclusion of a $\Gamma$ $\sim$ 1 power-law model component (i.e., VV~114 X-1) extended to higher energies is inconsistent with the data (see Figure~\ref{fig:xmmnu_spec_resid}). While AGN may exhibit spectral turnover features, the break energies typically occur at $\gtrsim$ 50 keV \citep[e.g.,][]{Molina2006,Winter2009}. The break energy at $\sim$ 4~keV found from our model is consistent with the spectral behavior measured from high-quality ULX spectra, indicative of disk and Comptonized corona components around accreting stellar mass compact objects \citep{Gladstone2009}.

Thus, the X-ray spectral analysis presented here demonstrates that VV~114 X-1 does not dominate the global 0.3--30 keV emission of VV~114, and that in fact the galaxy-integrated emission is dominated by emission from ULXs. Notably, this finding is corroborated by previous X-ray studies of VV 114: \citet{Grimes2006} showed that if VV~114 X-1 is an AGN it does not dominate the global emission of VV~114, and similarly \citet{Basu-Zych2016} concluded that the removal of VV~114 X-1 from the XLF results in a luminosity distribution consistent with a collection of blended HMXBs drawn from a ``standard" HMXB XLF. Both of these studies thus conclude that the galaxy-integrated X-ray emission from VV~114 is consistent with expectations for a galaxy with $\gtrsim$ 2 keV emission dominated by XRBs. In the following sections, we therefore discuss our results for VV~114 assuming the $\gtrsim$ 2 keV emission is dominated by such sources.  

\subsection{Comparison with SEDs in other Star-Forming Galaxies}\label{sec:sed_comp}

\begin{figure*}
\centering
\includegraphics[width=\textwidth,trim=0 0 0 0, clip]{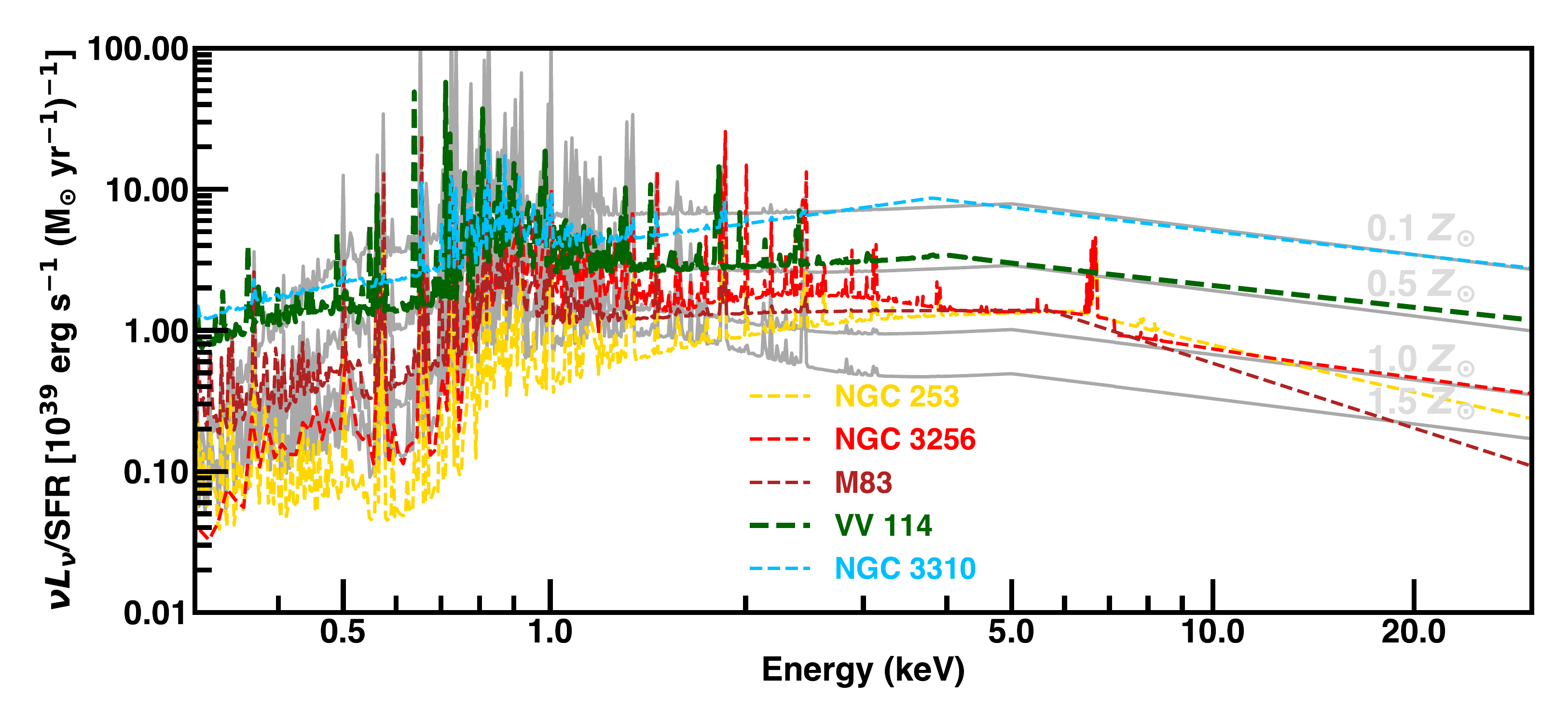}
\caption{The SFR-normalized 0.3--30 keV SED of VV~114 (green), based on the best-fit spectral model to the nearly-simultaneous \xmm\ and \nustar\ data. The light grey lines in the background are simulated SEDs for a star-forming galaxy at 1.5 $Z_{\odot}$, $Z_{\odot}$, 0.5 $Z_{\odot}$, and 0.1 $Z_{\odot}$, annotated going from low to high normalization (see Section~\ref{sec:sed_metal} for details). We display the observed SFR-normalized SEDs of four other star-forming galaxies at different metallicities for comparison: NGC 253 (yellow, $\sim$ 1.1 $Z_{\odot}$; \citealt{Zaritsky1994}), NGC 3256 (red, $\sim$ 1.5 $Z_{\odot}$; \citealt{MH1999}, \citealt{Trancho2007}), M83 (dark red, $\sim$ 0.96 $Z_{\odot}$; \citealt{Zaritsky1994}), and NGC 3310 (cyan, $\sim$ 0.30 $Z_{\odot}$; \citealt{Eng2008}). The SFR-normalized SED of VV~114 is notably elevated relative to the roughly solar metallicity star-forming galaxies (NGC 253, NGC 3256, and M83), in line with theoretical predictions.}\label{fig:sed_comp}
\end{figure*}

In this section we present the 0.3--30 keV SED of VV~114 relative to the X-ray SEDs from a small sample of star-forming galaxies at different metallicities already in the literature to discuss the effects of metallicity and SFH on the X-ray SED. In Figure~\ref{fig:sed_comp} we show the SFR-normalized SED of VV~114 (green) relative to the SFR-normalized SEDs of NGC 253 (yellow, \citealt{Wik2014}), NGC 3256 (red, \citealt{Lehmer2015}), M 83 (dark red, \citealt{Yukita2016}), and NGC 3310 (blue, \citealt{Lehmer2015}). All the galaxies in Figure~\ref{fig:sed_comp} have a similar overall SED shape, with spectral turnovers at energies between $\approx$ 3--8 keV, indicating that all these galaxies have substantial ULX populations. 

While the basic SED shape indicates the same class of source provides the bulk of the hard (2--30 keV) X-ray emission in all these galaxies, the normalizations for the sub-solar metallicity galaxy SEDs of VV~114 and NGC 3310 are noticeably elevated with respect to the normalizations of the solar to super-solar metallicity galaxies \citep[NGC 253, NGC 3256, and M 83;][]{Zaritsky1994,MH1999,Trancho2007}. In particular, the SED of VV~114 is elevated by a factor of $\sim$ 3--11 relative to the solar metallicity SEDs, a factor consistent with theoretical predictions for the enhancement in the bright XRB population at half solar metallicity from \citet{Fragos2013}. Furthermore, the SED of VV~114 appears elevated in the soft band relative to the solar metallicity galaxies, which, as will be discussed in Section~\ref{sec:sed_metal}, is possibly the result of different ISM conditions at lower metallicity. 

The SED of NGC 3310 likewise appears elevated in the soft band relative to the solar metallicity galaxies, and furthermore displays an even higher normalization than VV~114 at energies $>$ 1 keV (enhancement by a factor of $\sim$ 8--25 relative to solar). This enhancement factor for NGC 3310 is more consistent with theoretical predictions for an XRB population at 0.1 $Z_{\odot}$, as the theoretical simulations suggest a nearly order of magnitude difference between the galaxy-integrated \Lx/SFR between XRB populations at solar and 0.1 $Z_{\odot}$ \citep{Linden2010,Fragos2013}. However, the reported gas phase metallicity for NGC 3310 is closer to 0.3 $Z_{\odot}$ \citep{dG2003,Eng2008}, which is only slightly lower than the half solar value reported for VV~114. It is possible that the simulated SEDs based on theoretical predictions presented in Figure~\ref{fig:sed_comp}, although simplified, are actually constraining the metallicity of NGC 3310, indicating that the galaxy may have a gas phase metallicity closer to the low end ($\sim$ 0.2 $Z_{\odot}$) of the measured uncertainty range \citep{Eng2008}. 

Alternatively, statistical scatter due to XLF sampling can affect galaxy-integrated XRB \Lx\ and therefore SED normalization \citep[e.g.,][]{Lehmer2019}. Given $M_{\star}$ $\approx$ 9 $\times$ 10$^{9}$ $M_{\odot}$ and SFR $\approx$ 6 $M_{\odot}$ yr$^{-1}$ for NGC 3310 \citep{Lehmer2015}, and $M_{\star}$ $\approx$ 4 $\times$ 10$^{10}$ $M_{\odot}$ and SFR $\approx$ 38 $M_{\odot}$ yr$^{-1}$ for VV 114 \citep{Basu-Zych2016}, we would expect the statistical scatter in XRB \Lx\ for these galaxies to be on the order of 0.3 dex and 0.2 dex, respectively \citep{Lehmer2019}. The SED normalizations of VV 114 and NGC 3310 are elevated with respect to the SEDs of NGC 253, NGC 3256, and M 83 by more than 0.5 dex in all cases, and the measured difference in normalization between NGC 3310 and VV 114 is $\sim$ 0.4 dex. Thus, statistical scatter cannot explain the difference in SED normalization between the sub-solar and solar metallicity galaxies, and it is not likely to be responsible for the difference in normalization between NGC 3310 and VV 114.

Another possible factor affecting the normalization of the SED is the SFH of the galaxy. HMXBs and ULXs represent a snapshot in the evolution of massive stars in binaries, and thus appear at early times ($\lesssim$ 50 Myr) following a burst of star formation and evolve rapidly ($\sim$ Myrs) away from their X-ray bright phase following core-collapse of the secondary donor star. Binary population synthesis models thus predict that the underlying SFH, or age of the stellar population, will affect the integrated \Lx\ from such a population, in addition to the aforementioned metallicity effects. Such models predict a peak in the number of bright XRBs produced at $Z = 0.4 Z_{\odot}$ on timescales 5--10 Myr post-starburst \citep{Linden2010}. In these models the lowest metallicity populations ($Z < 0.1 Z_{\odot}$) produce vastly more HMXBs, and therefore higher \Lx/SFR, than solar metallicity environments, but only on timescales $>$ 10 Myr post-starburst. A number of recent observational studies have corroborated this SFH-dependence for HMXB production using spatially and temporally resolved SFHs in the vicinities of HMXBs \citep[e.g.,][]{Antoniou2010,Antoniou2016,Lehmer2017,Antoniou2019,Garofali2018}. 

As we do not have a detailed SFH of VV~114 or measurements of individual cluster ages, a joint analysis of the effect of both stellar population age and metallicity on the SED is beyond the scope of this work. However, we can make some conjectures as to the differences between the host environment in NGC 3310 and VV~114 using measured star cluster ages in NGC 3310. These cluster ages, derived from SED-fitting of \hst\ data, reveal a peak in the cluster age distribution at $\sim$ 30 Myr post-starburst \citep{dG2003SED,dG2003}, well beyond the most favorable timescale ($<$ 10 Myr post-starburst) for boosted HXMB or ULX production at  $Z = 0.4 Z_{\odot}$ discussed above. On its surface, this would seem to indicate that NGC 3310 does not have a more favorable SFH in terms of XRB production relative to VV~114, and that instead metallicity, perhaps as low as 0.1 $Z_{\odot}$, is the primary effect driving the enhanced SED normalization for NGC 3310. Of course, this analysis is highly simplified, as it assumes simple bursts of star formation when galaxies in fact have much more complex SFHs \citep[e.g.,][]{Eufrasio2017}. In fact, recent work exploring the \Lx-SFR scaling relation for XRBs using sub-galactic regions in NGC 3310 identified stellar population age as the likely dominant factor in driving the excess of XRB emission relative to galaxy-wide scaling relations \citep{An2019}. This highlights the need for further studies exploring {\it both} the age and metallicity effects on XRB production, ideally for a larger sample of galaxies across different metallicities, in order to provide improved empirical constraints for the scaling of XRB \Lx\ with these host galaxy properties.

\subsection{The Effect of Metallicity on the X-ray SED}\label{sec:sed_metal}

\begin{figure*}
\centering
\includegraphics[width=\textwidth,trim=0 0 0 0, clip]{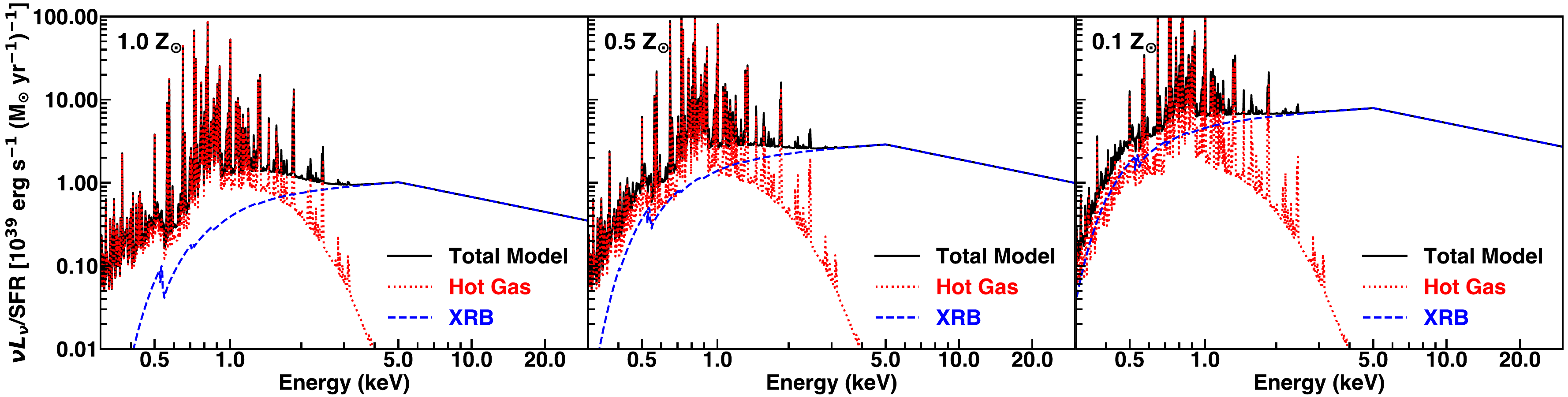}
\caption{Simulated SFR-normalized SEDs for a star-forming galaxy at three different metallicities, from left to right: $Z_{\odot}$, 0.5 $Z_{\odot}$, and 0.1 $Z_{\odot}$. In each panel the solid black line shows the total, galaxy-integrated SED, while the dotted red and dashed blue lines show the hot gas and XRB components, respectively. The $Z_{\odot}$ model is constructed to approximate an average of the SED from star-forming galaxies presented in \citet{Wik2014}, \citet{Lehmer2015}, and \citet{Yukita2016}. The 0.5 $Z_{\odot}$ and 0.1 $Z_{\odot}$ models are constructed by scaling the normalization of the XRB component of the solar metallicity SED in the left-hand panel following the theoretical predictions of \citet{Fragos2013}. The progression of panels from left to right illustrates the theoretical expectation for how the elevation of galaxy-integrated \Lx/SFR due to an enhanced XRB population with decreasing metallicity is reflected in the galaxy-wide SED.}\label{fig:sed_theory}
\end{figure*}

As demonstrated in Section~\ref{sec:sed_comp} for the small sample of nearby, star-forming galaxies, metallicity appears to be a key property affecting the emergent X-ray SED of a galaxy. Likewise, studies of nearby, star-forming galaxies have demonstrated that galaxy-integrated \Lx/SFR increases with decreasing metallicity, an effect which is corroborated by theoretical binary population synthesis work \citep[e.g.,][]{Mapelli2010,Linden2010,Fragos2013,Prestwich2013,Douna2015,Brorby2016,Basu-Zych2016,Wiktor2017,Wiktor2019}. This behavior can be attributed to the effects of metallicity on stellar and binary evolution and thus the characteristics of the resultant XRB population, namely the formation of more massive BHs at lower metallicities given weaker stellar winds \cite[e.g.,][]{Mapelli2010}, the formation of more high accretion rate Roche lobe overflow (RLOF) systems due to the more compact nature of the binaries at lower metallicities \citep[e.g.,][]{Linden2010}, and the wider parameter space leading to survivable common envelope phases and therefore production of RLOF systems at low metallicity \citep[e.g.,][]{Bel2010,Linden2010}. The net effect of metallicity on XRB production is therefore the appearance of not only {\it more} HMXBs with decreasing metallicity, but also possibly the presence of more {\it luminous} HMXBs, leading to the expectation of a population whose XLF has both a higher normalization and a flatter slope. 

While observational and theoretical studies alike suggest that there are more luminous XRBs per unit SFR at lower metallicity, the effect of this enhanced \Lx/SFR on the emergent X-ray SED is not yet constrained empirically. To understand where our newly measured low-metallicity SED for VV~114 fits in with theoretical expectations for the metallicity dependence of XRB populations we must first build up a theoretically-motivated picture of the changes to the X-ray SED with metallicity. 

To simulate X-ray SEDs for star-forming galaxies at different metallicities we begin with a baseline X-ray SED informed by the SED studies of the nearby star-forming galaxies shown in Figure~\ref{fig:sed_comp} from \citet{Wik2014}, \citet{Lehmer2015}, and \citet{Yukita2016}. Our baseline SED is constructed from a {\tt Tbabs*(apec + vphabs*apec + vphabs*bknpow)} model in {\tt XSPEC}. This model choice is empirically motivated: the hot gas component in star-forming galaxies has been shown to be well-fit by two-temperature thermal plasma models (e.g., {\tt apec + vphabs*apec}) across a range of SFRs as described in Section~\ref{sec:point_src_decomp} \citep{Strickland2004,Ott2005I,Ott2005II,Grimes2005,Tull2006I,Tull2006II,Li2013,MineoGas,Smith2018}, and studies of Milky Way XRBs and extragalactic ULXs, the dominant sources of compact emission in star-forming galaxies, show spectra well-described by broken power-laws \citep{McClintock2006,Gladstone2009,Fragos2013other}. In this model we set the column densities (both foreground and intrinsic), thermal plasma temperatures and normalizations, and broken power-law break energy, photon indices and normalization to reproduce the average SED of M83, NGC 3256, NGC 253, and NGC 3310 which represent the best current empirical constraints on the form of the X-ray SED for star-forming galaxies \citep{Wik2014,Lehmer2015,Yukita2016}. 

In what follows, we use this toy model to address how the emergent SED evolves away from the solar metallicity benchmark described above due to changes in gas-phase metallicity; however, our approach is simplified, as we cannot address all the complexities of the effect of metallicity on both the hot gas and XRB emission given the relative lack of observational constraints on the X-ray emission from star-forming galaxies across a range of metallicities. We thus account for metallicity effects on this baseline spectrum in two ways: (1) through the abundances of the {\tt vphabs} components; and (2) through a change to the {\tt bknpow} normalization. Altering the {\tt vphabs} abundances for a chosen metallicity is straightforward, where we use the \citet{Asplund2009} abundance tables in {\tt XSPEC} to set the abundances relative to solar. To account for the increase in galaxy-integrated \Lx/SFR with decreasing metallicity due to XRBs, we scale the {\tt bknpow} normalization from the baseline model by a factor determined from the theoretical scalings of galaxy-integrated XRB \Lx\ with metallicity from \citet{Fragos2013}. We choose to scale the XRB component normalization using theoretical scalings, as such scaling relations provide a physically-motivated estimate of XRB \Lx\ as a function of metallicity that is broadly consistent with empirical constraints. We leave the {\tt APEC} model parameters fixed with changes in metallicity, as we do not yet have strong observational or theoretical constraints to show how the underlying hot gas component varies with metallicity and SFR; however, we do not expect the {\it shape} of the intrinsic hot gas component to vary strongly with stellar mass or SFR of a star-forming galaxy \citep[e.g.,][]{Ott2005I,Grimes2005,Smith2005,MineoGas,An2016,Smith2018,An2019}. 

In the above described model, the {\tt APEC} abundances are thus fixed at solar metallicity for {\it all} simulated SEDs, regardless of assumed gas-phase metallicity. This is in contrast to our fits to the spectra of VV~114, where we set the {\tt APEC} abundances to the measured gas-phase metallicity of the galaxy. In the case of VV~114, we are able to constrain via fits to the observed spectra the characteristics of the hot gas (e.g., temperature and normalization) given the sub-solar abundance. In the case of our simulated SEDs, the {\tt APEC} component temperatures and normalizations are set based on observed constraints from largely solar metallicity galaxies, thus the adopted values in the toy model are appropriate assuming solar metallicity abundances. Because the emission from hot gas in star-forming galaxies may be a complex function of metallicity, we choose not to change the {\tt APEC} abundances in the simulated SEDs in order to keep the shape of the {\it intrinsic} hot gas component fixed as a function of metallicity. We stress that these are simplifying assumptions, meant to produce toy models of the X-ray SED for star-forming galaxies at different metallicities for the purposes of comparison with observed SEDs, as described below. A much larger sample of star-forming galaxies across a range of metallicities would be required to produce a more universal model for the X-ray SED on the basis of host galaxy properties such as metallicity, SFR, and stellar mass.

In Figure~\ref{fig:sed_theory}, we show our simulated X-ray SEDs for star-forming galaxies at $Z_{\odot}$, 0.5 $Z_{\odot}$, and 0.1 $Z_{\odot}$. By construction, the shape of neither the intrinsic XRB nor the intrinsic hot gas component changes in our simulated SEDs from $Z_{\odot}$ to 0.1 $Z_{\odot}$; however, the overall normalization of the SED changes with metallicity. In particular, the flux from the XRB component at 0.1 $Z_{\odot}$ is $\sim$ 10$\times$ higher than at $Z_{\odot}$ in the 0.5--8 keV band. The flux due to hot gas likewise increases from $Z_{\odot}$ to 0.1 $Z_{\odot}$, albeit by a factor of $\sim$ 3$\times$ in the 0.5--8 keV band. The nearly order of magnitude change in the normalization of the XRB component from $Z_{\odot}$ to 0.1 $Z_{\odot}$ is due to the theoretical increase in XRB \Lx\ per unit SFR with decreasing metallicity, while the increase in normalization for the hot gas component at 0.1 $Z_{\odot}$ relative to solar can be ascribed to decreased absorption, particularly of soft band photons, given the sub-solar metallicity assumed for the ISM.

We show the simulated SED at 0.5 $Z_{\odot}$ as a light grey labelled line relative to the observed SED of VV~114 (green line) in Figure~\ref{fig:sed_unfolded}, and similarly display simulated SEDs at 1.5 $Z_{\odot}$, $Z_{\odot}$, 0.5 $Z_{\odot}$, and 0.1 $Z_{\odot}$ relative to other star-forming galaxies in Figure~\ref{fig:sed_comp}. The measured SED of VV~114 becomes ULX-dominated at energies $\gtrsim$ 1.5 keV and is entirely consistent with the simulated SED at 0.5 $Z_{\odot}$ in this energy range. In other words, with the newly measured 0.3--30 keV SED of VV~114 we confirm theoretical predictions for the effect of metallicity on the high energy SED of a star-forming galaxy, namely an elevated normalization relative to solar indicative of an enhanced XRB population at lower metallicity.

By contrast, the soft band (0.5--2 keV) portion of the SED of VV~114 does not agree well with the soft band of the simulated SED at 0.5 $Z_{\odot}$. In particular, the emergent 0.5--2 keV flux of VV~114 is $\sim$ 20$\times$ higher than the flux of our simulated 0.5 $Z_{\odot}$ SED in this same band. This discrepancy is further highlighted when comparing the hot gas versus XRB population contributions to the total soft band emission for VV~114 relative to the simulated SED. From Table~\ref{tab:comp_lx}, we find that $L^{\rm XRB}_{\rm 0.5-2~keV}$ is $\sim$ 40\% of the {\it total} emergent soft band luminosity for VV~114, while it is $\sim$ 20\% of the total for the simulated SED at 0.5 $Z_{\odot}$. Similarly, the soft band portion of the SED of NGC 3310 (blue line, Figure~\ref{fig:sed_comp}) appears elevated relative to the simulated soft band SED at 0.5 $Z_{\odot}$. Recently, \citet{An2019} measured that the hot diffuse component of NGC 3310 constitutes $\sim$ 57\% of the soft band emission in NGC 3310, implying that the XRB component in this galaxy likewise provides a larger share of the total soft band emission relative to expectations from the simulated SED at similar metallicity.

These results imply that the disagreement between the measured and simulated soft band SEDs at low metallicity may stem from incorrect assumptions about ISM properties and, notably, the level of intrinsic absorption, which is most important at energies $\lesssim$ 1.5 keV and likely varies galaxy to galaxy. In the low-metallicity simulated SEDs, the hot gas component is modeled with the simplified assumption that it approximates the hot ISM of a solar metallicity galaxy, where metallicity is only accounted for in the abundances set in the {\tt vphabs}, or intrinsic absorption, component modifying the hot gas model. As noted above, this assumption is made in order to keep the intrinsic hot gas shape constant as a function of metallicity, and because there is a lack of observational constraints in the literature on how detailed ISM abundance patterns affect the hot gas emission. Additionally, the broken power-law normalization representing the XRB contribution to the simulated SED is scaled by the theoretical predictions for the change in XRB \Lx\ with metallicity from \citet{Fragos2013}; however, these theoretical scalings for XRB \Lx\ with metallicity are predicated on X-ray SEDs modeled after Milky Way XRBs, and therefore assume Milky Way-like intrinsic absorption modifying the XRB flux. 

It is quite possible that the ISM properties of a solar metallicity galaxy are different from those of a lower metallicity galaxy such as VV~114 or NGC 3310 at the same SFR. Comparing the measured SED of VV~114 to our toy model for the X-ray SED at 0.5 $Z_{\odot}$ suggests that the assumption of intrinsic absorption measured primarily from solar metallicity galaxies (e.g., M83, NGC 3256, and NGC 253) may be inappropriate for lower metallicity galaxies. This is possibly because more metal-poor galaxies have lower intrinsic column densities, an effect which is manifested most strongly in the soft band portion of the SED. As we do not yet have strong observational constraints on how the ISM properties change as a function of host galaxy properties (e.g., metallicity and SFR) we leave it to a future work to provide more rigorous investigation of the origins of the differences in the soft band SED in star-forming galaxies across different metallicities.

\subsection{Comparison of Galaxy-Integrated Properties with Empirical and Theoretical Scaling Relations}\label{sec:compare_prev}

As discussed in Section~\ref{sec:sed_comp}, the 0.3--30 keV SED of VV~114 displays a clearly elevated normalization relative to the X-ray SEDs of solar metallicity star-forming galaxies. In this section, we focus on comparing our results for the galaxy-integrated \Lx/SFR of VV~114 with results from previous works, as well as expected theoretical and empirical scalings of XRB \Lx/SFR for star-forming galaxies as a function of metallicity.  

As reported in Table~\ref{tab:gal_wide_fits}, we measured a galaxy-integrated total luminosity of log $L^{\rm gal}_{\rm 2-10~keV}$ = 41.27 $\pm$ 0.04 and log $L^{\rm gal}_{\rm 2-10~keV}$ = 41.41 $\pm$ 0.04 from the {\tt bknpow$_{\rm ULX}$} model fit to the \xmm\ and \nustar\ spectra and the {\tt pow$_{\rm XRB}$} model fit to the \chandra\ spectrum, respectively. The only other comparable spectral analysis in the literature for VV~114 was performed by \citet{Grimes2006}, using the same \chandra\ data as presented in this work, in addition to using shallow, archival \xmm\ data (not used here). Their analysis found log $L^{\rm gal}_{\rm 2-10~keV}$ = 41.38 for the galaxy-wide emission from fits to the \chandra\ spectra of the eastern and western components of the galaxy, consistent within the uncertainties of our \chandra-derived total X-ray luminosity (log $L^{\rm gal}_{\rm 2-10~keV}$ = 41.41 $\pm$ 0.04). They do not report a luminosity from their best-fit model to the archival \xmm\ data, so we input the best-fit parameters from their model (Table~5 in \citealt{Grimes2006}) in {\tt XSPEC} and use the {\tt flux} command to derive log~$L^{\rm gal}_{\rm 2-10~keV}$ = 41.23 (corrected for foreground Galactic absorption), consistent with our log~$L^{\rm gal}_{\rm 2-10~keV}$~=~41.27~$\pm$~0.04 for the newly obtained \xmm\ and \nustar\ observations. We note that this agreement is significant, as the luminosity derived from the \citet{Grimes2006} \xmm\ model is based on archival \xmm\ data taken at a different epoch than the \xmm\ and \nustar\ observations presented here. This agreement therefore suggests that variability, in a galaxy-integrated sense, is not significant between the epochs at which the archival and new \xmm\ observations were obtained. 

We next turn to the comparison of XRB \Lx/SFR for VV~114 with recent theoretical and empirical constraints on the scaling of these quantities with metallicity. To get a ``clean" estimate of XRB \Lx\ a thorough accounting of the different contributions, including resolved point sources (ULXs), unresolved XRBs, the hot ISM contribution, and any possible AGN contamination, is important. In this work, we measure XRB \Lx\ from the broken power-law component of the {\tt bknpow$_{\rm ULX}$} model fit to the \xmm\ and \nustar\ spectra. We convert luminosities reported in different energy bands from other works to the 0.5--8 keV band using conversion factors determined from webPIMMs under the assumption of a simple power-law spectrum with $\Gamma$ = 1.7, appropriate for XRBs.

\begin{figure}
\centering
\includegraphics[width=0.5\textwidth,trim=0 0 0 0, clip]{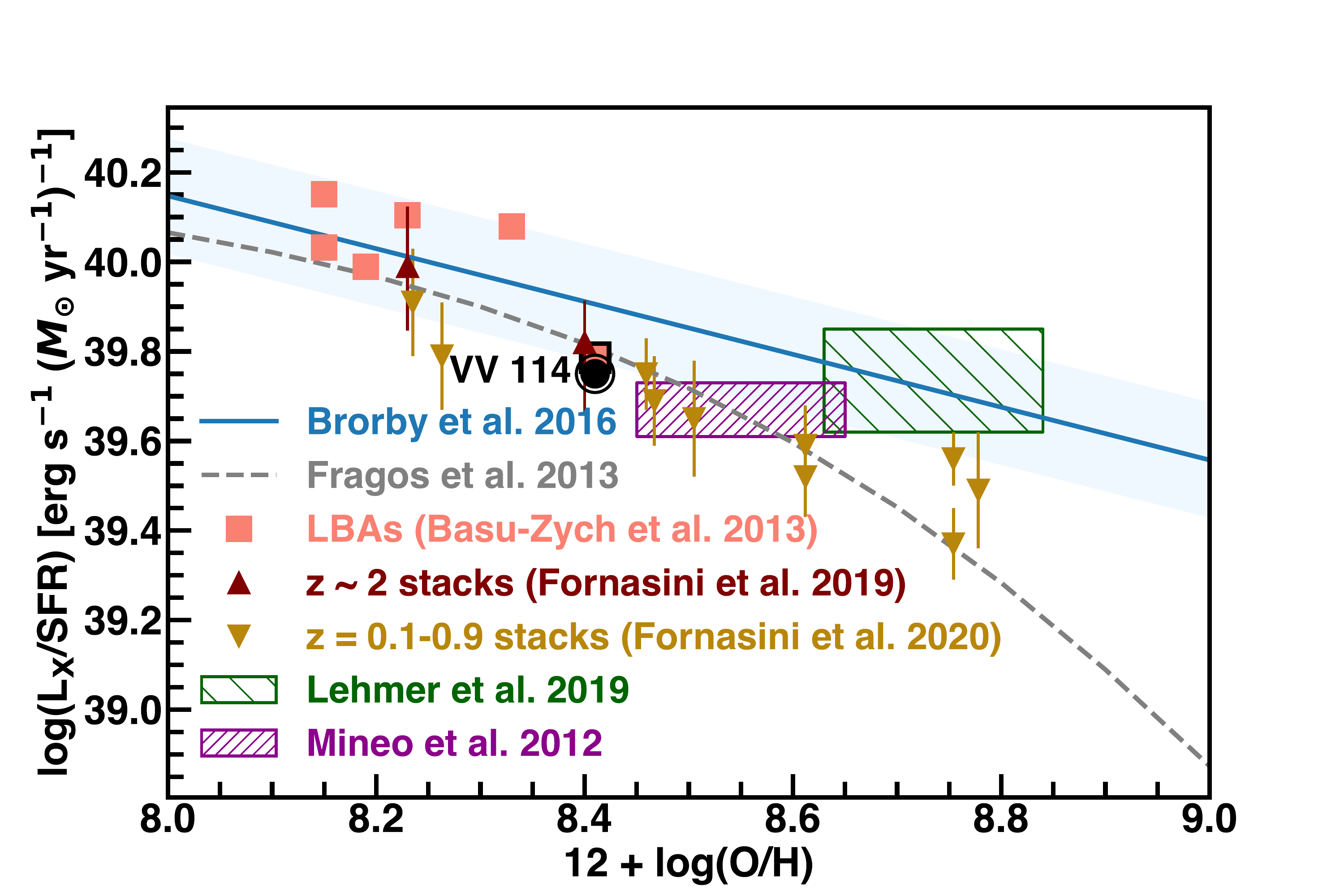}
\caption{Galaxy-integrated XRB \Lx\ (0.5--8 keV) per unit SFR as a function of gas-phase metallicity for a selection of empirical and theoretical studies. We show the results for VV~114 from this work as the black, labelled circle, as well as from a sample of LBAs \citep[salmon squares;][]{BZ13LBG}, $z = 0.1-0.9$ galaxy stacks \citep[yellow triangles;][]{Fornasini2020}, and $z = 2$ galaxy stacks \citep[maroon triangles;][]{Fornasini2019}. The salmon square outlined in black is the value for VV~114 from \citet{BZ13LBG}. We likewise show the canonical \Lx/SFR values derived from XLF-fitting of resolved XRB populations in star-forming galaxies from \citet{Mineo2012} and \citet{Lehmer2019} as the hashed purple and green regions, respectively, where the extent of the regions shows denotes the mean metallicity and standard deviations of the galaxies in each sample. Finally, we show the empirically derived scaling for the \Lx-SFR-$Z$ plane from \citet{Brorby2016} as the solid blue line, with the dispersion in the relation shown in light blue, and the theoretical scaling of galaxy-integrated XRB \Lx/SFR with metallicity from \citet{Fragos2013} as the dashed grey line. The observed galaxy-integrated XRB \Lx/SFR value for VV~114 is consistent with the empirical and theoretical scalings of \Lx/SFR which account for the effects of metallicity on XRB populations.}\label{fig:scale_comp}
\end{figure}

In Figure~\ref{fig:scale_comp} we show the $L^{\rm XRB}_{\rm 0.5-8~keV}$/SFR of VV~114, assuming a SFR of 38 $M_{\odot}$ yr$^{-1}$, (black circle) relative to the best-fit \Lx/SFR in the same band based on fits to the XLF of star-forming galaxies from \citealt{Lehmer2019} (hashed green region) and \citealt{Mineo2012} (hashed purple region). The horizontal extents of each region show the mean metallicity of all galaxies in each sample and the standard deviations, which for the \citet{Lehmer2019} sample come from their Table~1 and for the \citet{Mineo2012} sample are taken from the calculations of \citet{Fornasini2019}. The vertical extents of the hashed regions represent the 1$\sigma$ uncertainties on both scalings. Both the \citet{Mineo2012} and \citet{Lehmer2019} scalings should be taken as appropriate for roughly solar metallicity environments. We find the measured $L^{\rm XRB}_{\rm 0.5-8~keV}$/SFR of VV~114 from the {\tt bknpow$_{\rm ULX}$} model fit to the \xmm\ + \nustar\ spectra is a 1$\sigma$ outlier from the \citet{Mineo2012} relation, but consistent with the \citet{Lehmer2019} relation within the 1$\sigma$ uncertainties on their XLF-derived scaling. 

We next compare our derived \Lx/SFR for VV~114 to theoretical and empirical scalings for the dependence of \Lx/SFR on metallicity from \citet{Fragos2013} and \citet{Brorby2016}, respectively. In Figure~\ref{fig:scale_comp} we show the theoretical \citet{Fragos2013} evolution of \Lx/SFR for XRBs with metallicity from binary population synthesis models as a dashed grey line (absorbed SED model from \citet{Fragos2013}), and the empirical findings of \citet{Brorby2016} based on analysis of the X-ray emission from a sample of Lyman break galaxies as a solid blue line, with the dispersion in their relation shown as light blue. Our measured value for VV~114 is consistent with the \citet{Brorby2016} empirical relation, though this is expected as their \Lx-SFR-$Z$ scaling is derived from a sample of highly star-forming galaxies similar to VV~114. Likewise, VV~114 is consistent with the theoretical predictions from \citet{Fragos2013}, despite the different assumptions for the underlying SED in the \citet{Fragos2013} models, which come from analyzing the spectra of Galactic BH and NS XRBs in different accretion states \citep{Fragos2013other}. 

We also show three observational samples for comparison in Figure~\ref{fig:scale_comp}, namely a selection of LBAs from \citet{BZ13LBG}, the high-sSFR sample of $z = 2$ galaxy stacks binned by metallicity from \citet{Fornasini2019}, and the sample of $z = 0.1-0.9$ galaxy stacks binned by metallicity from \citet{Fornasini2020}, all of which are selected to be HMXB-dominated samples. VV~114 is included in the original \citet{BZ13LBG} sample, so we plot VV~114's XRB \Lx/SFR from \citet{BZ13LBG} as the salmon square with black outline. The value for VV~114 from \citet{BZ13LBG} is taken from galaxy-wide 2--10 keV luminosity from the \citet{Grimes2006} analysis. To make an appropriate comparison, we therefore correct the \citet{BZ13LBG} \Lx\ value for VV~114 to the 0.5--8 keV band, and subtract the contribution from the hot gas component based on the galaxy-wide spectral model presented in \citet{Grimes2006}. With this correction, we find the values from this work and \citet{Grimes2006} and \citet{BZ13LBG} are in good agreement. Notably, all observed samples (VV~114, LBAs, $z = 2$ stacks, and $z = 0.1-0.9$ stacks) are elevated with respect to the scalings derived from XLF fitting for nearby star-forming galaxies \citep{Mineo2012,Lehmer2019}, but are consistent with the theoretical and empirical scalings which account for the metallicity dependence of XRB populations. These results underscore the necessity of accounting for metallicity effects in studies of XRBs in star-forming environments, including future empirical constraints on the dependence of the XRB XLF on metallicity.

\subsection{Soft Band SED and Relevance to IGM Thermal History}\label{sec:gas_igm}

In Section~\ref{sec:sed_comp}, we showed that the normalization of the hard band (2--30 keV) SED of VV~114 agrees with theoretical predictions for enhanced XRB \Lx/SFR at lower metallicity. There are no such theoretical predictions currently in the literature for how the shape or normalization of the soft band (0.5--2 keV) SED scales with metallicity; however, there are past observational works which have investigated the scaling of the hot gas \Lx\ (as measured in the 0.5--2 keV band) in star-forming galaxies with SFR \citep[e.g.,][]{Strickland2004,Ott2005I,Ott2005II,Grimes2005,MineoGas,Smith2018}. In particular, the \citet{MineoGas} study used \chandra\ data for a sample of 21 star-forming galaxies, covering a range of SFRs from 0.1--20 $M_{\odot}$~yr$^{-1}$, to parametrize the linear scaling between $L^{\rm gas}_{\rm X}$ and SFR. Below, we discuss our results for the hot gas \Lx\ of VV~114 derived from our best-fit SED relative to the \citet{MineoGas} empirical scaling, and connect this discussion to the importance of the soft band SED to the thermal history of the IGM. 

From the {\tt APEC} components of the {\tt bknpow$_{\rm ULX}$} model fit to the \nustar\ + \xmm\ spectra (Table~\ref{tab:gal_wide_fits}), we find  $L^{\rm gas}_{\rm X}$(0.5--2 keV) = 1.23 $\times$ 10$^{41}$ erg~s$^{-1}$ for VV~114. By contrast, the linear $L^{\rm gas}_{\rm X}$--SFR scaling derived from the \citet{MineoGas} sample predicts $L^{\rm gas}_{\rm X}$(0.5--2 keV) $\sim$ 3 $\times$ 10$^{40}$ erg~s$^{-1}$ given the SFR for VV~114, a difference of $\sim$ 0.6 dex from our measured hot gas \Lx. It is possible the lower metallicity of VV 14 is what results in this discrepancy, as the \citet{MineoGas} study does not explicitly account for metallicity in deriving the hot gas \Lx--SFR scaling and quotes a dispersion of only $\sigma$ = 0.34 dex in the relation. 

Several previous X-ray studies with \chandra\ have also investigated the scaling of $L^{\rm gas}_{\rm X}$ with SFR across a range of star-forming galaxies, from ULIRGs to dwarf starbursts \citep{Grimes2005,Ott2005I,Ott2005II}. Notably, \citet{Ott2005I,Ott2005II} studied a sample of eight dwarf starbursts using \chandra, of which the majority were low-metallicity ($Z < 0.3 Z_{\odot}$), finding that the diffuse gas \Lx\ was linearly correlated with ``current" (i.e., H$\alpha$-based) SFR. While \citet{Ott2005I, Ott2005II} {\it did} consider the gas-phase metallicity of the dwarf starbursts in their sample in the context of their study, they did not find any explicit correlation between the diffuse gas \Lx\ per unit SFR as a function of metallicity. Two of their lowest metallicity galaxies did not have any detected diffuse component, though they note this may because such emission was below their detection thresholds, while four of their low-metallicity  ($Z <$ 0.3 $Z_{\odot}$)  dwarf starbursts had substantial diffuse emission detected. Interestingly, they found that the diffuse emission in these dwarf starburst galaxies constituted 60--80\% of the 0.3--8 keV photons from the galaxy \citep{Ott2005I}, a much higher percentage contribution from hot gas \Lx\ than predicted from our simulated low-metallicity SEDs presented in Section~\ref{sec:sed_metal}, but in line with the measured percentage contribution from diffuse emission to the soft band for the low-metallicity galaxies VV~114 and NGC 3310 ($\sim$ 60\%, this work and \citealt{An2019}, respectively). 

The scaling of $L^{\rm gas}_{\rm X}$(0.5--2 keV) with SFR and metallicity has implications for the importance of X-ray photons from star-forming galaxies at high redshift, and their effect on the thermal history of the IGM. In particular, the soft band (0.5--2 keV) portion of the X-ray SED is most important for the epoch of heating, prior to the epoch of reionization \citep[e.g.,][]{Pacucci2014,Das2017}. The mean free paths of photons at $z \sim$10--20 approach the Hubble length at energies $\gtrsim$ 2 keV, thus photons with energies $\gtrsim$ 2 keV will effectively ``free stream'' through the IGM during this epoch \citep[e.g.,][]{McQuinn2012}. 

Theoretical studies have shown that the cosmic 21-cm signal, which should be measurable with second generation interferometers such as HERA and SKA, will therefore be sensitive to the shape of the soft band SED of the first galaxies and its scaling with SFR \citep[e.g.,][]{Mesinger2014,Greig2017}. Likewise, theoretical work indicates that the {\it timing} of IGM heating is affected by the shape of the X-ray SED, with ``early" heating which precedes reionization predicted for softer SEDs, and ``late" heating which occurs during reionization predicted for harder SEDs \citep[e.g.,][]{Fialkov2014}. Critically, these theoretical studies often assume that the X-ray photons which heat the IGM are produced solely by XRBs in the first star-forming galaxies, without accounting for the hot gas emission from such galaxies \citep[e.g.,][]{Fialkov2014,Madau2017}. As we have shown here, the hot gas emission can be substantial in the soft band, especially for low-metallicity galaxies. Thus, work such as this, which constrains both the form of the X-ray SED as well as its component parts, is critical to constraining {\it when} significant IGM heating may occur and predicting the 21-cm fluctuations measurable by next generation interferometers.

The best current empirical constraints on the soft band X-ray SED from star-forming galaxies do not account for the host galaxy metallicity and its effect on the hot gas versus XRB contribution to the emergent soft band flux \citep[e.g.,][]{Grimes2005,MineoGas}. Theoretical studies based on these empirical results show that there may be a factor of three difference in the 21-cm power on large scales between assuming hot gas versus XRBs dominate the soft band emission (soft versus hard spectrum, respectively) \citep[e.g.,][]{Pacucci2014}. As we might expect the pristine, low-metallicity galaxies in the early Universe to have different ISM properties than local galaxies, nearby low-metallicity galaxies such as VV~114 serve as better analogs for the first galaxies when it comes to constraining the form of the X-ray SED as it applies to the epoch of heating and the cosmic 21-cm signal. Modeling of the expected 21-cm signal shows that tuning model predictions to constraints based on local star-forming galaxies (ostensibly at solar metallicity) can lead to estimates of 5$\times$ fewer soft photons escaping the galaxy compared to a ``metal-free" ISM, which is more transparent \citep[e.g.,][]{Das2017}. This, in turn, affects the thermal history of the IGM; if the ISM conditions in the early Universe are similar to star-forming galaxies today, this implies fewer soft photons escape galaxies at high redshift, leading to inefficient heating of the IGM which occurs closer to reionization.

We present evidence that VV~114 has a higher than expected elevation of \Lx\ per unit SFR in the soft band relative to other highly star-forming galaxies at solar metallicity \citep[e.g.,][]{MineoGas,Wik2014,Lehmer2015,Yukita2016}. This may imply more soft photons escape galaxies at low metallicity at high redshift, ultimately leading to larger fluctuations in the IGM temperature, and therefore higher amplitude for the large-scale 21-cm power spectrum \citep{Pacucci2014}. As noted above, the soft band portion of the SED is in general not well calibrated down to low metallicities, but VV~114 offers tantalizing evidence that X-rays from star-forming galaxies may play a critical role in heating the IGM. 

\section{Summary \& Future Work}\label{sec:conclude}

Here we have measured, for the first time, the 0.3--30 keV SED of the low-metallicity, star-forming galaxy VV~114. Through detailed spectral fitting of archival \chandra\ as well as the newly obtained, near simultaneous \xmm\ and \nustar\ observations we showed that the SED of VV~114 has (1) an elevated normalization relative to the X-ray SEDs of solar metallicity galaxies; and (2) a characteristic break at high energies ($\sim$ 4 keV). These SED characteristics are indicative of an enhanced ULX population, which dominates the global X-ray emission from VV~114. Our findings for VV~114 are consistent with theoretical expectations, which predict a factor of at least two enhancement in the galaxy-integrated \Lx\ from XRBs at 0.5 $Z_{\odot}$ relative to production of XRBs at solar metallicity. 

We further show that the X-ray SED has a similar {\it shape} for star-forming galaxies of different metallicities, namely that the SED is ULX-dominated at high energies with a substantial hot gas contribution in the soft band. We also present evidence, for the first time, that VV~114 has an elevated soft band (0.5--2 keV) luminosity relative to predictions for the scaling of diffuse gas emission with SFR from previous empirical studies. This elevated soft band \Lx\ for VV~114 is due possibly to the more pristine ISM conditions in the galaxy given its lower metallicity. This work underlines the importance of broadening the sample of low-metallicity galaxies across a range of SFRs for which there are measured X-ray SEDs, with constraints on the contribution from both hot, diffuse gas, as well as compact sources of emission such as XRBs, to offer the best possible empirical framework for interpreting future high redshift measurements and informing binary population synthesis work. 
 
\section*{Acknowledgements}

We thank the referee for very helpful comments, which greatly improved the quality of the manuscript. K.G. and B.D.L. gratefully acknowledge financial support from NASA grant 80NSSC18K1605.

{\it Facilities:} {\it XMM-Newton} (EPIC-MOS and EPIC-pn), {\it NuSTAR} (FPMA and FPMB), {\it Chandra} (ACIS-S)
\software{Astropy \citep{astropy2013,astropy2018},
Matplotlib \citep{matplotlib}, XSPEC \citep{Arnaud1996}}, MARX \citep{marx2017}, SAOImage DS9 \citep{ds9}, CIAO \citep{ciao}, HEASoft \citep{heasoft2014}
%%%%%%%%%%%%%%%%%%%%%%%%%%%%%%%%%%%%%%%%%%%%%%%%%%


\newpage
\bibliographystyle{aasjournal}

\begin{thebibliography}{}
\expandafter\ifx\csname natexlab\endcsname\relax\def\natexlab#1{#1}\fi
\providecommand{\url}[1]{\href{#1}{#1}}
\providecommand{\dodoi}[1]{doi:~\href{http://doi.org/#1}{\nolinkurl{#1}}}
\providecommand{\doeprint}[1]{\href{http://ascl.net/#1}{\nolinkurl{http://ascl.net/#1}}}
\providecommand{\doarXiv}[1]{\href{https://arxiv.org/abs/#1}{\nolinkurl{https://arxiv.org/abs/#1}}}

\bibitem[{{Aird} {et~al.}(2017){Aird}, {Coil}, \& {Georgakakis}}]{Aird2017}
{Aird}, J., {Coil}, A.~L., \& {Georgakakis}, A. 2017, \mnras, 465, 3390,
  \dodoi{10.1093/mnras/stw2932}

\bibitem[{{Anastasopoulou} {et~al.}(2016){Anastasopoulou}, {Zezas}, {Ballo}, \&
  {Della Ceca}}]{An2016}
{Anastasopoulou}, K., {Zezas}, A., {Ballo}, L., \& {Della Ceca}, R. 2016,
  \mnras, 460, 3570, \dodoi{10.1093/mnras/stw1200}

\bibitem[{{Anastasopoulou} {et~al.}(2019){Anastasopoulou}, {Zezas}, {Gkiokas},
  \& {Kovlakas}}]{An2019}
{Anastasopoulou}, K., {Zezas}, A., {Gkiokas}, V., \& {Kovlakas}, K. 2019,
  \mnras, 483, 711, \dodoi{10.1093/mnras/sty3131}

\bibitem[{{Anders} \& {Grevesse}(1989)}]{Anders1989}
{Anders}, E., \& {Grevesse}, N. 1989, \gca, 53, 197,
  \dodoi{10.1016/0016-7037(89)90286-X}

\bibitem[{{Antoniou} \& {Zezas}(2016)}]{Antoniou2016}
{Antoniou}, V., \& {Zezas}, A. 2016, \mnras, 459, 528,
  \dodoi{10.1093/mnras/stw167}

\bibitem[{{Antoniou} {et~al.}(2010){Antoniou}, {Zezas}, {Hatzidimitriou}, \&
  {Kalogera}}]{Antoniou2010}
{Antoniou}, V., {Zezas}, A., {Hatzidimitriou}, D., \& {Kalogera}, V. 2010,
  \apjl, 716, L140, \dodoi{10.1088/2041-8205/716/2/L140}

\bibitem[{{Antoniou} {et~al.}(2019){Antoniou}, {Zezas}, {Drake}, {Badenes},
  {Haberl}, {Wright}, {Hong}, {Di Stefano}, {Gaetz}, {Long}, {Plucinsky},
  {Sasaki}, {Williams}, {Winkler}, \& {SMC XVP collaboration}}]{Antoniou2019}
{Antoniou}, V., {Zezas}, A., {Drake}, J.~J., {et~al.} 2019, \apj, 887, 20,
  \dodoi{10.3847/1538-4357/ab4a7a}

\bibitem[{{Arnaud}(1996)}]{Arnaud1996}
{Arnaud}, K.~A. 1996, in Astronomical Society of the Pacific Conference Series,
  Vol. 101, Astronomical Data Analysis Software and Systems V, ed. G.~H.
  {Jacoby} \& J.~{Barnes}, 17

\bibitem[{{Asplund} {et~al.}(2009){Asplund}, {Grevesse}, {Sauval}, \&
  {Scott}}]{Asplund2009}
{Asplund}, M., {Grevesse}, N., {Sauval}, A.~J., \& {Scott}, P. 2009, \araa, 47,
  481, \dodoi{10.1146/annurev.astro.46.060407.145222}

\bibitem[{{Astropy Collaboration} {et~al.}(2013){Astropy Collaboration},
  {Robitaille}, {Tollerud}, {Greenfield}, {Droettboom}, {Bray}, {Aldcroft},
  {Davis}, {Ginsburg}, {Price-Whelan}, {Kerzendorf}, {Conley}, {Crighton},
  {Barbary}, {Muna}, {Ferguson}, {Grollier}, {Parikh}, {Nair}, {Unther},
  {Deil}, {Woillez}, {Conseil}, {Kramer}, {Turner}, {Singer}, {Fox}, {Weaver},
  {Zabalza}, {Edwards}, {Azalee Bostroem}, {Burke}, {Casey}, {Crawford},
  {Dencheva}, {Ely}, {Jenness}, {Labrie}, {Lim}, {Pierfederici}, {Pontzen},
  {Ptak}, {Refsdal}, {Servillat}, \& {Streicher}}]{astropy2013}
{Astropy Collaboration}, {Robitaille}, T.~P., {Tollerud}, E.~J., {et~al.} 2013,
  \aap, 558, A33, \dodoi{10.1051/0004-6361/201322068}

\bibitem[{{Astropy Collaboration} {et~al.}(2018){Astropy Collaboration},
  {Price-Whelan}, {Sip{\H o}cz}, {G{\"u}nther}, {Lim}, {Crawford}, {Conseil},
  {Shupe}, {Craig}, {Dencheva}, {Ginsburg}, {VanderPlas}, {Bradley},
  {P{\'e}rez-Su{\'a}rez}, {de Val-Borro}, {Aldcroft}, {Cruz}, {Robitaille},
  {Tollerud}, {Ardelean}, {Babej}, {Bach}, {Bachetti}, {Bakanov}, {Bamford},
  {Barentsen}, {Barmby}, {Baumbach}, {Berry}, {Biscani}, {Boquien}, {Bostroem},
  {Bouma}, {Brammer}, {Bray}, {Breytenbach}, {Buddelmeijer}, {Burke},
  {Calderone}, {Cano Rodr{\'{\i}}guez}, {Cara}, {Cardoso}, {Cheedella},
  {Copin}, {Corrales}, {Crichton}, {D'Avella}, {Deil}, {Depagne}, {Dietrich},
  {Donath}, {Droettboom}, {Earl}, {Erben}, {Fabbro}, {Ferreira}, {Finethy},
  {Fox}, {Garrison}, {Gibbons}, {Goldstein}, {Gommers}, {Greco}, {Greenfield},
  {Groener}, {Grollier}, {Hagen}, {Hirst}, {Homeier}, {Horton}, {Hosseinzadeh},
  {Hu}, {Hunkeler}, {Ivezi{\'c}}, {Jain}, {Jenness}, {Kanarek}, {Kendrew},
  {Kern}, {Kerzendorf}, {Khvalko}, {King}, {Kirkby}, {Kulkarni}, {Kumar},
  {Lee}, {Lenz}, {Littlefair}, {Ma}, {Macleod}, {Mastropietro}, {McCully},
  {Montagnac}, {Morris}, {Mueller}, {Mumford}, {Muna}, {Murphy}, {Nelson},
  {Nguyen}, {Ninan}, {N{\"o}the}, {Ogaz}, {Oh}, {Parejko}, {Parley}, {Pascual},
  {Patil}, {Patil}, {Plunkett}, {Prochaska}, {Rastogi}, {Reddy Janga},
  {Sabater}, {Sakurikar}, {Seifert}, {Sherbert}, {Sherwood-Taylor}, {Shih},
  {Sick}, {Silbiger}, {Singanamalla}, {Singer}, {Sladen}, {Sooley},
  {Sornarajah}, {Streicher}, {Teuben}, {Thomas}, {Tremblay}, {Turner},
  {Terr{\'o}n}, {van Kerkwijk}, {de la Vega}, {Watkins}, {Weaver}, {Whitmore},
  {Woillez}, {Zabalza}, \& {Astropy Contributors}}]{astropy2018}
{Astropy Collaboration}, {Price-Whelan}, A.~M., {Sip{\H o}cz}, B.~M., {et~al.}
  2018, \aj, 156, 123, \dodoi{10.3847/1538-3881/aabc4f}

\bibitem[{{Baldwin} {et~al.}(1981){Baldwin}, {Phillips}, \& {Terlevich}}]{BPT}
{Baldwin}, J.~A., {Phillips}, M.~M., \& {Terlevich}, R. 1981, \pasp, 93, 5,
  \dodoi{10.1086/130766}

\bibitem[{{Bartalucci} {et~al.}(2014){Bartalucci}, {Mazzotta}, {Bourdin}, \&
  {Vikhlinin}}]{Bartalucci2014}
{Bartalucci}, I., {Mazzotta}, P., {Bourdin}, H., \& {Vikhlinin}, A. 2014, \aap,
  566, A25, \dodoi{10.1051/0004-6361/201423443}

\bibitem[{{Basu-Zych} {et~al.}(2016){Basu-Zych}, {Lehmer}, {Fragos},
  {Hornschemeier}, {Yukita}, {Zezas}, \& {Ptak}}]{Basu-Zych2016}
{Basu-Zych}, A.~R., {Lehmer}, B., {Fragos}, T., {et~al.} 2016, \apj, 818, 140,
  \dodoi{10.3847/0004-637X/818/2/140}

\bibitem[{{Basu-Zych} {et~al.}(2009){Basu-Zych}, {Gon{\c{c}}alves}, {Overzier},
  {Law}, {Schiminovich}, {Heckman}, {Martin}, {Wyder}, \& {O'Dowd}}]{BZ2009}
{Basu-Zych}, A.~R., {Gon{\c{c}}alves}, T.~S., {Overzier}, R., {et~al.} 2009,
  \apjl, 699, L118, \dodoi{10.1088/0004-637X/699/2/L118}

\bibitem[{{Basu-Zych} {et~al.}(2013{\natexlab{a}}){Basu-Zych}, {Lehmer},
  {Hornschemeier}, {Bouwens}, {Fragos}, {Oesch}, {Belczynski}, {Brandt},
  {Kalogera}, {Luo}, {Miller}, {Mullaney}, {Tzanavaris}, {Xue}, \&
  {Zezas}}]{BZ13LBG}
{Basu-Zych}, A.~R., {Lehmer}, B.~D., {Hornschemeier}, A.~E., {et~al.}
  2013{\natexlab{a}}, \apj, 762, 45, \dodoi{10.1088/0004-637X/762/1/45}

\bibitem[{{Basu-Zych} {et~al.}(2013{\natexlab{b}}){Basu-Zych}, {Lehmer},
  {Hornschemeier}, {Gon{\c{c}}alves}, {Fragos}, {Heckman}, {Overzier}, {Ptak},
  \& {Schiminovich}}]{Basu-Zych2013}
---. 2013{\natexlab{b}}, \apj, 774, 152, \dodoi{10.1088/0004-637X/774/2/152}

\bibitem[{{Belczynski} {et~al.}(2010){Belczynski}, {Dominik}, {Bulik},
  {O'Shaughnessy}, {Fryer}, \& {Holz}}]{Bel2010}
{Belczynski}, K., {Dominik}, M., {Bulik}, T., {et~al.} 2010, \apjl, 715, L138,
  \dodoi{10.1088/2041-8205/715/2/L138}

\bibitem[{{Berghea} {et~al.}(2008){Berghea}, {Weaver}, {Colbert}, \&
  {Roberts}}]{Berghea2008}
{Berghea}, C.~T., {Weaver}, K.~A., {Colbert}, E.~J.~M., \& {Roberts}, T.~P.
  2008, \apj, 687, 471, \dodoi{10.1086/591722}

\bibitem[{{Boroson} {et~al.}(2011){Boroson}, {Kim}, \&
  {Fabbiano}}]{Boroson2011}
{Boroson}, B., {Kim}, D.-W., \& {Fabbiano}, G. 2011, \apj, 729, 12,
  \dodoi{10.1088/0004-637X/729/1/12}

\bibitem[{{Brorby} {et~al.}(2014){Brorby}, {Kaaret}, \&
  {Prestwich}}]{Brorby2014}
{Brorby}, M., {Kaaret}, P., \& {Prestwich}, A. 2014, \mnras, 441, 2346,
  \dodoi{10.1093/mnras/stu736}

\bibitem[{{Brorby} {et~al.}(2016){Brorby}, {Kaaret}, {Prestwich}, \&
  {Mirabel}}]{Brorby2016}
{Brorby}, M., {Kaaret}, P., {Prestwich}, A., \& {Mirabel}, I.~F. 2016, \mnras,
  457, 4081, \dodoi{10.1093/mnras/stw284}

\bibitem[{{Cash}(1979)}]{Cash1979}
{Cash}, W. 1979, \apj, 228, 939, \dodoi{10.1086/156922}

\bibitem[{{Colbert} {et~al.}(2004){Colbert}, {Heckman}, {Ptak}, {Strickland},
  \& {Weaver}}]{Colbert2004}
{Colbert}, E. J.~M., {Heckman}, T.~M., {Ptak}, A.~F., {Strickland}, D.~K., \&
  {Weaver}, K.~A. 2004, \apj, 602, 231, \dodoi{10.1086/380899}

\bibitem[{{Dahlem} {et~al.}(2000){Dahlem}, {Parmar}, {Oosterbroek}, {Orr},
  {Weaver}, \& {Heckman}}]{Dahlem2000}
{Dahlem}, M., {Parmar}, A., {Oosterbroek}, T., {et~al.} 2000, \apj, 538, 555,
  \dodoi{10.1086/309186}

\bibitem[{{Das} {et~al.}(2017){Das}, {Mesinger}, {Pallottini}, {Ferrara}, \&
  {Wise}}]{Das2017}
{Das}, A., {Mesinger}, A., {Pallottini}, A., {Ferrara}, A., \& {Wise}, J.~H.
  2017, \mnras, 469, 1166, \dodoi{10.1093/mnras/stx943}

\bibitem[{{de Grijs} {et~al.}(2003{\natexlab{a}}){de Grijs}, {Anders},
  {Bastian}, {Lynds}, {Lamers}, \& {O'Neil}}]{dG2003}
{de Grijs}, R., {Anders}, P., {Bastian}, N., {et~al.} 2003{\natexlab{a}},
  \mnras, 343, 1285, \dodoi{10.1046/j.1365-8711.2003.06777.x}

\bibitem[{{de Grijs} {et~al.}(2003{\natexlab{b}}){de Grijs}, {Fritze-v.
  Alvensleben}, {Anders}, {Gallagher}, {Bastian}, {Taylor}, \&
  {Windhorst}}]{dG2003SED}
{de Grijs}, R., {Fritze-v. Alvensleben}, U., {Anders}, P., {et~al.}
  2003{\natexlab{b}}, \mnras, 342, 259,
  \dodoi{10.1046/j.1365-8711.2003.06536.x}

\bibitem[{{Douna} {et~al.}(2015){Douna}, {Pellizza}, {Mirabel}, \&
  {Pedrosa}}]{Douna2015}
{Douna}, V.~M., {Pellizza}, L.~J., {Mirabel}, I.~F., \& {Pedrosa}, S.~E. 2015,
  \aap, 579, A44, \dodoi{10.1051/0004-6361/201525617}

\bibitem[{{Engelbracht} {et~al.}(2008){Engelbracht}, {Rieke}, {Gordon},
  {Smith}, {Werner}, {Moustakas}, {Willmer}, \& {Vanzi}}]{Eng2008}
{Engelbracht}, C.~W., {Rieke}, G.~H., {Gordon}, K.~D., {et~al.} 2008, \apj,
  678, 804, \dodoi{10.1086/529513}

\bibitem[{{Eufrasio} {et~al.}(2017){Eufrasio}, {Lehmer}, {Zezas}, {Dwek},
  {Arendt}, {Basu-Zych}, {Wiklind}, {Yukita}, {Fragos}, {Hornschemeier},
  {Markwardt}, {Ptak}, \& {Tzanavaris}}]{Eufrasio2017}
{Eufrasio}, R.~T., {Lehmer}, B.~D., {Zezas}, A., {et~al.} 2017, \apj, 851, 10,
  \dodoi{10.3847/1538-4357/aa9569}

\bibitem[{{Fabbiano}(2006)}]{Fabbiano2006}
{Fabbiano}, G. 2006, \araa, 44, 323,
  \dodoi{10.1146/annurev.astro.44.051905.092519}

\bibitem[{{Fialkov} {et~al.}(2014){Fialkov}, {Barkana}, \&
  {Visbal}}]{Fialkov2014}
{Fialkov}, A., {Barkana}, R., \& {Visbal}, E. 2014, \nat, 506, 197,
  \dodoi{10.1038/nature12999}

\bibitem[{{Fialkov} {et~al.}(2017){Fialkov}, {Cohen}, {Barkana}, \&
  {Silk}}]{Fialkov2017}
{Fialkov}, A., {Cohen}, A., {Barkana}, R., \& {Silk}, J. 2017, \mnras, 464,
  3498, \dodoi{10.1093/mnras/stw2540}

\bibitem[{{Fornasini} {et~al.}(2020){Fornasini}, {Civano}, \&
  {Suh}}]{Fornasini2020}
{Fornasini}, F.~M., {Civano}, F., \& {Suh}, H. 2020, \mnras, 495, 771,
  \dodoi{10.1093/mnras/staa1211}

\bibitem[{{Fornasini} {et~al.}(2019){Fornasini}, {Kriek}, {Sanders}, {Shivaei},
  {Civano}, {Reddy}, {Shapley}, {Coil}, {Mobasher}, {Siana}, {Aird}, {Azadi},
  {Freeman}, {Leung}, {Price}, {Fetherolf}, {Zick}, \& {Barro}}]{Fornasini2019}
{Fornasini}, F.~M., {Kriek}, M., {Sanders}, R.~L., {et~al.} 2019, \apj, 885,
  65, \dodoi{10.3847/1538-4357/ab4653}

\bibitem[{{Fragos} {et~al.}(2013{\natexlab{a}}){Fragos}, {Lehmer}, {Naoz},
  {Zezas}, \& {Basu-Zych}}]{Fragos2013}
{Fragos}, T., {Lehmer}, B.~D., {Naoz}, S., {Zezas}, A., \& {Basu-Zych}, A.
  2013{\natexlab{a}}, \apjl, 776, L31, \dodoi{10.1088/2041-8205/776/2/L31}

\bibitem[{{Fragos} {et~al.}(2013{\natexlab{b}}){Fragos}, {Lehmer}, {Tremmel},
  {Tzanavaris}, {Basu-Zych}, {Belczynski}, {Hornschemeier}, {Jenkins},
  {Kalogera}, {Ptak}, \& {Zezas}}]{Fragos2013other}
{Fragos}, T., {Lehmer}, B., {Tremmel}, M., {et~al.} 2013{\natexlab{b}}, \apj,
  764, 41, \dodoi{10.1088/0004-637X/764/1/41}

\bibitem[{{Fruscione} {et~al.}(2006){Fruscione}, {McDowell}, {Allen},
  {Brickhouse}, {Burke}, {Davis}, {Durham}, {Elvis}, {Galle}, {Harris},
  {Huenemoerder}, {Houck}, {Ishibashi}, {Karovska}, {Nicastro}, {Noble},
  {Nowak}, {Primini}, {Siemiginowska}, {Smith}, \& {Wise}}]{ciao}
{Fruscione}, A., {McDowell}, J.~C., {Allen}, G.~E., {et~al.} 2006, in
  \procspie, Vol. 6270, Society of Photo-Optical Instrumentation Engineers
  (SPIE) Conference Series, 62701V, \dodoi{10.1117/12.671760}

\bibitem[{{Garofali} {et~al.}(2018){Garofali}, {Williams}, {Hillis}, {Gilbert},
  {Dolphin}, {Eracleous}, \& {Binder}}]{Garofali2018}
{Garofali}, K., {Williams}, B.~F., {Hillis}, T., {et~al.} 2018, \mnras, 479,
  3526, \dodoi{10.1093/mnras/sty1612}

\bibitem[{{Garofali} {et~al.}(2017){Garofali}, {Williams}, {Plucinsky},
  {Gaetz}, {Wold}, {Haberl}, {Long}, {Blair}, {Pannuti}, {Winkler}, \&
  {Gross}}]{Garofali2017}
{Garofali}, K., {Williams}, B.~F., {Plucinsky}, P.~P., {et~al.} 2017, \mnras,
  472, 308, \dodoi{10.1093/mnras/stx1905}

\bibitem[{{Gilfanov}(2004)}]{Gilfanov2004}
{Gilfanov}, M. 2004, \mnras, 349, 146, \dodoi{10.1111/j.1365-2966.2004.07473.x}

\bibitem[{{Gladstone} {et~al.}(2009){Gladstone}, {Roberts}, \&
  {Done}}]{Gladstone2009}
{Gladstone}, J.~C., {Roberts}, T.~P., \& {Done}, C. 2009, \mnras, 397, 1836,
  \dodoi{10.1111/j.1365-2966.2009.15123.x}

\bibitem[{{Greig} \& {Mesinger}(2017)}]{Greig2017}
{Greig}, B., \& {Mesinger}, A. 2017, \mnras, 472, 2651,
  \dodoi{10.1093/mnras/stx2118}

\bibitem[{{Grimes} {et~al.}(2006){Grimes}, {Heckman}, {Hoopes}, {Strickland},
  {Aloisi}, {Meurer}, \& {Ptak}}]{Grimes2006}
{Grimes}, J.~P., {Heckman}, T., {Hoopes}, C., {et~al.} 2006, \apj, 648, 310,
  \dodoi{10.1086/505680}

\bibitem[{{Grimes} {et~al.}(2005){Grimes}, {Heckman}, {Strickland}, \&
  {Ptak}}]{Grimes2005}
{Grimes}, J.~P., {Heckman}, T., {Strickland}, D., \& {Ptak}, A. 2005, \apj,
  628, 187, \dodoi{10.1086/430692}

\bibitem[{{Grimm} {et~al.}(2003){Grimm}, {Gilfanov}, \& {Sunyaev}}]{Grimm2003}
{Grimm}, H.~J., {Gilfanov}, M., \& {Sunyaev}, R. 2003, \mnras, 339, 793,
  \dodoi{10.1046/j.1365-8711.2003.06224.x}

\bibitem[{{G{\"u}nther} {et~al.}(2017){G{\"u}nther}, {Frost}, \&
  {Theriault-Shay}}]{marx2017}
{G{\"u}nther}, H.~M., {Frost}, J., \& {Theriault-Shay}, A. 2017, \aj, 154, 243,
  \dodoi{10.3847/1538-3881/aa943b}

\bibitem[{{Harrison} {et~al.}(2013){Harrison}, {Craig}, {Christensen},
  {Hailey}, {Zhang}, {Boggs}, {Stern}, {Cook}, {Forster}, {Giommi},
  {Grefenstette}, {Kim}, {Kitaguchi}, {Koglin}, {Madsen}, {Mao}, {Miyasaka},
  {Mori}, {Perri}, {Pivovaroff}, {Puccetti}, {Rana}, {Westergaard}, {Willis},
  {Zoglauer}, {An}, {Bachetti}, {Barri{\`e}re}, {Bellm}, {Bhalerao},
  {Brejnholt}, {Fuerst}, {Liebe}, {Markwardt}, {Nynka}, {Vogel}, {Walton},
  {Wik}, {Alexander}, {Cominsky}, {Hornschemeier}, {Hornstrup}, {Kaspi},
  {Madejski}, {Matt}, {Molendi}, {Smith}, {Tomsick}, {Ajello}, {Ballantyne},
  {Balokovi{\'c}}, {Barret}, {Bauer}, {Blandford}, {Brandt}, {Brenneman},
  {Chiang}, {Chakrabarty}, {Chenevez}, {Comastri}, {Dufour}, {Elvis}, {Fabian},
  {Farrah}, {Fryer}, {Gotthelf}, {Grindlay}, {Helfand}, {Krivonos}, {Meier},
  {Miller}, {Natalucci}, {Ogle}, {Ofek}, {Ptak}, {Reynolds}, {Rigby},
  {Tagliaferri}, {Thorsett}, {Treister}, \& {Urry}}]{Harrison2013}
{Harrison}, F.~A., {Craig}, W.~W., {Christensen}, F.~E., {et~al.} 2013, \apj,
  770, 103, \dodoi{10.1088/0004-637X/770/2/103}

\bibitem[{{Hartwell} {et~al.}(2004){Hartwell}, {Stevens}, {Strickland},
  {Heckman}, \& {Summers}}]{Hartwell2004}
{Hartwell}, J.~M., {Stevens}, I.~R., {Strickland}, D.~K., {Heckman}, T.~M., \&
  {Summers}, L.~K. 2004, \mnras, 348, 406,
  \dodoi{10.1111/j.1365-2966.2004.07375.x}

\bibitem[{HEASARC(2014)}]{heasoft2014}
HEASARC. 2014, {HEAsoft: Unified Release of FTOOLS and XANADU}.
\newblock \doeprint{1408.004}

\bibitem[{{Heckman} {et~al.}(2005){Heckman}, {Hoopes}, {Seibert}, {Martin},
  {Salim}, {Rich}, {Kauffmann}, {Charlot}, {Barlow}, {Bianchi}, {Byun},
  {Donas}, {Forster}, {Friedman}, {Jelinsky}, {Lee}, {Madore}, {Malina},
  {Milliard}, {Morrissey}, {Neff}, {Schiminovich}, {Siegmund}, {Small},
  {Szalay}, {Welsh}, \& {Wyder}}]{Heckman2005}
{Heckman}, T.~M., {Hoopes}, C.~G., {Seibert}, M., {et~al.} 2005, \apjl, 619,
  L35, \dodoi{10.1086/425979}

\bibitem[{{HI4PI Collaboration} {et~al.}(2016){HI4PI Collaboration}, {Ben
  Bekhti}, {Fl{\"o}er}, {Keller}, {Kerp}, {Lenz}, {Winkel}, {Bailin},
  {Calabretta}, {Dedes}, {Ford}, {Gibson}, {Haud}, {Janowiecki}, {Kalberla},
  {Lockman}, {McClure-Griffiths}, {Murphy}, {Nakanishi}, {Pisano}, \&
  {Staveley-Smith}}]{nHpimms}
{HI4PI Collaboration}, {Ben Bekhti}, N., {Fl{\"o}er}, L., {et~al.} 2016, \aap,
  594, A116, \dodoi{10.1051/0004-6361/201629178}

\bibitem[{{Hoopes} {et~al.}(2007){Hoopes}, {Heckman}, {Salim}, {Seibert},
  {Tremonti}, {Schiminovich}, {Rich}, {Martin}, {Charlot}, {Kauffmann},
  {Forster}, {Friedman}, {Morrissey}, {Neff}, {Small}, {Wyder}, {Bianchi},
  {Donas}, {Lee}, {Madore}, {Milliard}, {Szalay}, {Welsh}, \&
  {Yi}}]{Hoopes2007}
{Hoopes}, C.~G., {Heckman}, T.~M., {Salim}, S., {et~al.} 2007, \apjs, 173, 441,
  \dodoi{10.1086/516644}

\bibitem[{Hunter(2007)}]{matplotlib}
Hunter, J.~D. 2007, Computing In Science \& Engineering, 9, 90,
  \dodoi{10.1109/MCSE.2007.55}

\bibitem[{{Iono} {et~al.}(2013){Iono}, {Saito}, {Yun}, {Kawabe}, {Espada},
  {Hagiwara}, {Imanishi}, {Izumi}, {Kohno}, {Motohara}, {Nakanishi}, {Sugai},
  {Tateuchi}, {Tamura}, {Ueda}, \& {Yoshii}}]{Iono2013}
{Iono}, D., {Saito}, T., {Yun}, M.~S., {et~al.} 2013, \pasj, 65, L7,
  \dodoi{10.1093/pasj/65.3.L7}

\bibitem[{{Joye} \& {Mandel}(2003)}]{ds9}
{Joye}, W.~A., \& {Mandel}, E. 2003, in Astronomical Society of the Pacific
  Conference Series, Vol. 295, Astronomical Data Analysis Software and Systems
  XII, ed. H.~E. {Payne}, R.~I. {Jedrzejewski}, \& R.~N. {Hook}, 489

\bibitem[{{Kouroumpatzakis} {et~al.}(2020){Kouroumpatzakis}, {Zezas}, {Sell},
  {Kovlakas}, {Bonfini}, {Willner}, {Ashby}, {Maragkoudakis}, \&
  {Jarrett}}]{Kou2020}
{Kouroumpatzakis}, K., {Zezas}, A., {Sell}, P., {et~al.} 2020, \mnras, 494,
  5967, \dodoi{10.1093/mnras/staa1063}

\bibitem[{{Kroupa}(2001)}]{Kroupa}
{Kroupa}, P. 2001, \mnras, 322, 231, \dodoi{10.1046/j.1365-8711.2001.04022.x}

\bibitem[{{Kuntz} \& {Snowden}(2000)}]{Kuntz2000}
{Kuntz}, K.~D., \& {Snowden}, S.~L. 2000, \apj, 543, 195,
  \dodoi{10.1086/317071}

\bibitem[{{Le Floc'h} {et~al.}(2002){Le Floc'h}, {Charmandaris}, {Laurent},
  {Mirabel}, {Gallais}, {Sauvage}, {Vigroux}, \& {Cesarsky}}]{LeFloch2002}
{Le Floc'h}, E., {Charmandaris}, V., {Laurent}, O., {et~al.} 2002, \aap, 391,
  417, \dodoi{10.1051/0004-6361:20020784}

\bibitem[{{Lehmer} {et~al.}(2010){Lehmer}, {Alexander}, {Bauer}, {Brand t},
  {Goulding}, {Jenkins}, {Ptak}, \& {Roberts}}]{Lehmer2010}
{Lehmer}, B.~D., {Alexander}, D.~M., {Bauer}, F.~E., {et~al.} 2010, \apj, 724,
  559, \dodoi{10.1088/0004-637X/724/1/559}

\bibitem[{{Lehmer} {et~al.}(2013){Lehmer}, {Wik}, {Hornschemeier}, {Ptak},
  {Antoniou}, {Argo}, {Bechtol}, {Boggs}, {Christensen}, {Craig}, {Hailey},
  {Harrison}, {Krivonos}, {Leyder}, {Maccarone}, {Stern}, {Venters}, {Zezas},
  \& {Zhang}}]{Lehmer2013}
{Lehmer}, B.~D., {Wik}, D.~R., {Hornschemeier}, A.~E., {et~al.} 2013, \apj,
  771, 134, \dodoi{10.1088/0004-637X/771/2/134}

\bibitem[{{Lehmer} {et~al.}(2015){Lehmer}, {Tyler}, {Hornschemeier}, {Wik},
  {Yukita}, {Antoniou}, {Boggs}, {Christensen}, {Craig}, {Hailey}, {Harrison},
  {Maccarone}, {Ptak}, {Stern}, {Zezas}, \& {Zhang}}]{Lehmer2015}
{Lehmer}, B.~D., {Tyler}, J.~B., {Hornschemeier}, A.~E., {et~al.} 2015, \apj,
  806, 126, \dodoi{10.1088/0004-637X/806/1/126}

\bibitem[{{Lehmer} {et~al.}(2016){Lehmer}, {Basu-Zych}, {Mineo}, {Brand t},
  {Eufrasio}, {Fragos}, {Hornschemeier}, {Luo}, {Xue}, {Bauer}, {Gilfanov},
  {Ranalli}, {Schneider}, {Shemmer}, {Tozzi}, {Trump}, {Vignali}, {Wang},
  {Yukita}, \& {Zezas}}]{Lehmer2016}
{Lehmer}, B.~D., {Basu-Zych}, A.~R., {Mineo}, S., {et~al.} 2016, \apj, 825, 7,
  \dodoi{10.3847/0004-637X/825/1/7}

\bibitem[{{Lehmer} {et~al.}(2017){Lehmer}, {Eufrasio}, {Markwardt}, {Zezas},
  {Basu-Zych}, {Fragos}, {Hornschemeier}, {Ptak}, {Tzanavaris}, \&
  {Yukita}}]{Lehmer2017}
{Lehmer}, B.~D., {Eufrasio}, R.~T., {Markwardt}, L., {et~al.} 2017, \apj, 851,
  11, \dodoi{10.3847/1538-4357/aa9578}

\bibitem[{{Lehmer} {et~al.}(2019){Lehmer}, {Eufrasio}, {Tzanavaris},
  {Basu-Zych}, {Fragos}, {Prestwich}, {Yukita}, {Zezas}, {Hornschemeier}, \&
  {Ptak}}]{Lehmer2019}
{Lehmer}, B.~D., {Eufrasio}, R.~T., {Tzanavaris}, P., {et~al.} 2019, \apjs,
  243, 3, \dodoi{10.3847/1538-4365/ab22a8}

\bibitem[{{Lehmer} {et~al.}(2020){Lehmer}, {Ferrell}, {Doore}, {Eufrasio},
  {Monson}, {Alexander}, {Basu-Zych}, {Brandt}, {Sivakoff}, {Tzanavaris},
  {Yukita}, {Fragos}, \& {Ptak}}]{Lehmer2020}
{Lehmer}, B.~D., {Ferrell}, A.~P., {Doore}, K., {et~al.} 2020, \apjs, 248, 31,
  \dodoi{10.3847/1538-4365/ab9175}

\bibitem[{{Li} \& {Wang}(2013)}]{Li2013}
{Li}, J.-T., \& {Wang}, Q.~D. 2013, \mnras, 435, 3071,
  \dodoi{10.1093/mnras/stt1501}

\bibitem[{{Linden} {et~al.}(2010){Linden}, {Kalogera}, {Sepinsky}, {Prestwich},
  {Zezas}, \& {Gallagher}}]{Linden2010}
{Linden}, T., {Kalogera}, V., {Sepinsky}, J.~F., {et~al.} 2010, \apj, 725,
  1984, \dodoi{10.1088/0004-637X/725/2/1984}

\bibitem[{{Lumb} {et~al.}(2002){Lumb}, {Warwick}, {Page}, \& {De
  Luca}}]{Lumb2002}
{Lumb}, D.~H., {Warwick}, R.~S., {Page}, M., \& {De Luca}, A. 2002, \aap, 389,
  93, \dodoi{10.1051/0004-6361:20020531}

\bibitem[{{Maccarone} {et~al.}(2016){Maccarone}, {Yukita}, {Hornschemeier},
  {Lehmer}, {Antoniou}, {Ptak}, {Wik}, {Zezas}, {Boyd}, {Kennea}, {Page},
  {Eracleous}, {Williams}, {Boggs}, {Christensen}, {Craig}, {Hailey},
  {Harrison}, {Stern}, \& {Zhang}}]{Maccarone2016}
{Maccarone}, T.~J., {Yukita}, M., {Hornschemeier}, A., {et~al.} 2016, \mnras,
  458, 3633, \dodoi{10.1093/mnras/stw530}

\bibitem[{{Madau} \& {Fragos}(2017)}]{Madau2017}
{Madau}, P., \& {Fragos}, T. 2017, \apj, 840, 39,
  \dodoi{10.3847/1538-4357/aa6af9}

\bibitem[{{Madsen} {et~al.}(2015){Madsen}, {Harrison}, {Markwardt}, {An},
  {Grefenstette}, {Bachetti}, {Miyasaka}, {Kitaguchi}, {Bhalerao}, {Boggs},
  {Christensen}, {Craig}, {Forster}, {Fuerst}, {Hailey}, {Perri}, {Puccetti},
  {Rana}, {Stern}, {Walton}, {J{\o}rgen Westergaard}, \& {Zhang}}]{Madsen2015}
{Madsen}, K.~K., {Harrison}, F.~A., {Markwardt}, C.~B., {et~al.} 2015, \apjs,
  220, 8, \dodoi{10.1088/0067-0049/220/1/8}

\bibitem[{{Mapelli} {et~al.}(2010){Mapelli}, {Ripamonti}, {Zampieri}, {Colpi},
  \& {Bressan}}]{Mapelli2010}
{Mapelli}, M., {Ripamonti}, E., {Zampieri}, L., {Colpi}, M., \& {Bressan}, A.
  2010, \mnras, 408, 234, \dodoi{10.1111/j.1365-2966.2010.17048.x}

\bibitem[{{Martin} {et~al.}(2002){Martin}, {Kobulnicky}, \&
  {Heckman}}]{Martin2002}
{Martin}, C.~L., {Kobulnicky}, H.~A., \& {Heckman}, T.~M. 2002, \apj, 574, 663,
  \dodoi{10.1086/341092}

\bibitem[{{Mas-Hesse} \& {Kunth}(1999)}]{MH1999}
{Mas-Hesse}, J.~M., \& {Kunth}, D. 1999, \aap, 349, 765.
\newblock \doarXiv{astro-ph/9812072}

\bibitem[{{McClintock} \& {Remillard}(2006)}]{McClintock2006}
{McClintock}, J.~E., \& {Remillard}, R.~A. 2006, {Black hole binaries},
  Vol.~39, 157--213

\bibitem[{{McQuinn}(2012)}]{McQuinn2012}
{McQuinn}, M. 2012, \mnras, 426, 1349, \dodoi{10.1111/j.1365-2966.2012.21792.x}

\bibitem[{{Mesinger} {et~al.}(2014){Mesinger}, {Ewall-Wice}, \&
  {Hewitt}}]{Mesinger2014}
{Mesinger}, A., {Ewall-Wice}, A., \& {Hewitt}, J. 2014, \mnras, 439, 3262,
  \dodoi{10.1093/mnras/stu125}

\bibitem[{{Mesinger} {et~al.}(2013){Mesinger}, {Ferrara}, \&
  {Spiegel}}]{Mesinger2013}
{Mesinger}, A., {Ferrara}, A., \& {Spiegel}, D.~S. 2013, \mnras, 431, 621,
  \dodoi{10.1093/mnras/stt198}

\bibitem[{{Mineo} {et~al.}(2012{\natexlab{a}}){Mineo}, {Gilfanov}, \&
  {Sunyaev}}]{Mineo2012}
{Mineo}, S., {Gilfanov}, M., \& {Sunyaev}, R. 2012{\natexlab{a}}, \mnras, 419,
  2095, \dodoi{10.1111/j.1365-2966.2011.19862.x}

\bibitem[{{Mineo} {et~al.}(2012{\natexlab{b}}){Mineo}, {Gilfanov}, \&
  {Sunyaev}}]{MineoGas}
---. 2012{\natexlab{b}}, \mnras, 426, 1870,
  \dodoi{10.1111/j.1365-2966.2012.21831.x}

\bibitem[{{Mirocha}(2014)}]{Mirocha2014}
{Mirocha}, J. 2014, \mnras, 443, 1211, \dodoi{10.1093/mnras/stu1193}

\bibitem[{{Molina} {et~al.}(2006){Molina}, {Malizia}, {Bassani}, {Bird},
  {Dean}, {Landi}, {de Rosa}, {Walter}, {Barlow}, {Clark}, {Hill}, \&
  {Sguera}}]{Molina2006}
{Molina}, M., {Malizia}, A., {Bassani}, L., {et~al.} 2006, \mnras, 371, 821,
  \dodoi{10.1111/j.1365-2966.2006.10715.x}

\bibitem[{{Moustakas} \& {Kennicutt}(2006)}]{Moustakas2006}
{Moustakas}, J., \& {Kennicutt}, Robert~C., J. 2006, \apjs, 164, 81,
  \dodoi{10.1086/500971}

\bibitem[{{Ott} {et~al.}(2005{\natexlab{a}}){Ott}, {Walter}, \&
  {Brinks}}]{Ott2005I}
{Ott}, J., {Walter}, F., \& {Brinks}, E. 2005{\natexlab{a}}, \mnras, 358, 1423,
  \dodoi{10.1111/j.1365-2966.2005.08862.x}

\bibitem[{{Ott} {et~al.}(2005{\natexlab{b}}){Ott}, {Walter}, \&
  {Brinks}}]{Ott2005II}
---. 2005{\natexlab{b}}, \mnras, 358, 1453,
  \dodoi{10.1111/j.1365-2966.2005.08863.x}

\bibitem[{{Pacucci} {et~al.}(2014){Pacucci}, {Mesinger}, {Mineo}, \&
  {Ferrara}}]{Pacucci2014}
{Pacucci}, F., {Mesinger}, A., {Mineo}, S., \& {Ferrara}, A. 2014, \mnras, 443,
  678, \dodoi{10.1093/mnras/stu1240}

\bibitem[{{Park} {et~al.}(2019){Park}, {Mesinger}, {Greig}, \&
  {Gillet}}]{Park2019}
{Park}, J., {Mesinger}, A., {Greig}, B., \& {Gillet}, N. 2019, \mnras, 484,
  933, \dodoi{10.1093/mnras/stz032}

\bibitem[{{Pettini} \& {Pagel}(2004)}]{PP04}
{Pettini}, M., \& {Pagel}, B. E.~J. 2004, \mnras, 348, L59,
  \dodoi{10.1111/j.1365-2966.2004.07591.x}

\bibitem[{{Prestwich} {et~al.}(2015){Prestwich}, {Jackson}, {Kaaret}, {Brorby},
  {Roberts}, {Saar}, \& {Yukita}}]{Prestwich2015}
{Prestwich}, A.~H., {Jackson}, F., {Kaaret}, P., {et~al.} 2015, \apj, 812, 166,
  \dodoi{10.1088/0004-637X/812/2/166}

\bibitem[{{Prestwich} {et~al.}(2013){Prestwich}, {Tsantaki}, {Zezas},
  {Jackson}, {Roberts}, {Foltz}, {Linden}, \& {Kalogera}}]{Prestwich2013}
{Prestwich}, A.~H., {Tsantaki}, M., {Zezas}, A., {et~al.} 2013, \apj, 769, 92,
  \dodoi{10.1088/0004-637X/769/2/92}

\bibitem[{{Rana} {et~al.}(2015){Rana}, {Harrison}, {Bachetti}, {Walton},
  {Furst}, {Barret}, {Miller}, {Fabian}, {Boggs}, {Christensen}, {Craig},
  {Grefenstette}, {Hailey}, {Madsen}, {Ptak}, {Stern}, {Webb}, \&
  {Zhang}}]{Rana2015}
{Rana}, V., {Harrison}, F.~A., {Bachetti}, M., {et~al.} 2015, \apj, 799, 121,
  \dodoi{10.1088/0004-637X/799/2/121}

\bibitem[{{Remillard} \& {McClintock}(2006)}]{Remillard2006}
{Remillard}, R.~A., \& {McClintock}, J.~E. 2006, \araa, 44, 49,
  \dodoi{10.1146/annurev.astro.44.051905.092532}

\bibitem[{{Saito} {et~al.}(2015){Saito}, {Iono}, {Yun}, {Ueda}, {Nakanishi},
  {Sugai}, {Espada}, {Imanishi}, {Motohara}, {Hagiwara}, {Tateuchi}, {Lee}, \&
  {Kawabe}}]{Saito2015}
{Saito}, T., {Iono}, D., {Yun}, M.~S., {et~al.} 2015, \apj, 803, 60,
  \dodoi{10.1088/0004-637X/803/2/60}

\bibitem[{{Smith} {et~al.}(2018){Smith}, {Campbell}, {Struck}, {Soria},
  {Swartz}, {Magno}, {Dunn}, \& {Giroux}}]{Smith2018}
{Smith}, B.~J., {Campbell}, K., {Struck}, C., {et~al.} 2018, \aj, 155, 81,
  \dodoi{10.3847/1538-3881/aaa1a6}

\bibitem[{{Smith} {et~al.}(2005){Smith}, {Struck}, \& {Nowak}}]{Smith2005}
{Smith}, B.~J., {Struck}, C., \& {Nowak}, M.~A. 2005, \aj, 129, 1350,
  \dodoi{10.1086/427858}

\bibitem[{{Strickland} {et~al.}(2004){Strickland}, {Heckman}, {Colbert},
  {Hoopes}, \& {Weaver}}]{Strickland2004}
{Strickland}, D.~K., {Heckman}, T.~M., {Colbert}, E. J.~M., {Hoopes}, C.~G., \&
  {Weaver}, K.~A. 2004, \apjs, 151, 193, \dodoi{10.1086/382214}

\bibitem[{{Str{\"u}der} {et~al.}(2001){Str{\"u}der}, {Briel}, {Dennerl},
  {Hartmann}, {Kendziorra}, {Meidinger}, {Pfeffermann}, {Reppin}, {Aschenbach},
  {Bornemann}, {Br{\"a}uninger}, {Burkert}, {Elender}, {Freyberg}, {Haberl},
  {Hartner}, {Heuschmann}, {Hippmann}, {Kastelic}, {Kemmer}, {Kettenring},
  {Kink}, {Krause}, {M{\"u}ller}, {Oppitz}, {Pietsch}, {Popp}, {Predehl},
  {Read}, {Stephan}, {St{\"o}tter}, {Tr{\"u}mper}, {Holl}, {Kemmer}, {Soltau},
  {St{\"o}tter}, {Weber}, {Weichert}, {von Zanthier}, {Carathanassis}, {Lutz},
  {Richter}, {Solc}, {B{\"o}ttcher}, {Kuster}, {Staubert}, {Abbey}, {Holland},
  {Turner}, {Balasini}, {Bignami}, {La Palombara}, {Villa}, {Buttler},
  {Gianini}, {Lain{\'e}}, {Lumb}, \& {Dhez}}]{Struder2001}
{Str{\"u}der}, L., {Briel}, U., {Dennerl}, K., {et~al.} 2001, \aap, 365, L18,
  \dodoi{10.1051/0004-6361:20000066}

\bibitem[{{Summers} {et~al.}(2003){Summers}, {Stevens}, {Strickland}, \&
  {Heckman}}]{Summers2003}
{Summers}, L.~K., {Stevens}, I.~R., {Strickland}, D.~K., \& {Heckman}, T.~M.
  2003, \mnras, 342, 690, \dodoi{10.1046/j.1365-8711.2003.06590.x}

\bibitem[{{Trancho} {et~al.}(2007){Trancho}, {Bastian}, {Miller}, \&
  {Schweizer}}]{Trancho2007}
{Trancho}, G., {Bastian}, N., {Miller}, B.~W., \& {Schweizer}, F. 2007, \apj,
  664, 284, \dodoi{10.1086/518886}

\bibitem[{{T{\"u}llmann} {et~al.}(2006{\natexlab{a}}){T{\"u}llmann},
  {Breitschwerdt}, {Rossa}, {Pietsch}, \& {Dettmar}}]{Tull2006II}
{T{\"u}llmann}, R., {Breitschwerdt}, D., {Rossa}, J., {Pietsch}, W., \&
  {Dettmar}, R.~J. 2006{\natexlab{a}}, \aap, 457, 779,
  \dodoi{10.1051/0004-6361:20054743}

\bibitem[{{T{\"u}llmann} {et~al.}(2006{\natexlab{b}}){T{\"u}llmann}, {Pietsch},
  {Rossa}, {Breitschwerdt}, \& {Dettmar}}]{Tull2006I}
{T{\"u}llmann}, R., {Pietsch}, W., {Rossa}, J., {Breitschwerdt}, D., \&
  {Dettmar}, R.~J. 2006{\natexlab{b}}, \aap, 448, 43,
  \dodoi{10.1051/0004-6361:20052936}

\bibitem[{{Turner} {et~al.}(2001){Turner}, {Abbey}, {Arnaud}, {Balasini},
  {Barbera}, {Belsole}, {Bennie}, {Bernard}, {Bignami}, {Boer}, {Briel},
  {Butler}, {Cara}, {Chabaud}, {Cole}, {Collura}, {Conte}, {Cros}, {Denby},
  {Dhez}, {Di Coco}, {Dowson}, {Ferrando}, {Ghizzardi}, {Gianotti}, {Goodall},
  {Gretton}, {Griffiths}, {Hainaut}, {Hochedez}, {Holland}, {Jourdain},
  {Kendziorra}, {Lagostina}, {Laine}, {La Palombara}, {Lortholary}, {Lumb},
  {Marty}, {Molendi}, {Pigot}, {Poindron}, {Pounds}, {Reeves}, {Reppin},
  {Rothenflug}, {Salvetat}, {Sauvageot}, {Schmitt}, {Sembay}, {Short},
  {Spragg}, {Stephen}, {Str{\"u}der}, {Tiengo}, {Trifoglio}, {Tr{\"u}mper},
  {Vercellone}, {Vigroux}, {Villa}, {Ward}, {Whitehead}, \&
  {Zonca}}]{Turner2001}
{Turner}, M.~J.~L., {Abbey}, A., {Arnaud}, M., {et~al.} 2001, \aap, 365, L27,
  \dodoi{10.1051/0004-6361:20000087}

\bibitem[{{Walton} {et~al.}(2013){Walton}, {Fuerst}, {Harrison}, {Stern},
  {Bachetti}, {Barret}, {Bauer}, {Boggs}, {Christensen}, {Craig}, {Fabian},
  {Grefenstette}, {Hailey}, {Madsen}, {Miller}, {Ptak}, {Rana}, {Webb}, \&
  {Zhang}}]{Walton2013}
{Walton}, D.~J., {Fuerst}, F., {Harrison}, F., {et~al.} 2013, \apj, 779, 148,
  \dodoi{10.1088/0004-637X/779/2/148}

\bibitem[{{Walton} {et~al.}(2015){Walton}, {Middleton}, {Rana}, {Miller},
  {Harrison}, {Fabian}, {Bachetti}, {Barret}, {Boggs}, {Christensen}, {Craig},
  {Fuerst}, {Grefenstette}, {Hailey}, {Madsen}, {Stern}, \&
  {Zhang}}]{Walton2015}
{Walton}, D.~J., {Middleton}, M.~J., {Rana}, V., {et~al.} 2015, \apj, 806, 65,
  \dodoi{10.1088/0004-637X/806/1/65}

\bibitem[{{Weaver} {et~al.}(2000){Weaver}, {Heckman}, \& {Dahlem}}]{Weaver2000}
{Weaver}, K.~A., {Heckman}, T.~M., \& {Dahlem}, M. 2000, \apj, 534, 684,
  \dodoi{10.1086/308786}

\bibitem[{{Wik} {et~al.}(2014{\natexlab{a}}){Wik}, {Hornstrup}, {Molendi},
  {Madejski}, {Harrison}, {Zoglauer}, {Grefenstette}, {Gastaldello}, {Madsen},
  {Westergaard}, {Ferreira}, {Kitaguchi}, {Pedersen}, {Boggs}, {Christensen},
  {Craig}, {Hailey}, {Stern}, \& {Zhang}}]{Wik2014back}
{Wik}, D.~R., {Hornstrup}, A., {Molendi}, S., {et~al.} 2014{\natexlab{a}},
  \apj, 792, 48, \dodoi{10.1088/0004-637X/792/1/48}

\bibitem[{{Wik} {et~al.}(2014{\natexlab{b}}){Wik}, {Lehmer}, {Hornschemeier},
  {Yukita}, {Ptak}, {Zezas}, {Antoniou}, {Argo}, {Bechtol}, {Boggs},
  {Christensen}, {Craig}, {Hailey}, {Harrison}, {Krivonos}, {Maccarone},
  {Stern}, {Venters}, \& {Zhang}}]{Wik2014}
{Wik}, D.~R., {Lehmer}, B.~D., {Hornschemeier}, A.~E., {et~al.}
  2014{\natexlab{b}}, \apj, 797, 79, \dodoi{10.1088/0004-637X/797/2/79}

\bibitem[{{Wiktorowicz} {et~al.}(2019){Wiktorowicz}, {Lasota}, {Middleton}, \&
  {Belczynski}}]{Wiktor2019}
{Wiktorowicz}, G., {Lasota}, J.-P., {Middleton}, M., \& {Belczynski}, K. 2019,
  \apj, 875, 53, \dodoi{10.3847/1538-4357/ab0f27}

\bibitem[{{Wiktorowicz} {et~al.}(2017){Wiktorowicz}, {Sobolewska}, {Lasota}, \&
  {Belczynski}}]{Wiktor2017}
{Wiktorowicz}, G., {Sobolewska}, M., {Lasota}, J.-P., \& {Belczynski}, K. 2017,
  \apj, 846, 17, \dodoi{10.3847/1538-4357/aa821d}

\bibitem[{{Winter} {et~al.}(2009){Winter}, {Mushotzky}, {Terashima}, \&
  {Ueda}}]{Winter2009}
{Winter}, L.~M., {Mushotzky}, R.~F., {Terashima}, Y., \& {Ueda}, Y. 2009, \apj,
  701, 1644, \dodoi{10.1088/0004-637X/701/2/1644}

\bibitem[{{Winter} {et~al.}(2008){Winter}, {Mushotzky}, {Tueller}, \&
  {Markwardt}}]{Winter2008}
{Winter}, L.~M., {Mushotzky}, R.~F., {Tueller}, J., \& {Markwardt}, C. 2008,
  \apj, 674, 686, \dodoi{10.1086/525274}

\bibitem[{{Yukita} {et~al.}(2016){Yukita}, {Hornschemeier}, {Lehmer}, {Ptak},
  {Wik}, {Zezas}, {Antoniou}, {Maccarone}, {Replicon}, {Tyler}, {Venters},
  {Argo}, {Bechtol}, {Boggs}, {Christensen}, {Craig}, {Hailey}, {Harrison},
  {Krivonos}, {Kuntz}, {Stern}, \& {Zhang}}]{Yukita2016}
{Yukita}, M., {Hornschemeier}, A.~E., {Lehmer}, B.~D., {et~al.} 2016, \apj,
  824, 107, \dodoi{10.3847/0004-637X/824/2/107}

\bibitem[{{Zaritsky} {et~al.}(1994){Zaritsky}, {Kennicutt}, \&
  {Huchra}}]{Zaritsky1994}
{Zaritsky}, D., {Kennicutt}, Robert~C., J., \& {Huchra}, J.~P. 1994, \apj, 420,
  87, \dodoi{10.1086/173544}

\end{thebibliography}

%%%%%%%%%%%%%%%%%%%%%%%%%%%%%%%%%%%%%%%%%%%%%%%%%%

% Don't change these lines
%\bsp	% typesetting comment
%\label{lastpage}
\end{document}